\documentclass[twocolumn,secnumarabic,amssymb, nobibnotes, aps, prd]{revtex4-2}

\usepackage{amsmath}
\usepackage{amsfonts}
\usepackage{array}
\usepackage{cancel}
\usepackage{caption}
\usepackage{comment}
\usepackage{etoolbox}
\usepackage{graphicx}
\usepackage[a4paper, portrait, margin=2cm]{geometry}
\usepackage{hyperref}
\usepackage{indentfirst} 
\usepackage{lineno,hyperref}
\usepackage{mathrsfs}
\usepackage{multirow}
\usepackage{ragged2e}
\usepackage{rotating}
\usepackage{siunitx}
\usepackage{soul}
\usepackage{subcaption}
\usepackage{tabularx}
\usepackage[normalem]{ulem}
\usepackage{url}
\usepackage{xcolor}

\usepackage[T1]{fontenc}
\usepackage[utf8]{inputenc}

\newcommand{\back}{\!\!\!\!}
\newcommand{\bsym}[1]{\boldsymbol{#1}}
\newcommand{\tc}[2]{\textcolor{#1}{#2}}

\definecolor{orange}{HTML}{FF5000}

\definecolor{changecolor}{HTML}{000000}
\newcommand{\add}[1]{\tc{changecolor}{#1}}
\newcommand{\del}[1]{}
\newtoggle{showdeleted}
\togglefalse{showdeleted}

\begin{document}
	
	\title{The role of porosity in the transition to inertial regime in porous media flows}%
	
	\author{Dawid Strzelczyk}%
	\email[Dawid Strzelczyk: ]{dawid.strzelczyk@ijs.si, dawid.strzelczyk@uwr.edu.pl}
	\affiliation{Institute of Theoretical Physics, Faculty of Physics and Astronomy, University of Wrocław, pl. M. Borna 9, 50-204, Wrocław, Poland}
	\affiliation{Institute Jo\v{z}ef Stefan, Jamova cesta 39, 1000, Ljubljana, Slovenia}
	
	\author{Gregor Kosec}%
	\affiliation{Institute Jo\v{z}ef Stefan, Jamova cesta 39, 1000, Ljubljana, Slovenia}
	
	\author{Maciej Matyka}%
	\affiliation{Institute of Theoretical Physics, Faculty of Physics and Astronomy, University of Wrocław, pl. M. Borna 9, 50-204, Wrocław, Poland}
	\affiliation{{\color{black}Institute Jo\v{z}ef Stefan, Jamova cesta 39, 1000, Ljubljana, Slovenia}}
	
	\date{November 2025}%
	\maketitle
	\tableofcontents
	
	%
	%
	%
	%
	
	\justifying
	
	\noindent \textbf{Abstract} 
	In this work, we investigate the fundamental physical mechanism of the transition from Darcy to inertial (Darcy-Forchheimer) regime in steady-state flows through porous media, with the focus on vortex formation. We investigate their influence on the tortuosity--Reynolds number relation during this transition for systems of various porosities. We do so by numerically solving the Navier-Stokes equations within the pore-scale of simple cubic systems and relating the observations made therein to stochastic systems of more complex geometry. We observe that the tortuosity defined by integrals over the whole fluid volume behaves similarly in both types of systems. At the same time, in simple cubic systems, the tortuosity based on averaging of the length of the streamlines diverges from the volume-integrated one when the inertia onset takes place. We show that the discrepancy between those two tortuosities at increasing Reynolds number carries information about the dynamics of the vortex growth in the system. We stipulate that those dynamics are directly governed by the porosity.
	Our results highlight the utility of various definitions of tortuosity as measures of inertia in porous media flows and explain the reasons for the differences between those definitions. This can lead to a more sensible choice of inertia indicators in more application-oriented problems.
	\vspace{0.25cm}
	
	\noindent \textbf{Keywords:} porous media, inertial flows, Darcy-Forchheimer, tortuosity
	\vspace{1cm}

	\section{Introduction}
	
	
	Many real-life flows through porous media occur at velocities at which the effects caused by fluid inertia cannot be omitted. Such scenarios are of much importance for gas extraction~\cite{Friedel2006,Mohan2009}, mixing and reactive processes~\cite{Yang2024,Lee2025}, or flow through network-like porous media~\cite{Andrade1998}, e.g. the circulatory system~\cite{Mesri2015,Bluestein2011}. It is thus crucial from the scientific and application point of view to well understand the physical principles of the transition of flow through porous media from the creeping flow (Darcy) regime to the inertial regime and the behavior of the flow within the latter. Of special interest are the relations between the pore-scale hydrodynamics and the effective (volume-averaged) hydrodynamic coefficients of a porous medium, which serve as a link connecting the fundamental fluid dynamical processes to the more practical quantities characterizing those systems.
	
	In the Darcy regime, the Stokes equations are used to model the mass and momentum transport within the pore space. In order to obtain the average picture of the transport in this regime, upscaling~\cite{Neuman1977,Whitaker1986} is performed which results in the Darcy's law~\cite{Darcy1856}
	\begin{equation}\label{eq:darcy}
		\bsym{U} = \frac{k_0}{\nu}\left( -\langle \nabla P \rangle + \rho\bsym{g} \right)
	\end{equation}
	stating the proportionality between the forcing term (mean pressure gradient $-\langle \nabla P\ \rangle$ and/or body force $\rho\bsym{g}$) and the superficial flow velocity $\bsym{U}$. The constant $k_0$ is the (intrinsic) permeability characteristic of an isotropic porous system as $-\langle \nabla P \rangle + \rho\bsym{g}$ term vanishes, and $\nu$ is the fluid kinematic viscosity. In the inertial regime, the contribution of the convective derivative in the pore-scale momentum transport becomes non-negligible. Thus, the full Navier-Stokes equations need to be considered, yielding the Darcy-Forchheimer law~\cite{Forchheimer1901,Whitaker1996}
	\begin{equation}\label{eq:darcy_forchhiemer}
		\bsym{U}  +
		\beta \rho \bsym{U} |\bsym{U}| + 
		\add{\gamma\bsym{U} |\bsym{U}|^2}
		=
		\frac{k_0}{\nu}\left( -\langle \nabla P \rangle + \rho\bsym{g} \right)
	\end{equation}
	where $|\bsym{U}|$ is the magnitude of the superficial velocity, $\beta$ \add{and $\gamma$ are} \del{is} the free coefficients and the porous medium is again assumed to be isotropic. The transition from the Darcy to the inertial regime is often associated with the significant reorganization of the velocity and pressure fields within the pore spaces. Among many aspects of such reorganization, it was observed that the fluid kinetic energy becomes more evenly distributed in the pore space~\cite{Andrade1999}, which has to do with significant mass flow through an increasing number of channels and confinement of more kinetic energy within the vortices~\cite{Sniezek2024,Agnaou2017,Naqvi2025}. Our recent studies on high-~\cite{Sniezek2024} and low-porosity~\cite{Naqvi2025} porous media suggest that the porosity is one of the factors determining the dynamics of the transition from Darcy to inertial regime. Nevertheless, the exact mechanism of how the geometry of the pores impacts the onset of inertia is still unknown.
	
	For the quantification of the inertial effects in porous media, a variety of measures are used. Fitting Eqs.~\eqref{eq:darcy} and  \eqref{eq:darcy_forchhiemer} to the experimental data allows one to assess the permeability and the inertial correction coefficient and to distinguish between various inertial subregimes~\cite{Agnaou2017,Lasseux2011,Andrade1999}. The measurements of the hydrodynamic drag force exerted on the solid fraction of the porous medium can also be used as the inertia onset indicator~\cite{Koch1997,Hill2001}. Another quantity, commonly used in porous media studies, is the tortuosity $T$, in certain cases related to the permeability through the Carman-Kozeny relation~\cite{Carman1997,Srisutthiyakorn2017} as $k_0 \propto T^{-1}$. Although originally introduced as an effective parameter describing solely the porous medium geometry, the notion of tortuosity was later extended to encompass factors other than geometry~\cite{Sivanesapillai2014}, and even to non-hydrodynamic phenomena~\cite{Sykova2008}. Due to this, in its nearly 100-year history, several definitions of the tortuosity have been proposed, and the discussion on their range of applicability and relation to each other is still ongoing~\cite{Koponen1996,Matyka2008,Zhang1995}. The majority of the hydrodynamic (hydraulic) tortuosity definitions define it as an average of the lengths of the streamlines obtained from the pore-scale velocity field. However, it was proven that one can calculate the tortuosity directly from the pore-scale velocity field, without the need to generate streamlines beforehand~\cite{Duda2011}. It becomes especially useful in low-porosity systems where even at low superficial velocities, recirculation zones are abundant and the automatic seeding of the streamlines poses a big challenge. However, this equivalence between the two approaches to calculate the tortuosity no longer holds in the inertial regime, where the impact of the recirculation zones on the pore-scale flow is much larger than in the Darcy regime~\cite{Duda2011}. This introduces even more ambiguity into the description and explanation of the onset of inertia in porous media.
	
	The aim of this work is to study the behavior of tortuosity during the transition from the Darcy to the inertial regime. We do so to explain how the interplay between the pores' geometry and the pore-scale velocity drives this transition in porous media flows. To describe the transition quantitatively, we use two definitions of the tortuosity -- one formulated using volume integrals of the velocity field, and the other, using the velocity streamlines. We explain why their dependence on Reynolds number changes with the porosity of the porous sample. We show that those differences provide information about the interaction between the geometry of the pores and the hydrodynamics in the pore-scale. In particular, we highlight that the main factor affecting the tortuosity during the Darcy-inertial transition is the growth of the recirculation zones in size and kinetic energy.

	\section{Methods}\label{sec:methods}
	
	\subsection{Physical and numerical setup}\label{ssec:setup}
	
	\subsubsection{Governing equations and boundary conditions}
	
	We solve the steady-state mass conservation (Eq.~\eqref{eq:mass}) and Navier-Stokes (Eq.~\eqref{eq:momentum}) equations for an incompressible fluid in the pore-scale of three-dimensional porous samples. We consider cubic samples of side length $L$. The fluid fully saturates the pore space $\Omega$. On the solid walls, $\partial\Omega_\text{solid}$, we impose Dirichlet boundary conditions for the velocity $\bsym{V}$ (Eq.~\eqref{eq:noslip}) and Neumann boundary conditions for the pressure $P$ (Eq.~\eqref{eq:pressure}). We also impose periodic boundary conditions along all three Cartesian coordinates (Eqs.~\eqref{eq:periodic_V} and \eqref{eq:periodic_P})
	\renewcommand{\arraystretch}{1.5}
	\begin{subequations}\label{eq:nse}
		\begin{align}
			\nabla \cdot \bsym{V} &= 0 & \quad\text{in }\Omega,\label{eq:mass}\\
			\bsym{V} \> (\nabla \cdot \bsym{V}) &= -\nabla \dfrac{P}{\rho} + \bsym{g} + \nu\nabla^2\bsym{V} & \text{in }\Omega,\label{eq:momentum}\\
			\bsym{V} & = \bsym{0} &  \text{on }\partial \Omega_\text{solid},\label{eq:noslip}\\
			\bsym{n} \cdot \nabla P &=0 &  \text{on }\partial \Omega_\text{solid},\label{eq:pressure}\\
			\bsym{V}(\bsym{x}) &= \bsym{V}(\bsym{x} + \{0,\pm L\}^3) & \text{in }\Omega,\label{eq:periodic_V}\\
			P(\bsym{x}) &= P(\bsym{x} + \{0,\pm L\}^3) & \text{in }\Omega.\label{eq:periodic_P}
		\end{align}
	\end{subequations}
	The acceleration is aligned with the $x$-axis, $\bsym{g} = [g,0,0]$, and $\bsym{n}$ is the local unit normal vector on the solid boundary. We adopt the boldcase notation for the vectors and refer to the Cartesian coordinates with indices $0,1,2$, e.g. $\bsym{V} \equiv [V_0,V_1,V_2]$. We denote the magnitude of a vector with a non-boldcase font, e.g. $|\bsym{V}| \equiv V$. \add{In the actual simulations, we set $\rho=1$ and $\nu=1$. We note that all the quantities in Eqs.~\eqref{eq:nse} are given in non-dimensionalized form and such will be presented in the further part of this work. To provide a better perspective on the real-life values of the solution fields in the studied systems, we provide exemplary conversion factors to the dimensionalized forms of the pressure and velocity in Appendix~\ref{subapp:dimensions}.}
	
\begin{figure}[!h]
		\centering
		
		\includegraphics[width=.7\linewidth]{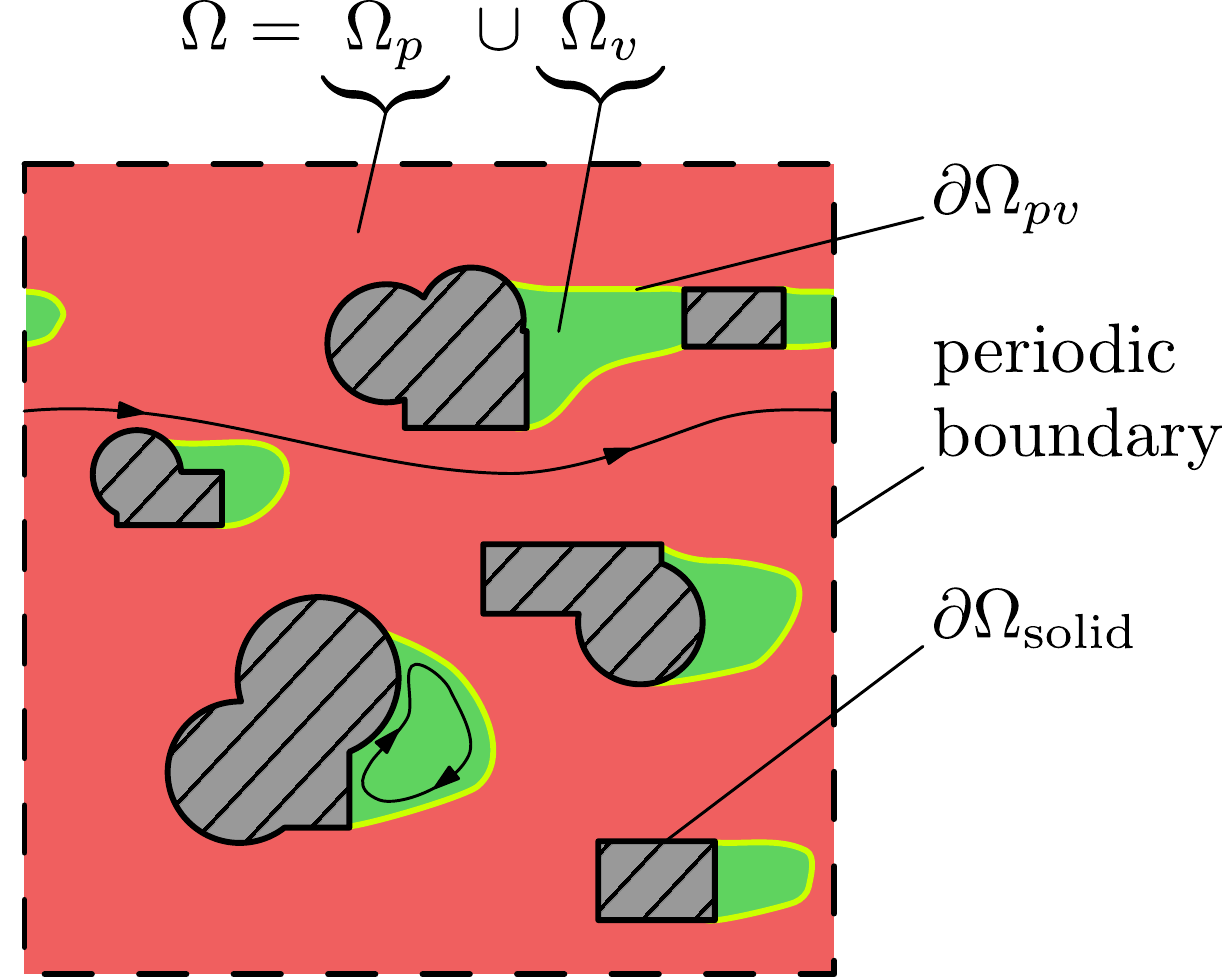}
		
		\caption{A schematic representation of the percolating volume ($\Omega_p$, \textit{red}), recirculation volume ($\Omega_v$, \textit{green}) and the boundary between the two ($\partial\Omega_{pv}$, \textit{yellow}) in a flow through a periodic porous sample (obstacles depicted striped gray with their boundary $\partial \Omega_\text{solid}$ plotted with solid black line). The curves with arrows along their length represent the percolating and the recirculating streamlines. The dashed edges of the large square represent the periodic boundary}
		\label{fig:volumes_defined_sketch}
	\end{figure}

	It is useful to define at this point the notion of the percolating volume $\Omega_p \subset \Omega$. It is the volume consisting of all the streamlines that connect any pair of the periodic boundaries of the fluid domain and are continuous within the porous sample (further called the \textit{percolating streamlines}). The volume consisting of all fluid points which are not in $\Omega_p$ will be called the recirculation (or vortex) volume $\Omega_v$, i.e. $\Omega_v = \Omega \> \backslash \> \Omega_p$. Such a definition of $\Omega_v$ is consistent with the definition of a steady-state vortex as the \textit{maximum set of nested closed streamlines} used in previous works~\cite{Lugt1979,Lagerstrom1975}. With such division of $\Omega$, the boundary between the percolating and the recirculation volume, $\partial\Omega_{pv}$ can be clearly defined. A schematic representation of the volumes and the boundaries considered in this work is shown in Fig.~\ref{fig:volumes_defined_sketch}.
	
	\subsubsection{Stochastic porous medium}\label{sssec:setup_stochastic_systems}
	
\begin{figure}[!ht]
		\centering
		\includegraphics[width=.45\linewidth]{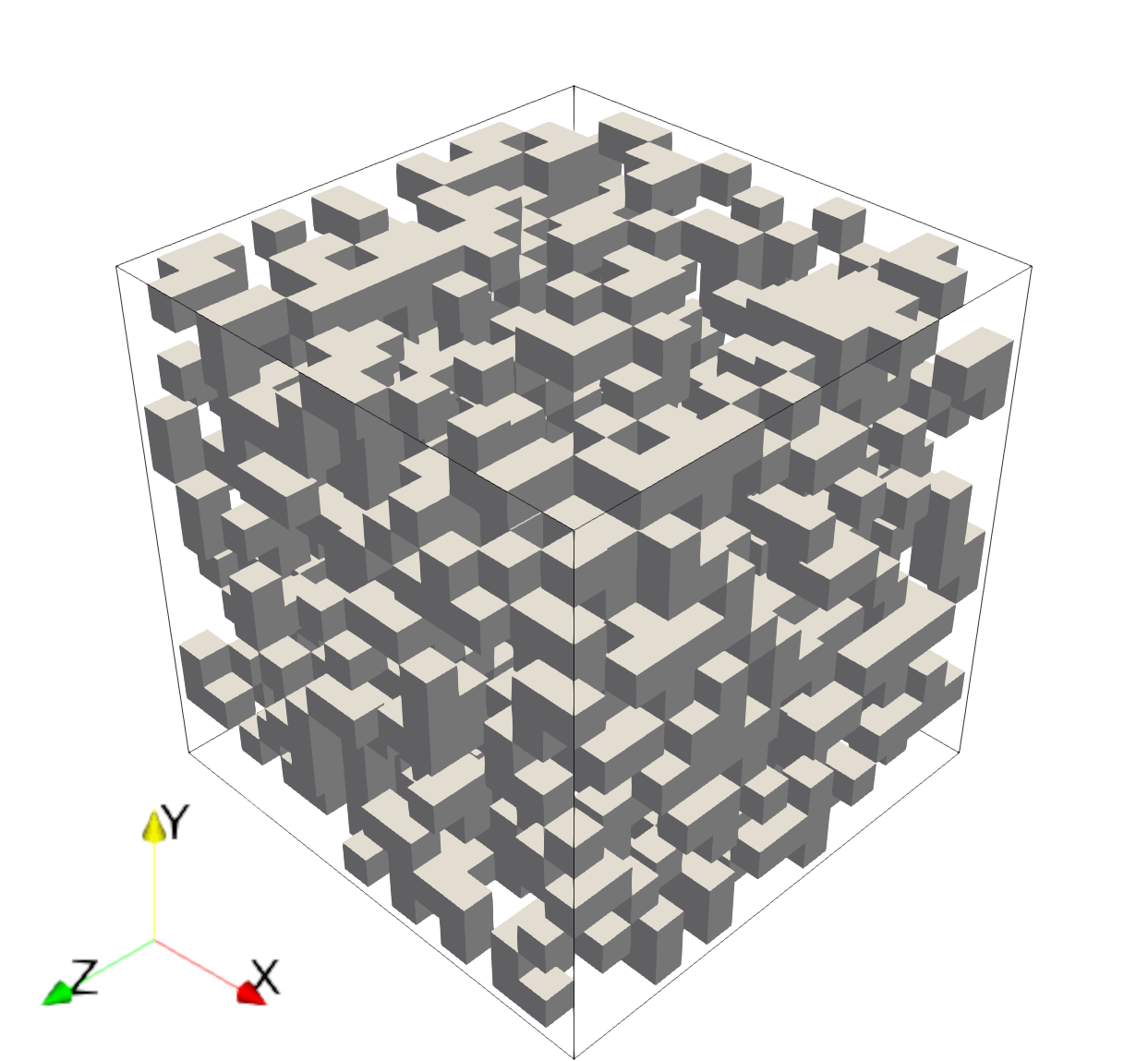}
		\includegraphics[width=.45\linewidth]{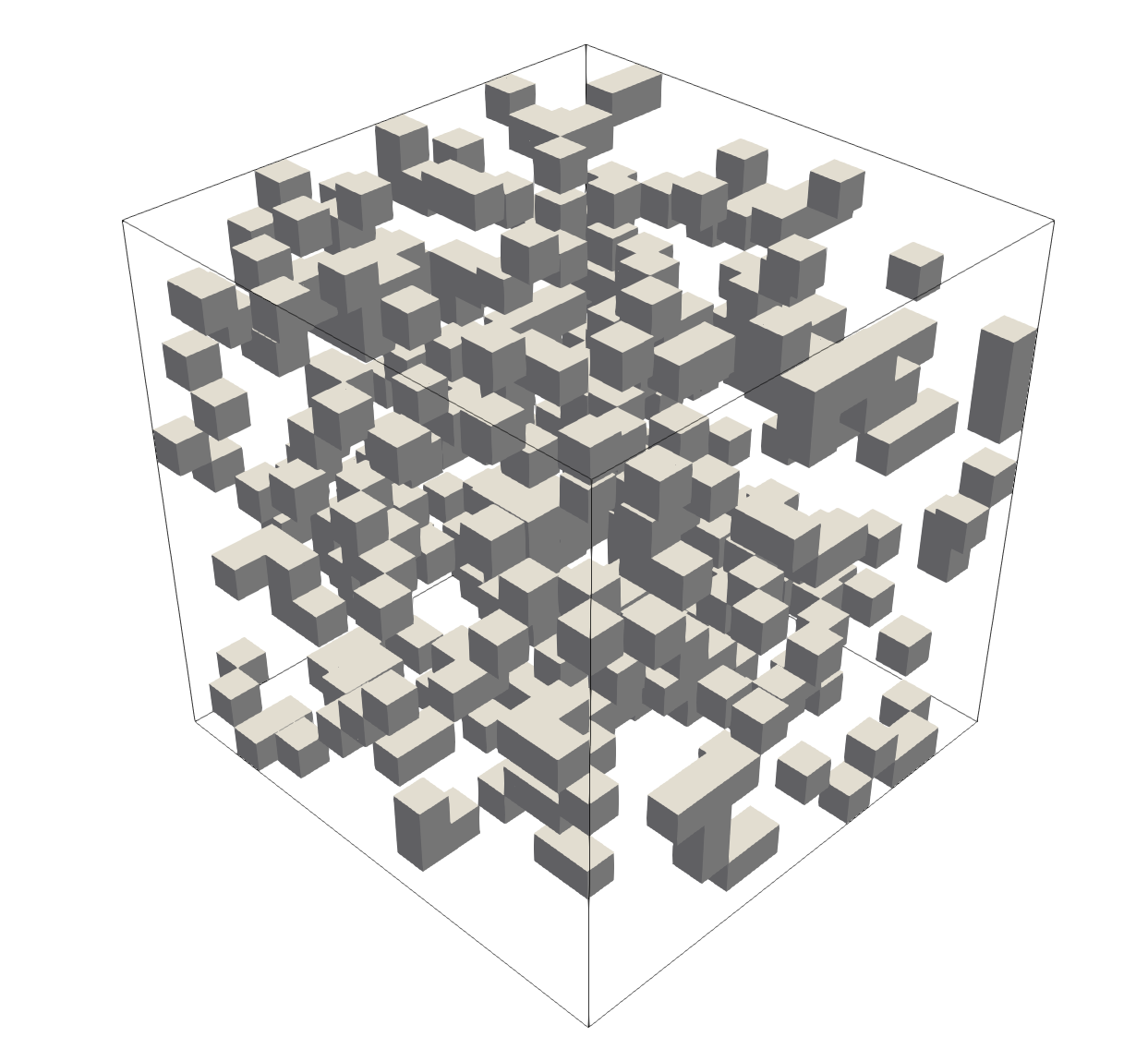}
		\caption{Visualization of the geometries of the solid phase of the two stochastic porous samples considered in this work. The target porosities are $\phi=0.7$ (\textit{left}) and $\phi=0.9$ (\textit{right}). The black-edged cube denotes the boundaries of the periodic cell.}
		\label{fig:stochastic_PM_geometry}
	\end{figure}

	The two stochastic porous samples considered in this work are confined within a cubic periodic cell of side length $L=1$. We generate the geometry of the solid phase by dividing the periodic cell into $16^3$ cubes and subsequently marking each of them as fluid with probability equal to the desired porosity $\phi$ of the sample. We consider values $\phi=0.7$ and $\phi=0.9$ and obtain the actual porosities of $0.696$ and $0.897$, respectively. The visualizations of the solid phase of the two samples are shown in Fig.~\ref{fig:stochastic_PM_geometry}.
	
	To solve the governing equations in the stochastic samples, we use Lattice Boltzmann Method (LBM)~\cite{Succi2018,Kruger2017} implemented for the GPU execution using Template Numerical Library~\cite{Klinkovsky2022,Oberhuber2021}. Due to the uniform discretization of space, LBM allows for simple treatment of complex boundaries and has been successfully used to model flows in porous media~\cite{Pan2006,Matyka2011,Foroughi2025}, We use D3Q19 velocity set for $\phi=0.7$ sample and D3Q27 set for $\phi=0.9$ sample, for better stability at higher velocities. Two-relaxation time (TRT) collision kernel~\cite{Ginzburg2018} with $\Lambda=1/4$ and the symmetric relaxation time $\tau^+=0.505$ is used, along with the lattice speed of sound $c_s=1/\sqrt{3}$. The discretization in space consists of $N=304$ lattice sites along each direction for $\phi=0.7$ sample and $N=160$ lattice sites along each direction for $\phi=0.9$ sample. The no-slip walls are implemented via the halfway-bounceback method~\cite{Succi2018,Kruger2017}. In the simulations with the highest Reynolds number, in the final timestep, the maximal value of any velocity component for either sample did not exceed $3.73\cdot 10^{-2}$ and the density variation $|1-\rho|$ did not exceed $1.73 \cdot 10^{-3}$, all in lattice units. \add{We provide a more detailed description of the used LBM model in Appendix~\ref{subapp:stochastic}.}
	
	The steady-state is assumed to be reached when the relative difference between the mass fluxes along $x_0$-direction per unit time measured \add{every $10^4$ timesteps} at $x_0=0.25$ defined as
	\begin{equation}\label{eq:u_mean_diff_lbm}
		\Delta Q_0 = \frac{|Q_0^t - Q_0^{t-10^4\Delta t}|}{Q_0^t} \cdot \frac{1}{10^4 \Delta t}
	\end{equation}
	\del{and measured every $10^4$ timesteps} falls below $10^{-2}$. In the above equation, $Q_0^t$ denotes the mass flux along $x_0$-direction at time $t$.
	
	\subsubsection{Simple cubic porous medium}\label{sssec:setup_sc_systems}
	
\begin{figure}[!ht]
		\centering
		\includegraphics[width=.45\linewidth]{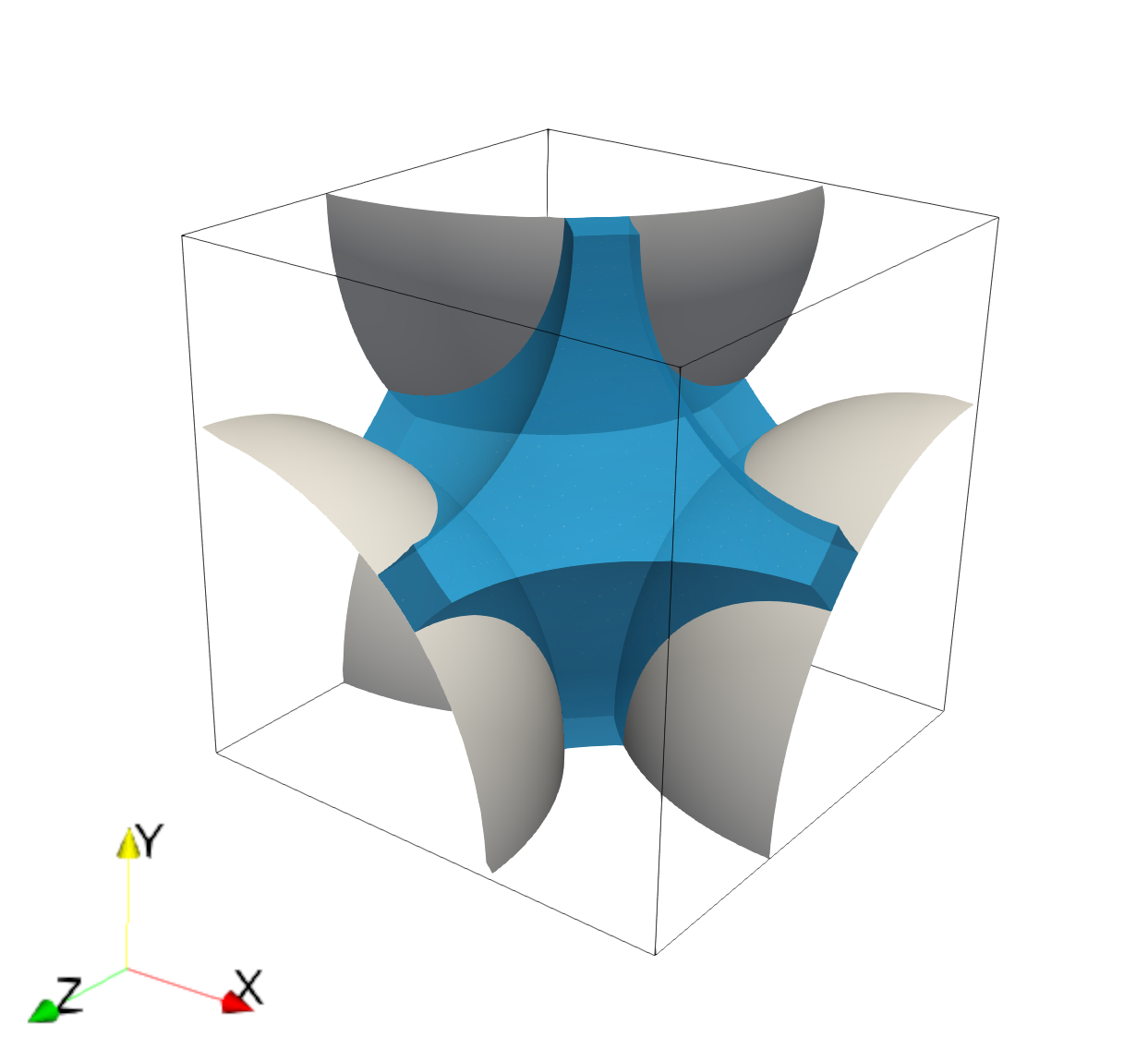}
		\includegraphics[width=.45\linewidth]{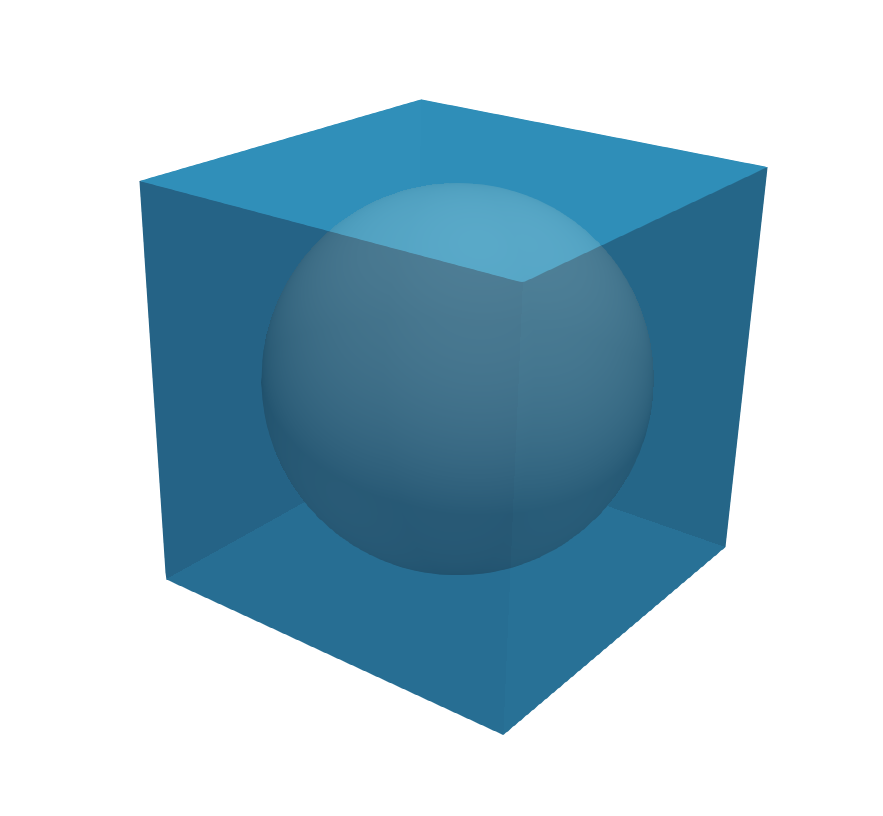}
		\includegraphics[width=.45\linewidth]{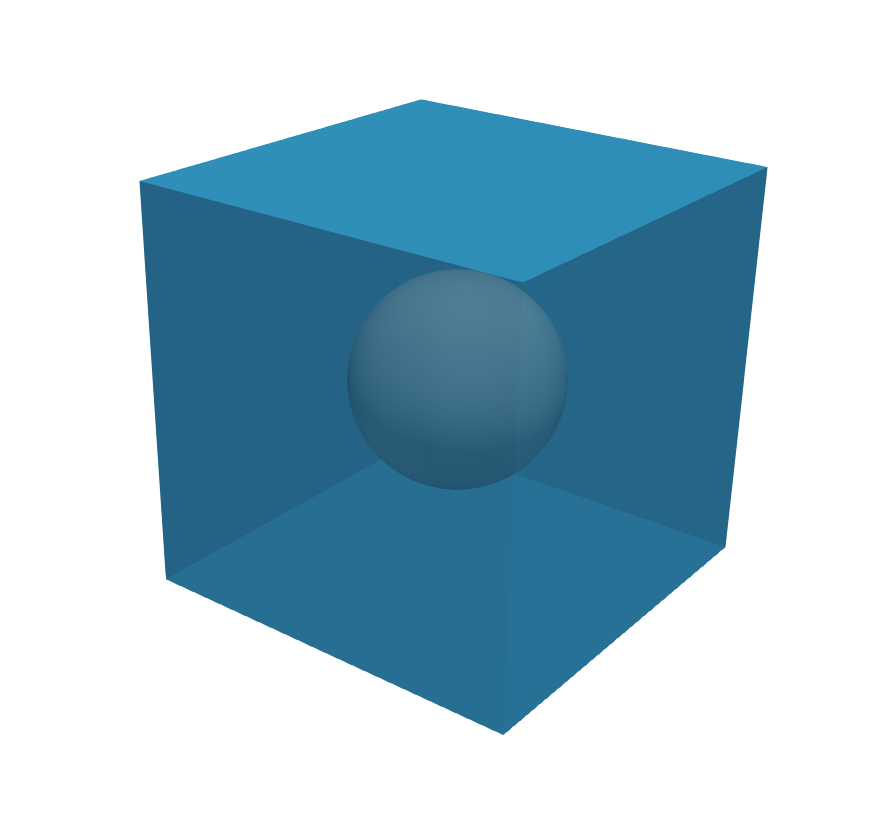}
		\includegraphics[width=.45\linewidth]{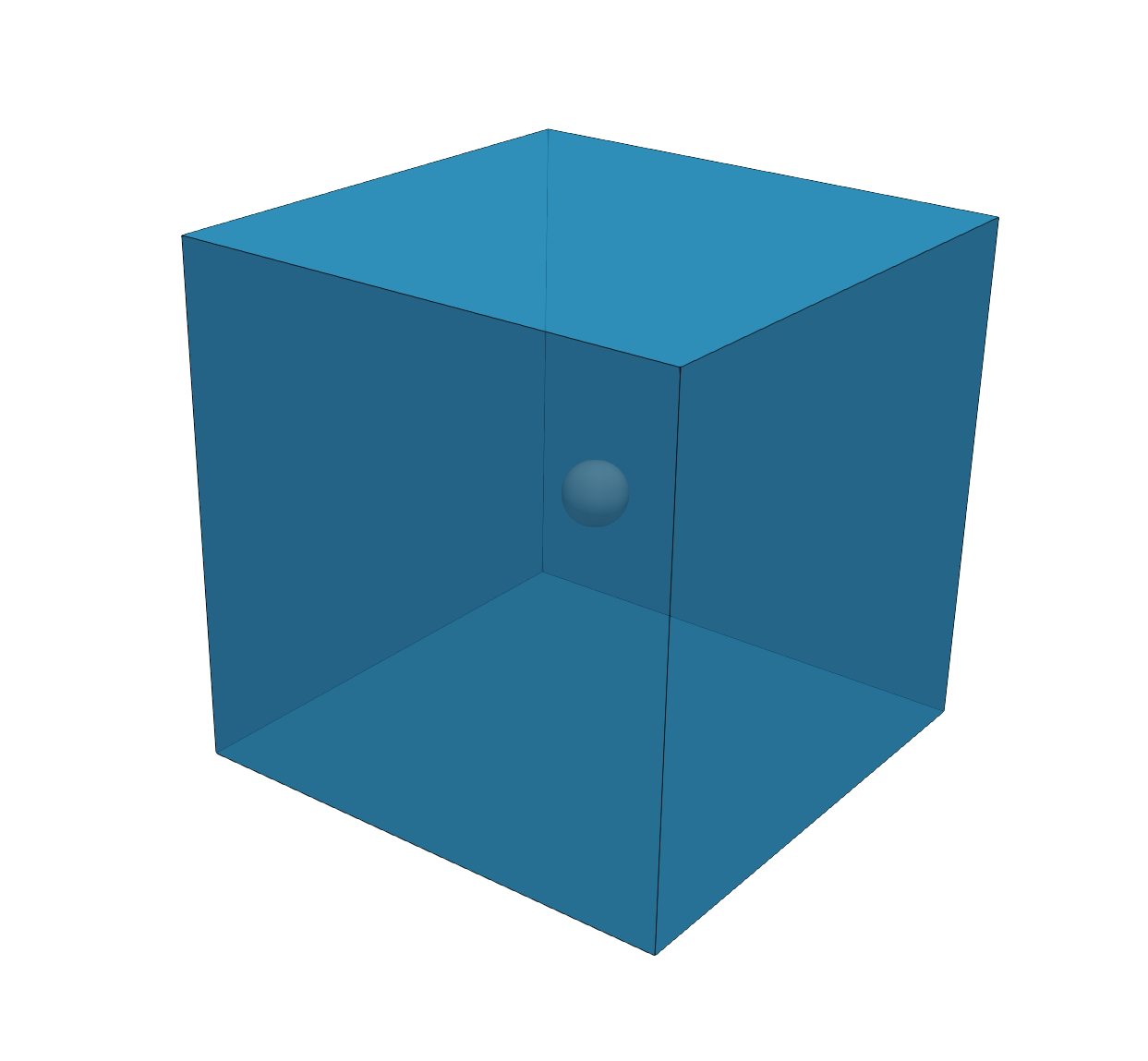}
		\caption{Visualization of the geometries of the solid (gray) and fluid (blue, opaque) phases of the two simple cubic porous samples considered in this work. The porosities are \del{$\phi=0.1$ (\textit{left}) and $\phi=0.999$ (\textit{right})} \add{$\phi=0.1, \> 0.59, \> 0.93, \> 0.999$, counting from the top-left to the bottom-right}. The black-edged cube denotes the boundaries of the periodic cell. In the case of the \del{low} \add{lowest}-porosity sample, only a few chosen obstacles residing in the periodic cell are shown for clarity.}
		\label{fig:SC_PM_geometry}
	\end{figure}

\begin{figure}[!ht]
		\centering
		\includegraphics[width=.45\linewidth]{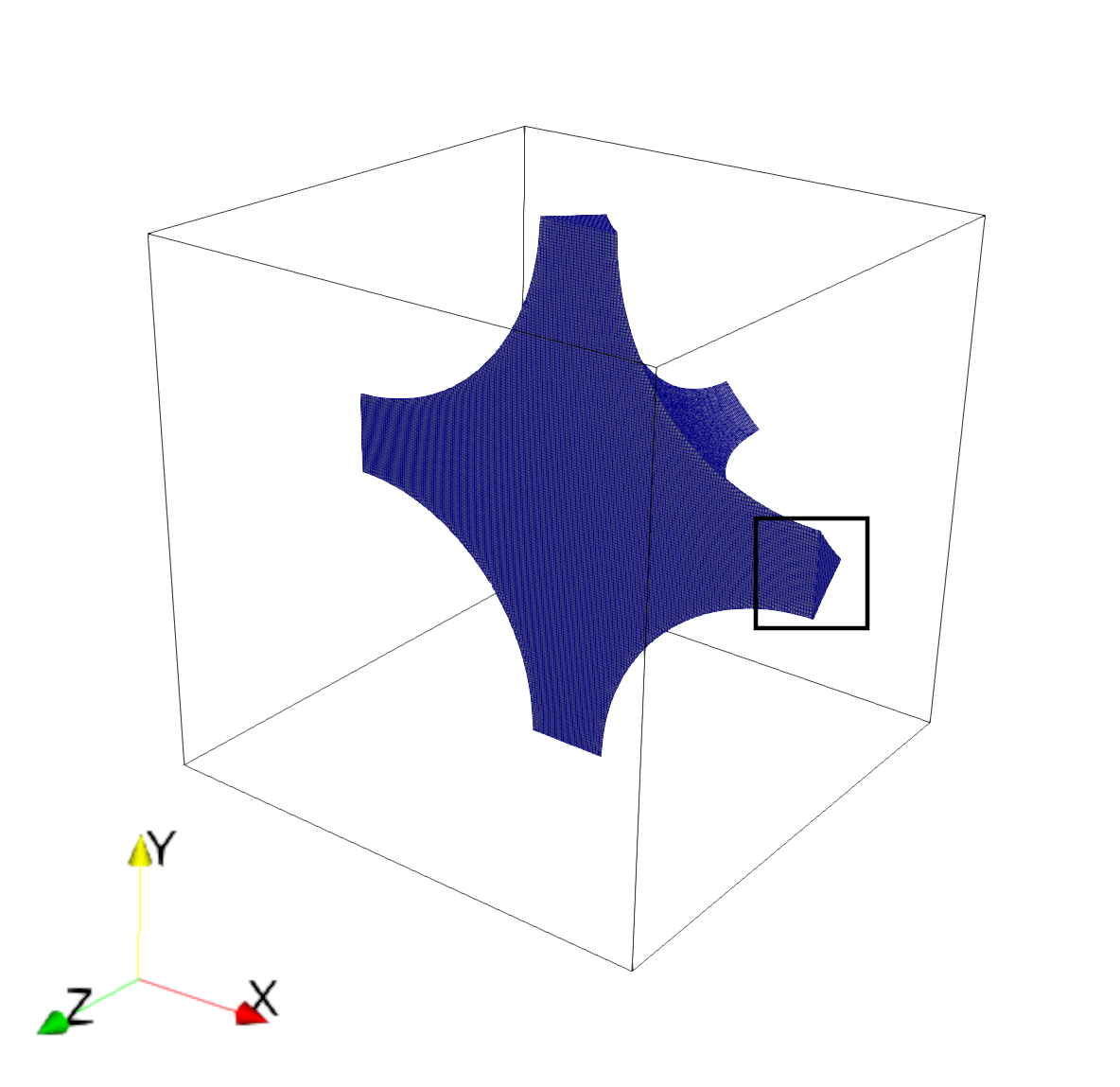}
		\includegraphics[width=.45\linewidth]{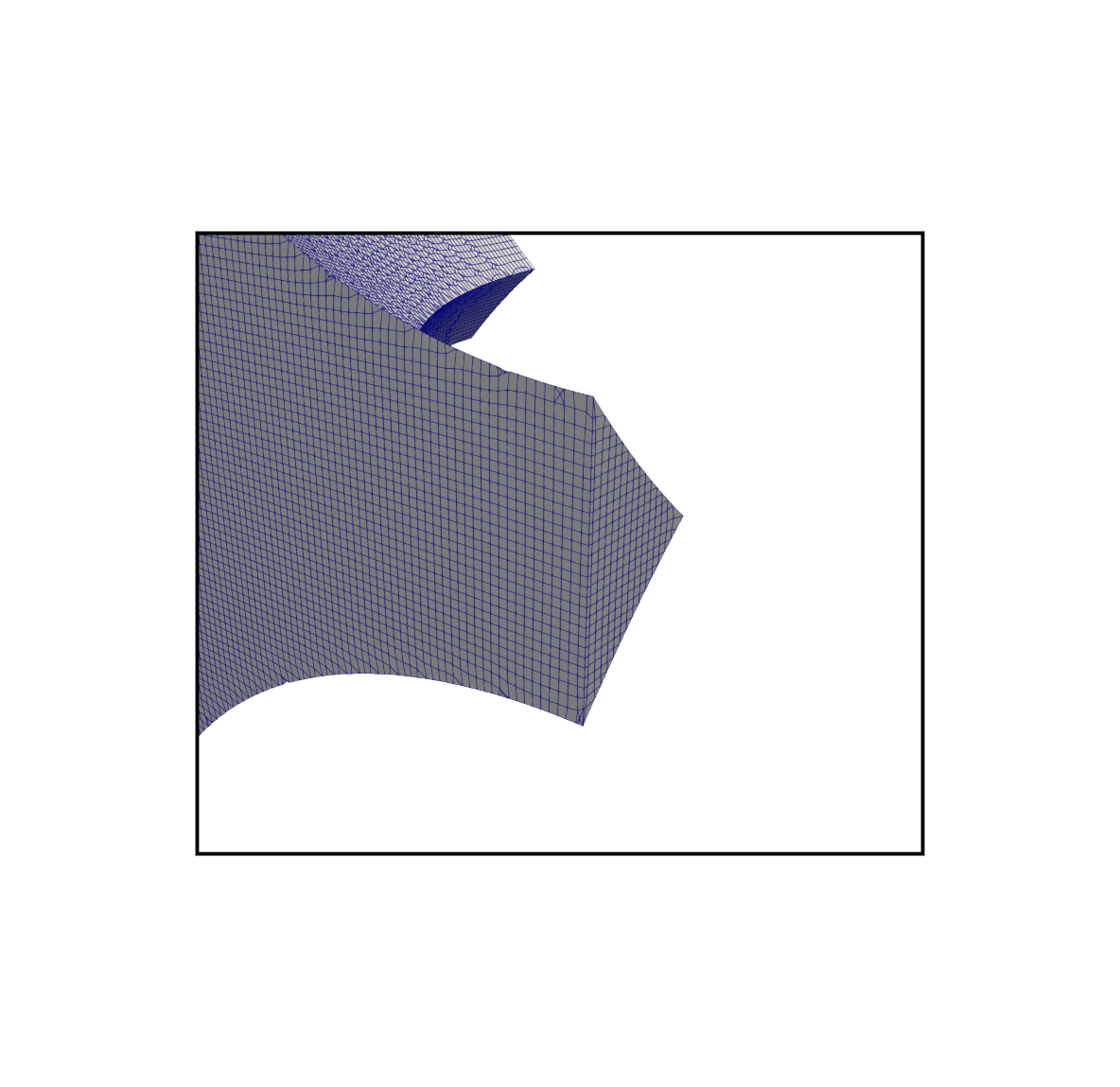}
		\\
		\includegraphics[width=.45\linewidth]{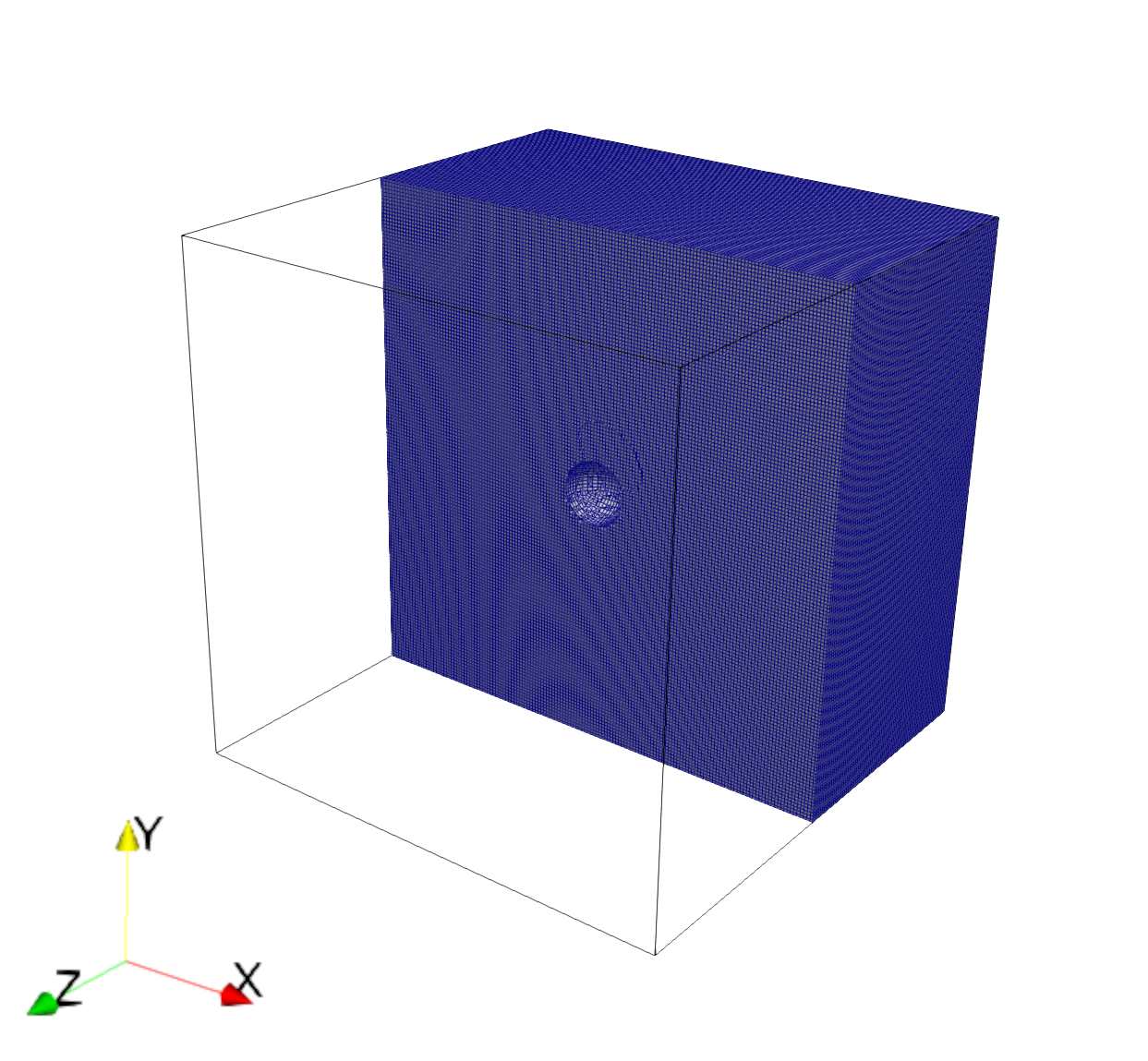}
		\includegraphics[width=0.45\linewidth]{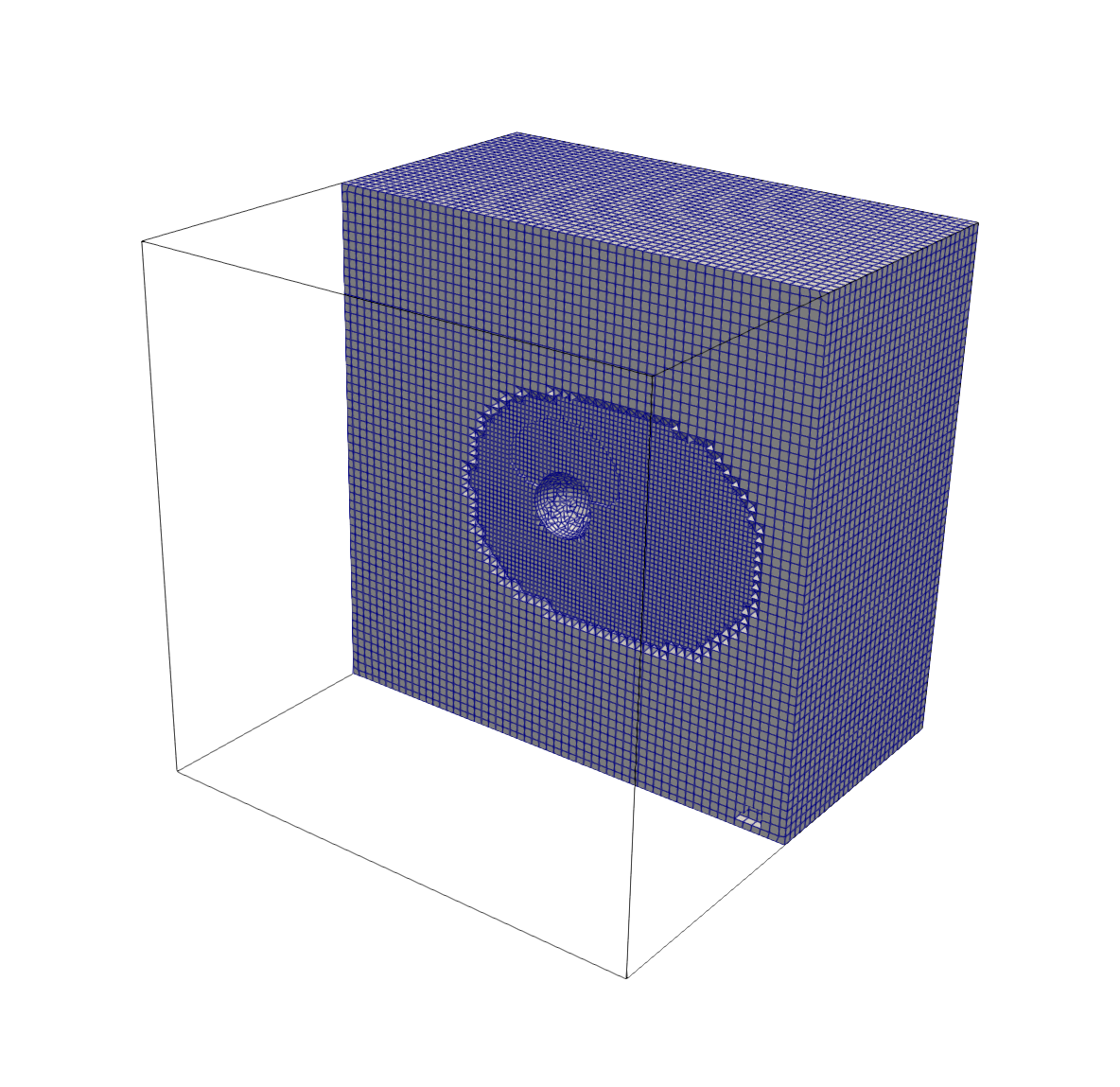}
		\caption{Visualization of the meshes of \del{the} two \add{chosen} simple cubic porous samples considered in this work. The porosities are $\phi=0.1$ (\textit{two top subplots}) and $\phi=0.999$ (\textit{two bottom subplots}, for the uniform and refined discretization). The {\itshape top-right} subplot shows the magnified view of the {\itshape top-left} mesh. The black-edged cube denotes the boundaries of the periodic cell. Only $z<0.5$ half of each mesh is shown.}
		\label{fig:sc_mesh}
	\end{figure}

	The \del{two} \add{four} simple cubic (SC) porous samples considered in this work are confined within a cubic periodic cell of side length $L=1$. We consider porosities \del{$\phi=0.1$ and $\phi=0.999$} \add{$\phi=0.1, \> 0.59, \> 0.93, \> 0.999$} corresponding to the obstacle radii \del{$r=0.6526$ and $r=0.062$} \add{$r=0.6526, \> 0.46, \> 0.26, \> 0.062$}, respectively. For the convenience of visualization and meshing, the computational geometry for the \del{lower} \add{lowest}-porosity sample considers the spherical obstacles placed in each of the periodic cell's corners, and for the \del{higher-porosity sample} \add{other samples}, we place a single obstacle in the center of the domain (see Fig.~\ref{fig:SC_PM_geometry}). We solve the governing equations using SIMPLE (Semi-Implicit Method for Pressure-Linked Equations) algorithm~\cite{patankar1980numerical} implemented in \texttt{simpleFOAM} solver of openFOAM~\cite{jasak2007openfoam} and use regular discretization with the mesh parameter $h=1/195$ for $\phi=0.1$ (see Fig.~\ref{fig:sc_mesh}, top row) and $h=1/150$ for \del{$\phi=0.999$} \add{the remaining samples} (see Fig.~\ref{fig:sc_mesh}, bottom-left \add{for the highest-porosity sample}). In contrast to LBM, SIMPLE allows to directly solve for the steady-state, reducing the simulation time. We use \texttt{pimpleFOAM} solver to confirm the steady state in several simulations of the highest Reynolds number. In the case of the \del{higher} \add{highest}-porosity sample, we do so on variable-density meshes with the target cell size ranging from $h_\text{bulk}=1/75$ to $h_\text{min}=1/150$ (see Fig.~\ref{fig:sc_mesh}, bottom-right). We include a more detailed description of the numerical setup for SC samples in Appendix~\ref{app:numerical_details}.\ref{subapp:sc}. We provide the analysis of the solution convergence with the refined mesh parameter ($h$ or \add{$h_\text{bulk}$ and} $h_\text{min}$) for SC systems in Appendix~\ref{app:convergence}. For all transient simulations, we choose the timestep $\Delta t$ such that the maximal Courant number in the system is below unity.
	
	
	\subsection{The used definitions of the tortuosity}\label{ssec:tortuosity}
	
\begin{figure}[!ht]
		\centering
		\includegraphics[width=.7\linewidth]{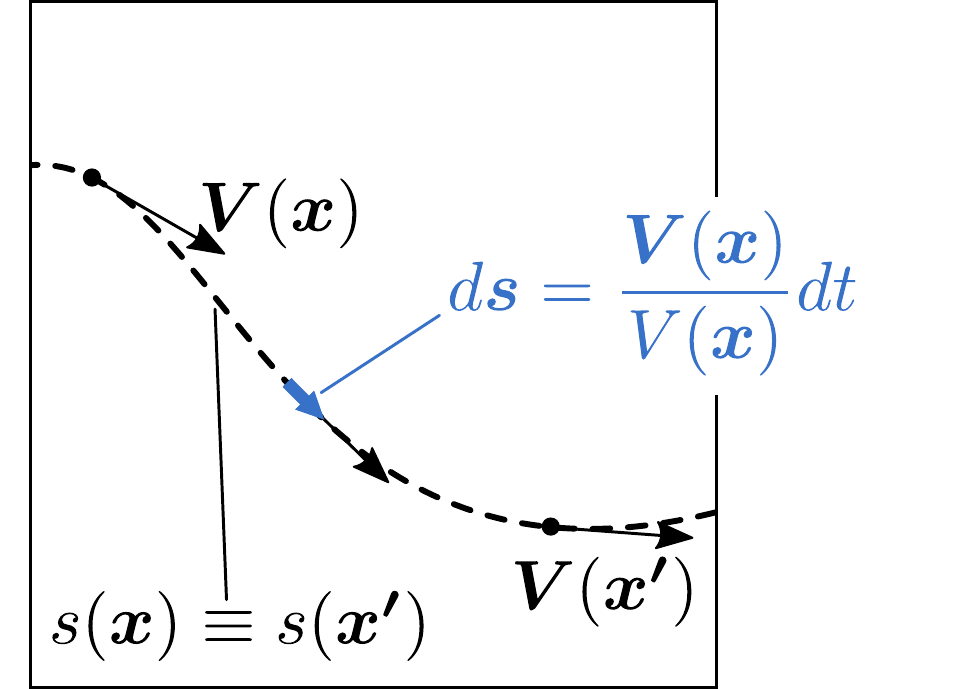}
		\caption{Schematic representation of a streamline $s$ ({\itshape dashed line}) as a curve tangent to the fluid velocity field vectors (black arrows) in each of the streamline's points (two chosen are shown here as black dots, $\bsym{x}$ and $\bsym{x'}$). An infinitesimal increment along a streamline, $d\bsym{s}$ (small blue arrow), is parallel to the local velocity vector, here normalized for convenience. Any of the streamline's points identifies it uniquely ($s(\bsym{x}) \equiv s(\bsym{x'})$).}
		\label{fig:streamline_definition}
	\end{figure}

	To quantify the inertial effects in the flow, let us introduce three definitions of tortuosity, two of which will be used further in this work. They mainly make use of the \textit{streamlines}, defined as curves $s$ tangent to the velocity field vectors at each of their points. Equivalently, for the steady state systems, a streamline can be defined as a path along which a massless particle is advected in the velocity field (see Fig.~\ref{fig:streamline_definition}). Since the velocity field has only one value in each fluid point $\bsym{x} \in \Omega$, the streamlines cannot cross each other, and thus, there exists a 1-to-1 mapping between the fluid points and the streamlines. In other words, a streamline can be identified uniquely by any of its points: $s \equiv s(\bsym{x})$ for any $\bsym{x} \in s(\bsym{x})$. We will use this fact when introducing the definitions of tortuosities in the following part of this section.
	
\begin{figure}[!ht]
		\raggedright
		\includegraphics[width=.45\linewidth]{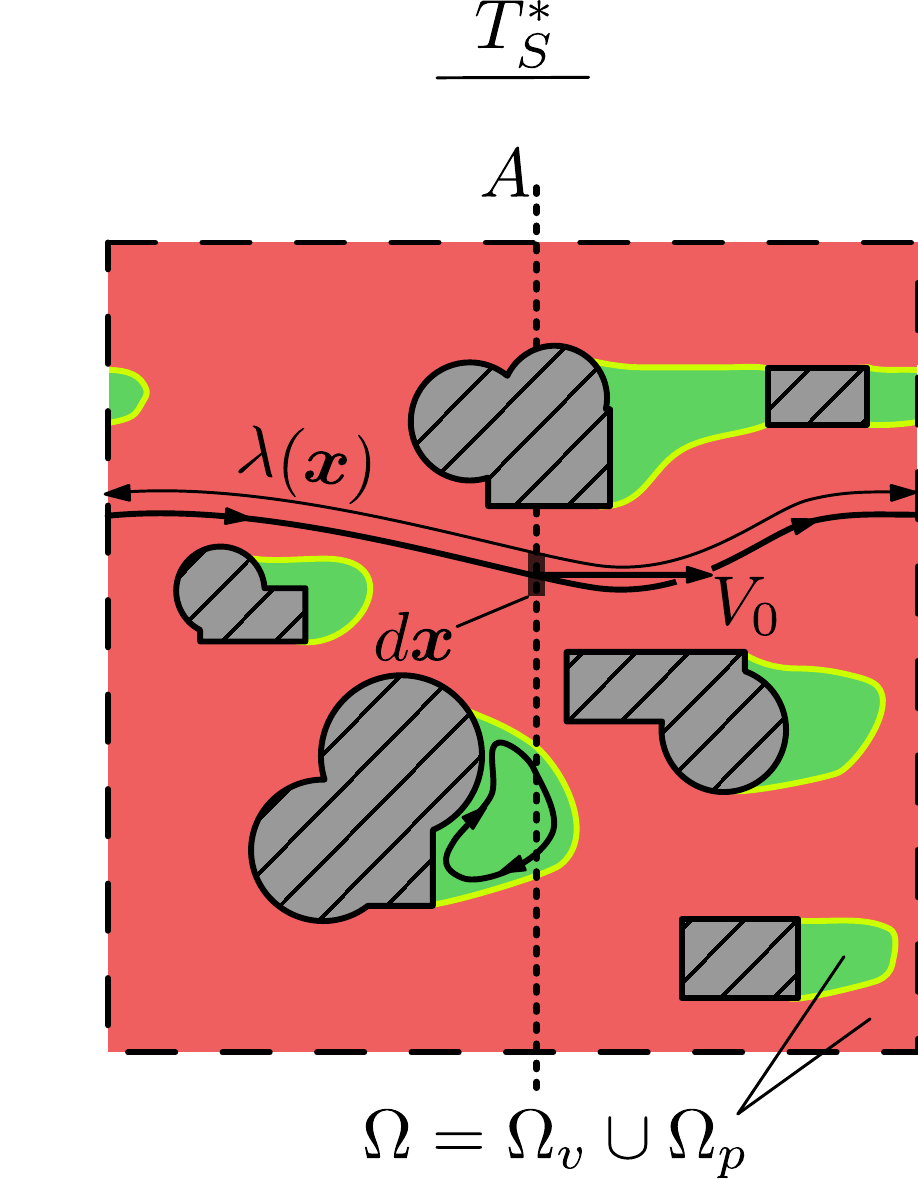}
		\includegraphics[width=.45\linewidth]{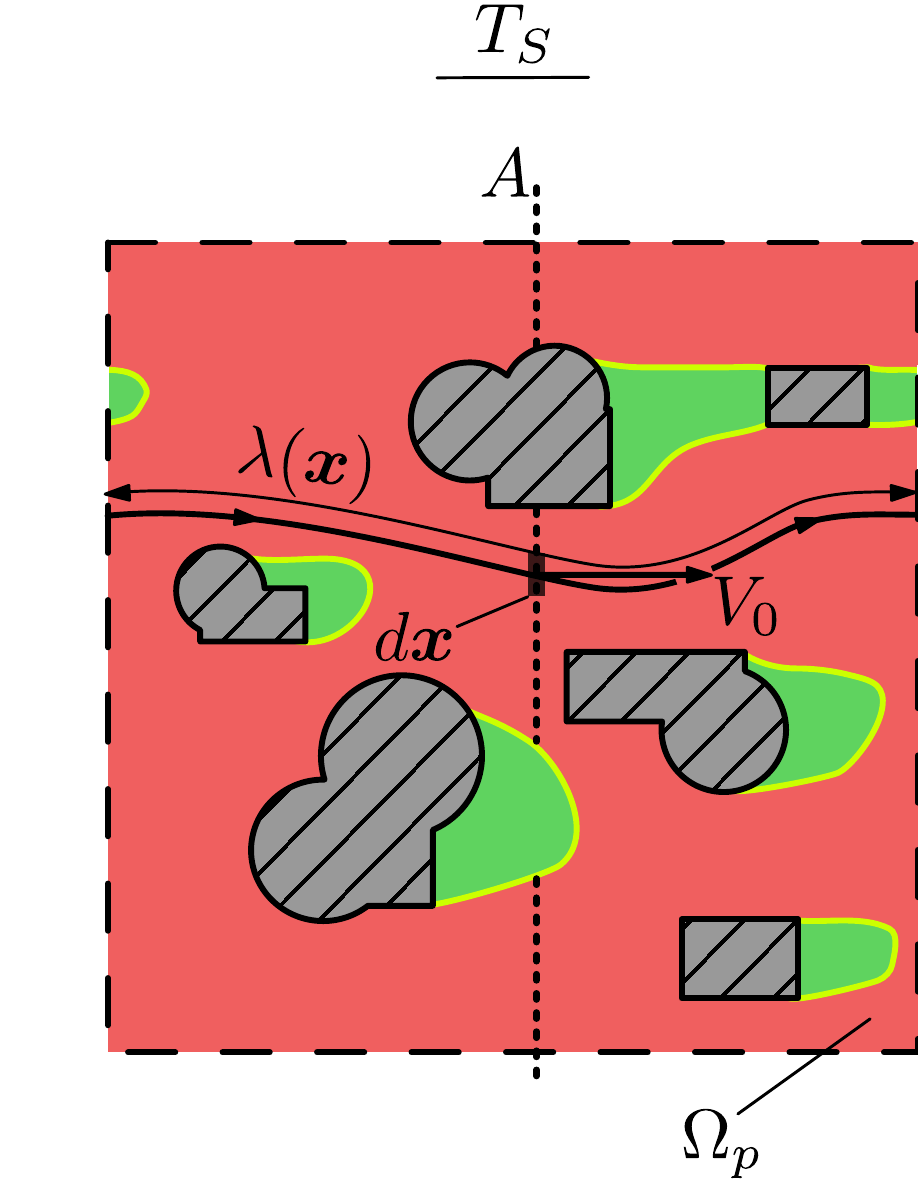}
		
		\vspace{.5cm}
		
		\includegraphics[width=.45\linewidth]{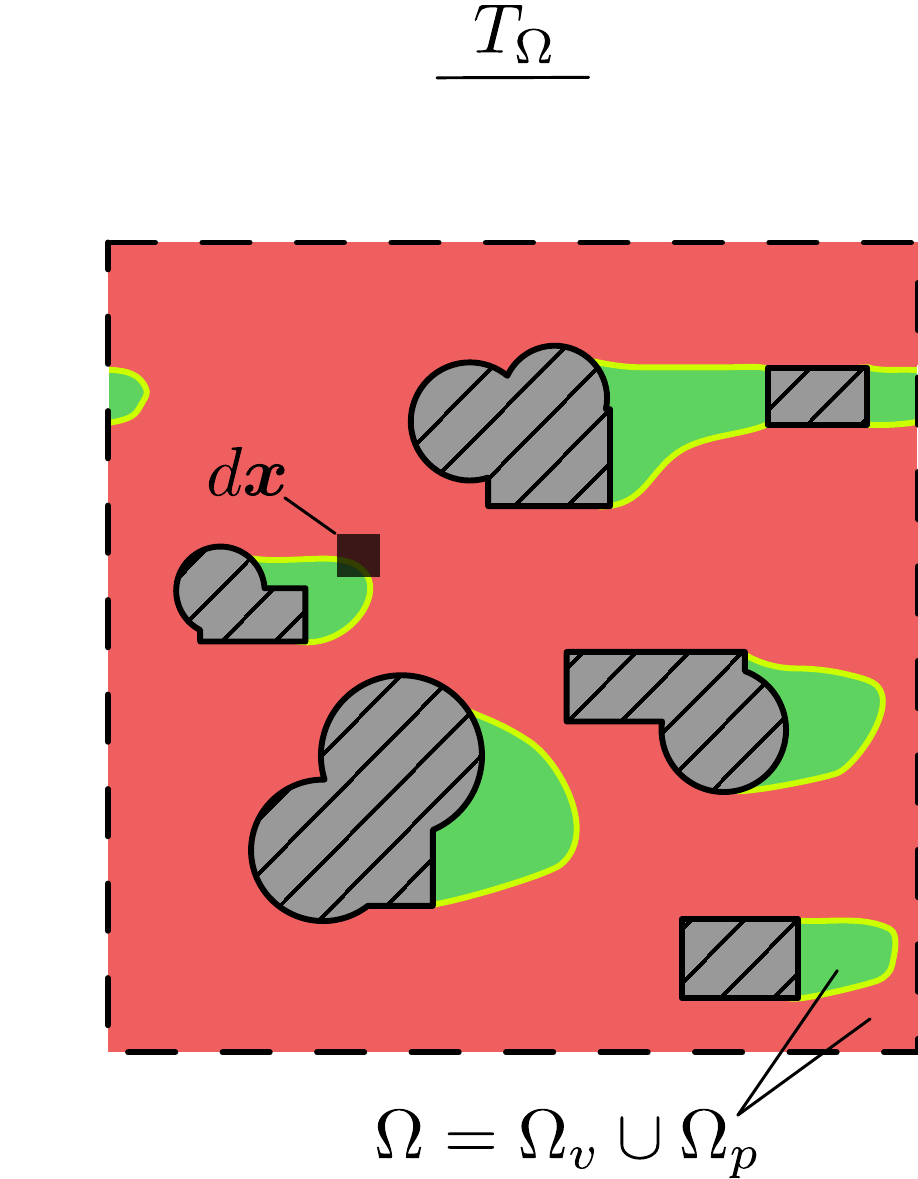}
		\caption{Schematic representation of the three definitions of tortuosities from Eqs.~\eqref{eq:T_sV0}, \eqref{eq:T_sV0_over_percolating_volume}, and \eqref{eq:TV}. $T_s^*$ (\textit{top left}) and $T_s$ (\textit{top right}) integrate over the surface elements $d\bsym{x}$ of the intersections of an arbitrary surface $A$ and the corresponding fluid volumes. $T_s^*$ considers all the fluid volume $\Omega$, while $T_s$ \add{--} only its percolating part, $\Omega_p$. $T_\Omega$ (\textit{bottom left}) integrates over the whole fluid volume $\Omega$. $T_s$ and $T_\Omega$ are used further in the present study.}
		\label{fig:T_definitions}
	\end{figure}

	The first presented definition of the tortuosity, $T_s^*$, was originally defined as a weighted average of the tortuosities of all the streamlines in the system~\cite{Matyka2008,Duda2011,Koponen1996}
	\begin{equation}\label{eq:T_sV0}
		T_s^* = \frac{\displaystyle\int\limits_{A \> \cap \> \Omega} d\bsym{x} \> V_0(\bsym{x}) \tilde\lambda(\bsym{x})}{\displaystyle\int\limits_{A \> \cap \> \Omega} d\bsym{x} \> V_0(\bsym{x})}.
	\end{equation}
	where $\tilde\lambda(\bsym{x})$ is the tortuosity of a streamline passing through a point $\bsym{x} \in A \cap \Omega$
	\begin{equation}\label{eq:single_streamline_tortuosity}
		\tilde\lambda(\bsym{x}) = \dfrac{\lambda(\bsym{x})}{L_\text{ref}}, \quad \lambda(\bsym{x}) \equiv \back \back \int\limits_{\text{along }s(\bsym{x})} \back ds
	\end{equation}
	with $\lambda(\bsym{x})$ denoting the (curvilinear) length of the streamline, $ds$ the length of the infinitesimal increment along the streamline, and $L_\text{ref}$ denoting the reference length. Unless otherwise stated, we take the reference length to be equal to the length of the porous sample's side, i.e., $L_\text{ref}=L$. The integration in Eq.~\eqref{eq:T_sV0} is performed over the intersection of \del{a plane $A$ perpendicular to the mean flow direction and} the fluid volume $\Omega$ \add{and plane $A$ perpendicular to the mean flow direction} (see Fig.~\ref{fig:T_definitions}, top-left plot). Also, $A$ must be intersected by all the streamlines in the system. As it is ambiguous to tell the length of a looping streamline inside a recirculation zone, the integration in Eq.~\eqref{eq:T_sV0} is usually limited only to the intersection of the plane perpendicular to the mean flow direction and the \textit{percolating} volume of the fluid (see Fig.~\ref{fig:T_definitions}, top-right plot)
	\begin{equation}\label{eq:T_sV0_over_percolating_volume}
		T_s = \frac{\displaystyle\int\limits_{A \> \cap \> \Omega_p} d\bsym{x} \> V_0(\bsym{x}) \tilde\lambda(\bsym{x})}{\displaystyle\int\limits_{A \> \cap \> \Omega_p} d\bsym{x} \> V_0(\bsym{x})},
	\end{equation}
	although in numerical evaluations of $T_s$ it is more convenient to place $A$ such that it intersects only the percolating volume. In the present work, we will use the definition from Eq.~\eqref{eq:T_sV0_over_percolating_volume} and will refer to it as the \textit{streamilne-based tortuosity}. In this, and all further integrations, one can equivalently assume that $\bsym{V}=\bsym{0}$ inside the solid phase and include the obstacles in the integrations.
	$T_s$ integrates the tortuosities of the percolating streamlines seeded from $A \cap \Omega_p$ with weighting proportional to the flux in the mean flow direction associated with each streamline. Such weighting considers the streamlines of large flux, and thus carrying more fluid mass in a unit time, as more important for the value of the tortuosity, which increases the impact of the preferential channels of the flow~\cite{Andrade1999}. It was proven in~\cite{Duda2011} that when there are no recirculation zones in the flow, $T_s$ is equivalent to the third tortuosity considered in the present work
	\begin{equation}\label{eq:TV}
		T_\Omega = \frac{\displaystyle\int\limits_{\Omega} d\bsym{x} \> V(\bsym{x})}{\displaystyle\int\limits_{\Omega} d\bsym{x} \> V_0(\bsym{x})} = \frac{I[V]}{I[V_0]}.
	\end{equation}
	where the integration goes over the whole fluid volume $\Omega$ (see Fig.~\ref{fig:T_definitions}, bottom-left plot). We will refer to this definition as the \textit{volume-integrated tortuosity}. The notation $I_\alpha[\> \cdot \>]$ is a shorthand for the integral of some quantity over the fluid volume or its subset, i.e.
	\begin{equation}
		I[f]_\alpha = \int\limits_{\Omega_\alpha} d\bsym{x} \> f(\bsym{x}), \quad \alpha \in \{\cdot,p,v\}
	\end{equation}
	and will be used further in the text. When the recirculation zones are present in the flow, the two latter tortuosities are not equal, rather the inequality holds~\cite{Duda2011}
	\begin{equation}\label{eq:Duda_inequality}
		T_s \le T_\Omega.
	\end{equation}
	
	{\color{changecolor}
		Splitting the integrals in the numerator and denominator of the volume-integrated tortuosity $T_\Omega$ into the contributions coming from the percolating volume $\Omega_p$ and the recirculation volume $\Omega_v$ gives
		\begin{equation}\label{eq:TV_split}
			T_\Omega = \frac{I[V]}{I[V_0]} = 
			\frac{I[V]_p + I[V]_v}{I[V_0]_p + I[V_0]_v}.
		\end{equation}
		According to the definition of the vortex provided earlier in this Section, the mean velocity of the vortex in steady-state flow is zero, i.e., $I[\bsym{V}]_v=\bsym{0}$, so the term $I[V_0]_v$ in the denominator can be omitted. Thus, we can rewrite Eq.~\eqref{eq:TV_split} as
		\begin{equation}\label{eq:TV_split_reduced}
			T_\Omega = \frac{I[V]_p}{I[V_0]_p} + \frac{I[V]_v}{I[V_0]_p}.
		\end{equation}
		Now, using the same procedure as in~\cite{Duda2011}, it is possible to show that the surface integrals over $A \cap \Omega_p$ in Eq.~\eqref{eq:T_sV0_over_percolating_volume} can be transformed into volume integrals over the entire percolating volume $\Omega_p$, i.e., $T_s = I[V]_p/I[V_0]_p$. This allows to rewrite Eq.~\eqref{eq:TV_split_reduced} as
		\begin{equation}\label{eq:TV_split_reduced_using_Ts}
			T_\Omega = T_s + \frac{I[V]_v}{I[V_0]_p}.
		\end{equation}
		Such a form of the volume-integrated tortuosity allows one to clearly distinguish between the contributions from the percolating and the recirculation zones and directly points to the sources of inequality in Eq.~\eqref{eq:Duda_inequality}. The denominator, $I[V_0]_p$, is proportional to the Reynolds number and does not introduce any new unknown variable into our considerations.
	}
	
	Out of the two tortuosities $T_s$ and $T_\Omega$, the latter turns out to be significantly easier to calculate numerically. This is because of its locality -- it does not make use of streamlines, which, due to numerical errors during their generation, can easily cross the boundary $\partial\Omega_{pv}$ between the percolating and recirculation volume during the integration along the fluid velocity field characteristics. This problem becomes especially pronounced in low-porosity porous media and at high Reynolds numbers, where a considerable portion of the fluid is occupied by the recirculation zones.
	
	\subsection{Results analysis}\label{ssec:postprocessing}
	
	\subsubsection{Generation of streamlines and determination of the recirculation and percolating volumes}\label{sssec:sl_generation}
	
	To generate the streamlines in SC samples, we numerically integrate the advection equation for the massless tracers transported in the fluid velocity field. We use the \del{first-order explicit Euler} \add{second- or third-order explicit} method for the timestepping and meshless RBF-FD approximation~\cite{tolstykh_using_2003} for the interpolation in space. We use the C++ implementation of meshless approximation routines provided by Medusa library~\cite{Slak2021}. We use streamlines to calculate the tortuosity and to identify the percolating/recirculation volumes in SC samples. Due to this, the initial positions of the tracers, as well as the timestep length depend on the task at hand and the considered system. The details of the streamline generation procedure and the algorithm used for the determination of percolating/recirculation volumes are delivered in Appendices~\ref{app:numerical_details}.\ref{subapp:streamlines} and \ref{app:numerical_details}.\ref{subapp:volumes_histogram}, respectively. The convergence of the value of the streamline-based tortuosity $T_s$ with the decreasing separation of the starting points of the streamlines $h_s$ and advection timestep length $\Delta t_s$ is provided in Appendix~\ref{app:convergence}.
	
	\subsubsection{Calculation of \add{the constituents of the volume-integrated tortuosity} \del{$T_\Omega$ constituents}}\label{sssec:T_Omega_calculation}
	
	To calculate the integrals of $V_0$ and $V$ over the whole volume of fluid in Eq.~\eqref{eq:TV}, in the case of stochastic porous samples, the first-order method, i.e., the summing of the values of the hydrodynamic fields from each lattice point, is used. In the case of SC samples, we use the ParaView utility \texttt{IntegrateVariables}~\cite{Ahrens2005,ParaView_Python}. For the integration of $V_0$ and $V$ in the recirculation or the percolating volume in SC samples, we interpolate $V$ and $V_0$ to the centers of the cells used for the identification of $\Omega_p$ and $\Omega_v$. We use the same RBF-FD interpolation as in the generation of streamlines. Finally, we use the first-order method for the integration of the fields in each of the volumes.
	
	\section{Results}
	
	\subsection{Difference in tortuosity--Reynolds number relation for porous samples of different porosities.}
	
\begin{figure}[!ht]
		\centering
		\includegraphics[width=\linewidth]{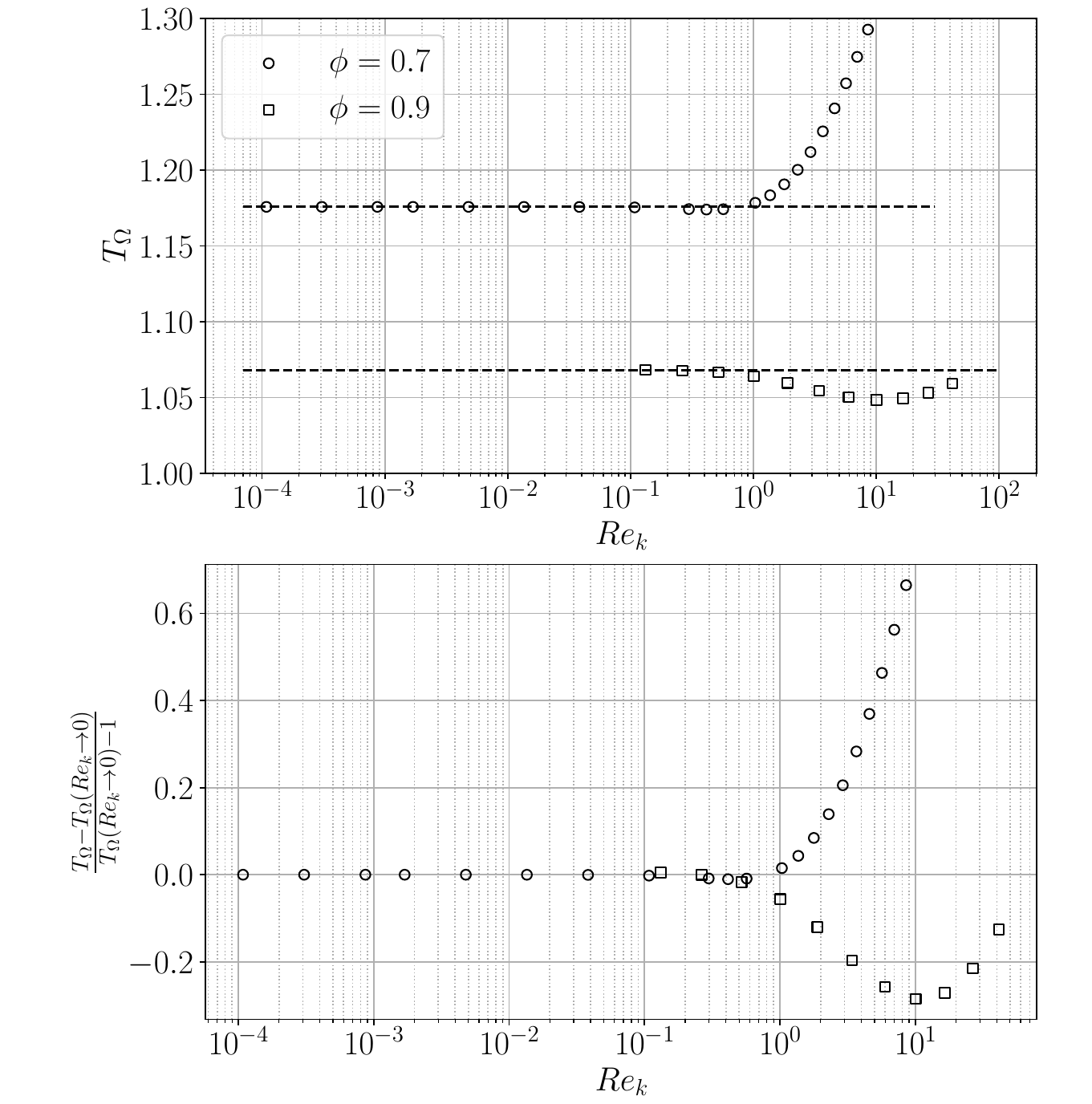}
		\caption{\textit{Top}: the values of $T_\Omega$ for stochastic porous media of porosity $\phi=0.7$ (\textit{circles})  and $\phi=0.9$ (\textit{squares}) during the Darcy-inertial transition. The horizontal dashed lines denote the values of $T_\Omega$ in the limit of vanishingly small $Re_k$ ($Re_k \rightarrow 0$). \textit{Bottom}: the same data as in the left subplot but rescaled by $T_\Omega$ at vanishingly small $Re_k$.}
		\label{fig:TV-Re_sotchastic_PM}
	\end{figure}

\begin{figure}[!h]
		\centering
		
		\includegraphics[width=.49\linewidth]{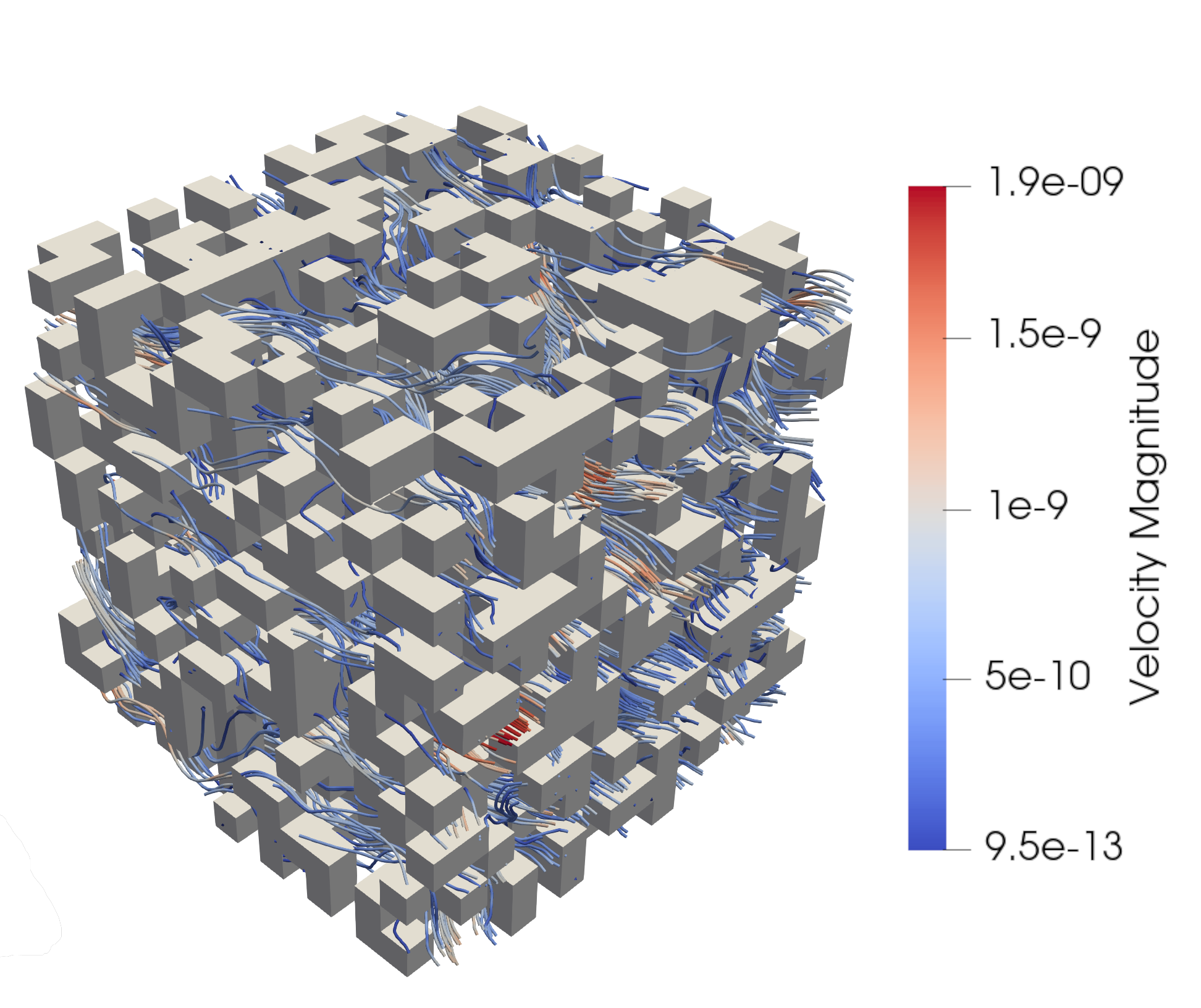}
		\includegraphics[width=.49\linewidth]{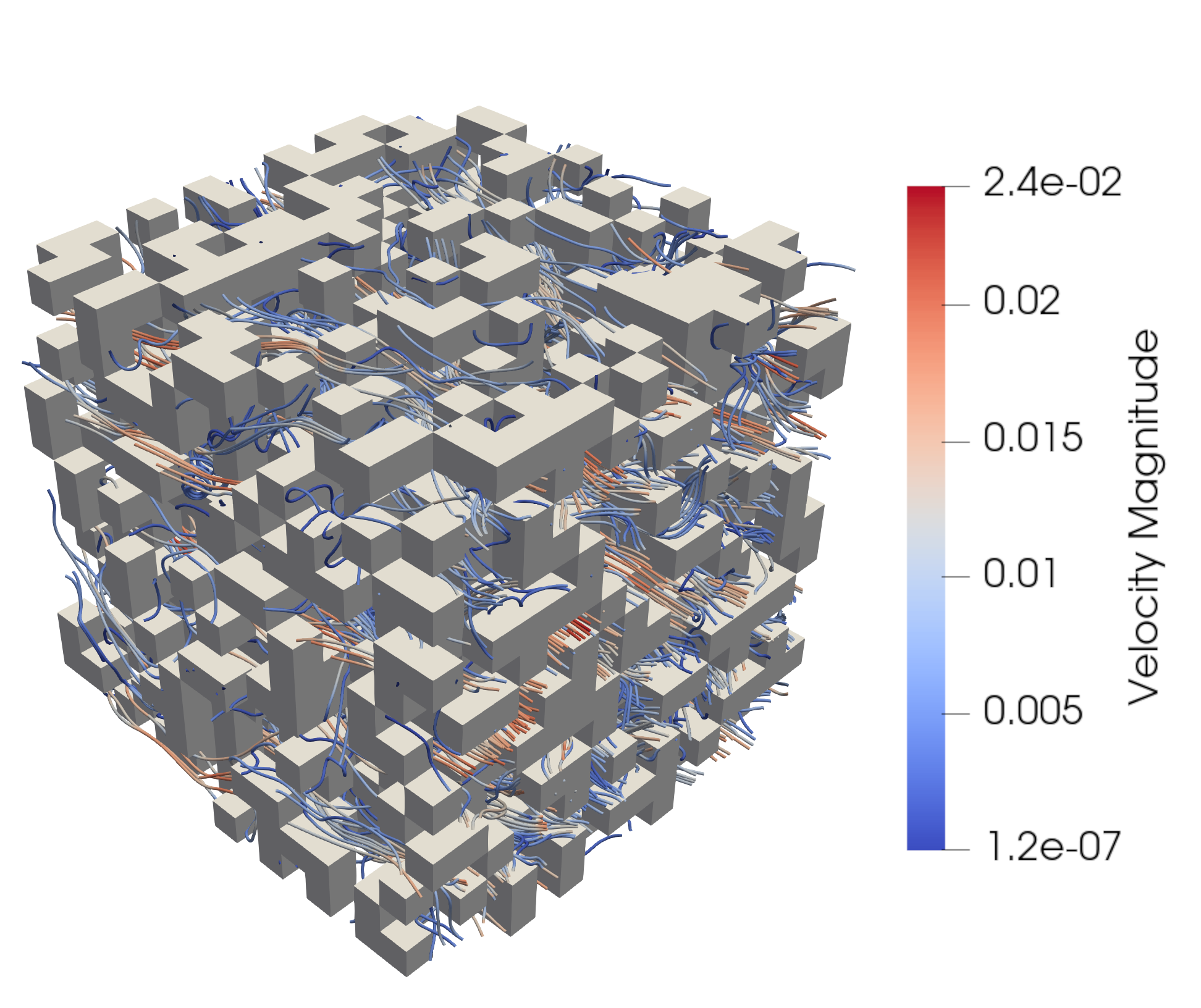}
		
		
		\includegraphics[width=.49\linewidth]{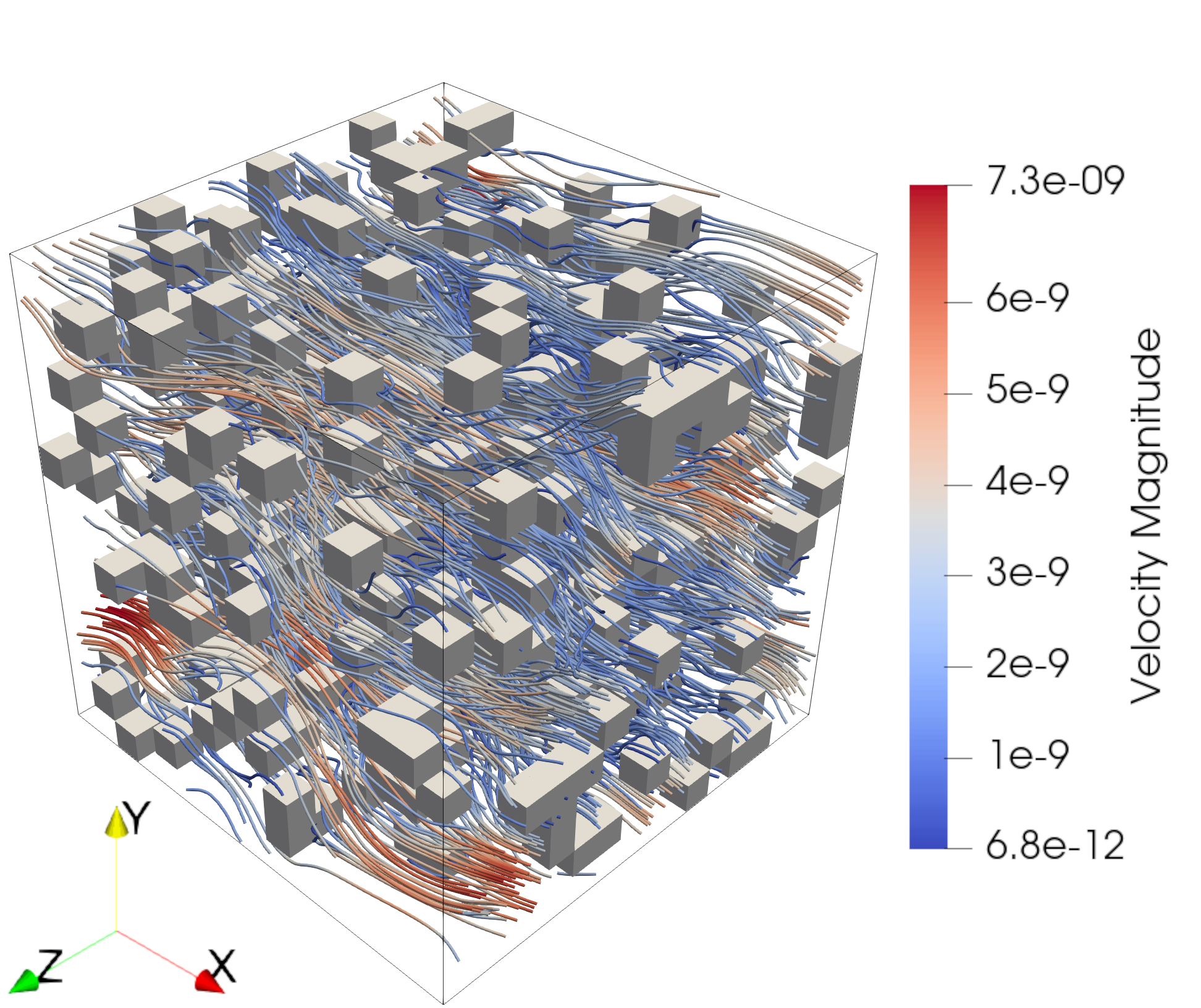}
		\includegraphics[width=.49\linewidth]{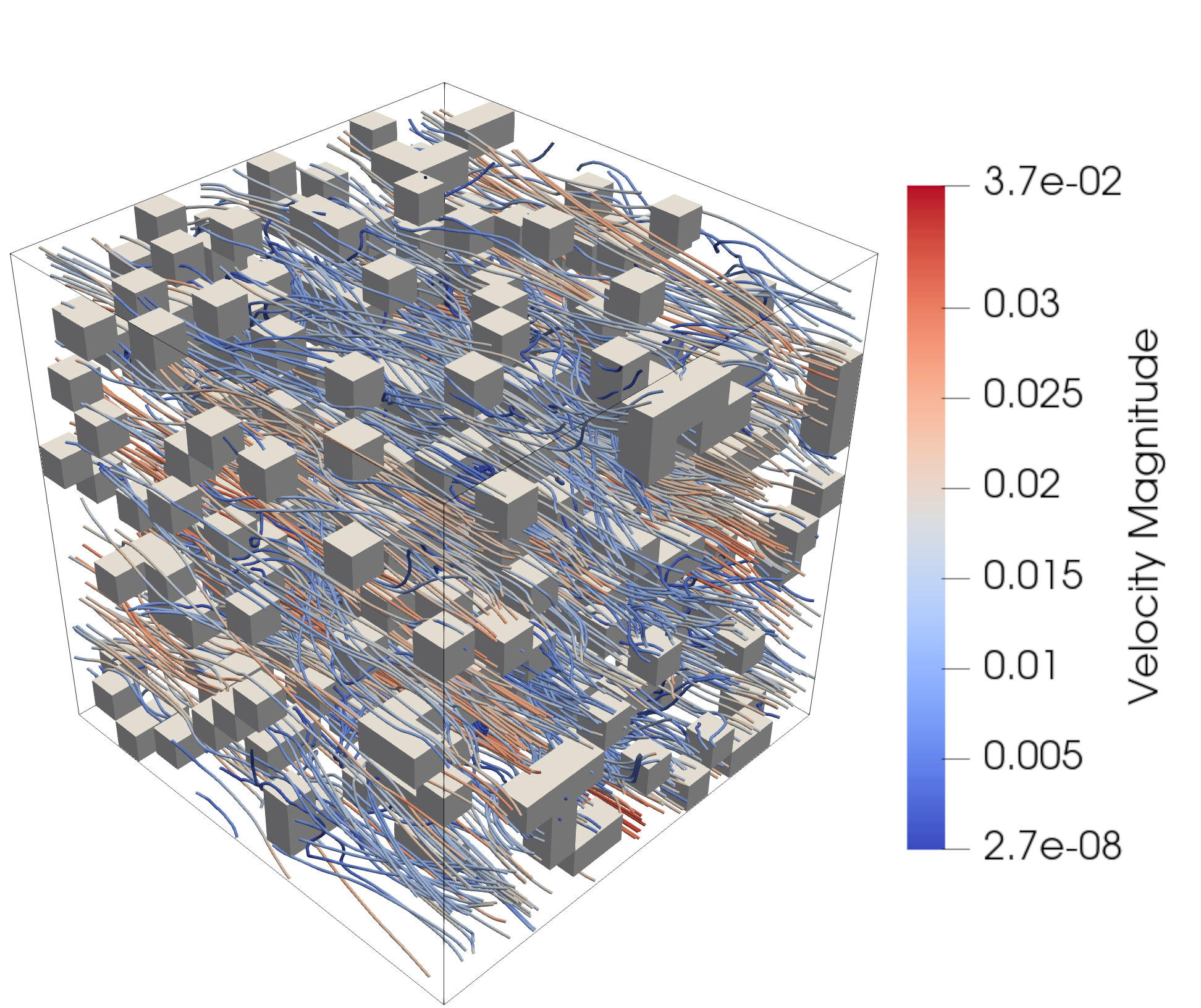}
		
		\caption{Streamlines of the velocity field for the stochastic porous media of porosity $\phi=0.7$ (\textit{top row}) and $\phi=0.9$ (\textit{bottom row}) at low (\textit{left column}) and high (\textit{right column}) $Re_k$. The flow is along $x_0$ axis. The shown streamlines are generated using the \texttt{streamTracer} utility of ParaView for the sake of the visualization only and are not used for the calculation of $T_s$ or the identification of percolating/recirculation volumes.}
		\label{fig:stochastic_PM_viz_3d}
	\end{figure}

\begin{figure}[!h]
		\centering
		
		\includegraphics[width=.49\linewidth]{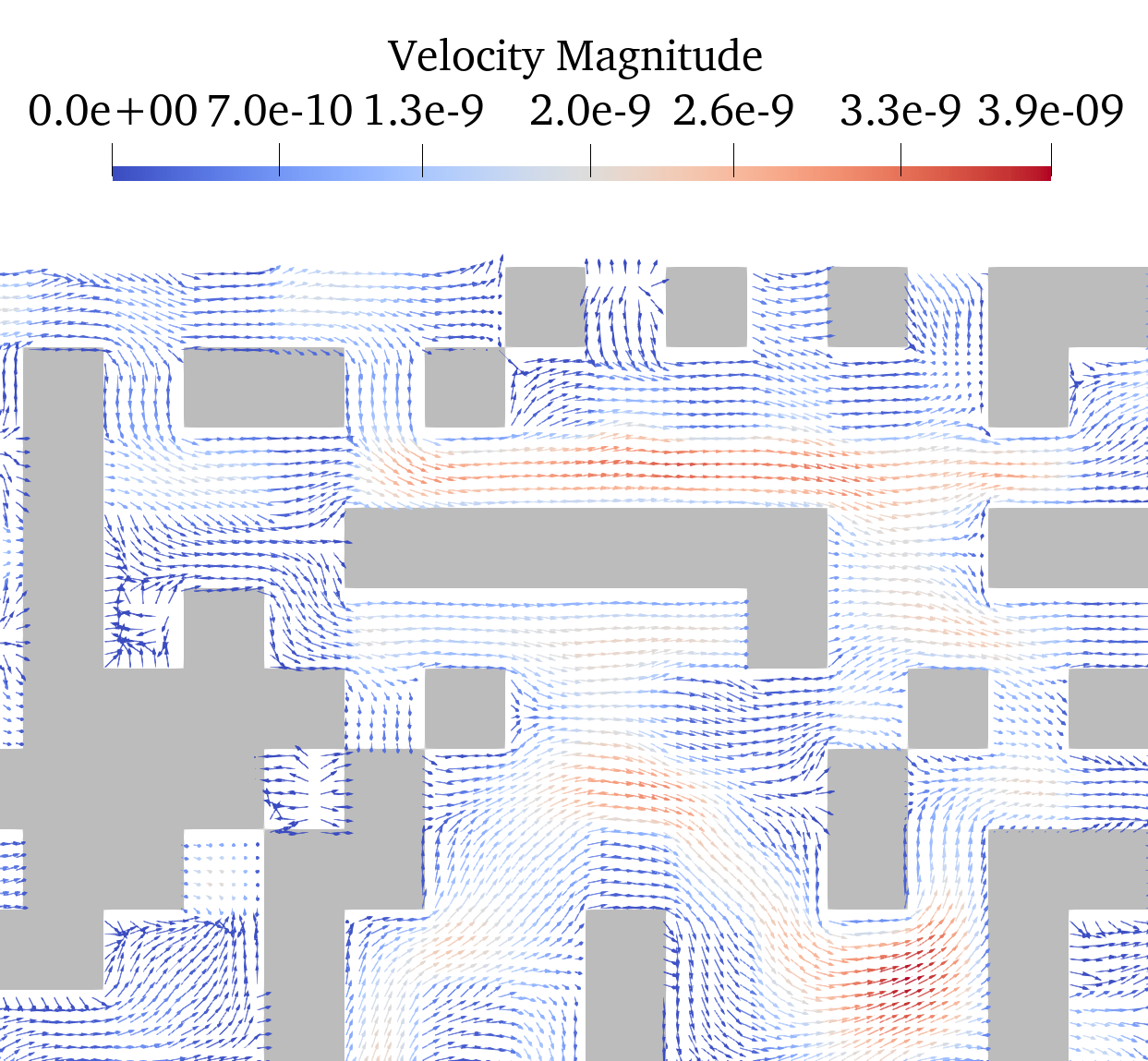}
		\includegraphics[width=.49\linewidth]{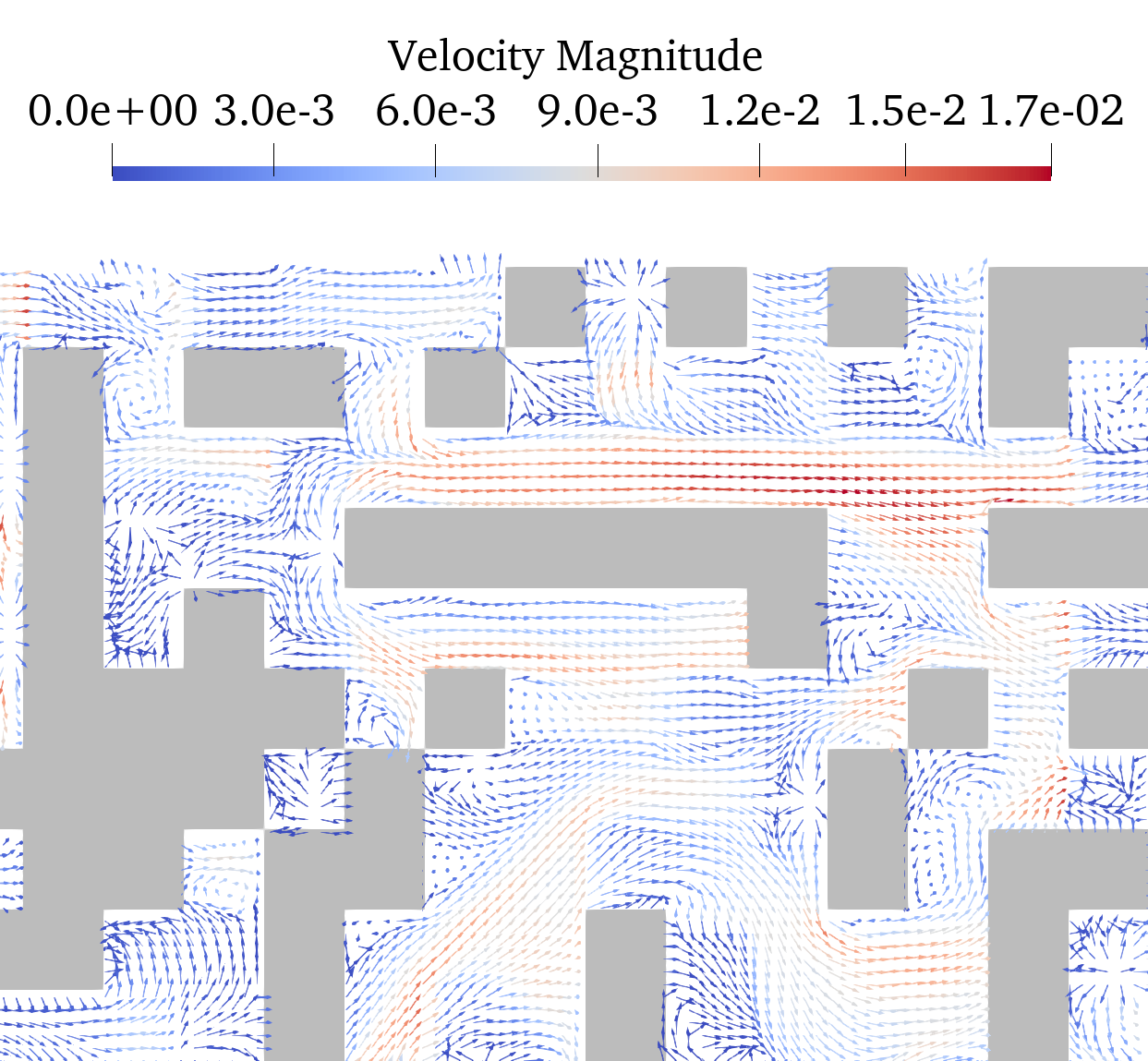}
		
		\vspace{1cm}
		
		\includegraphics[width=.49\linewidth]{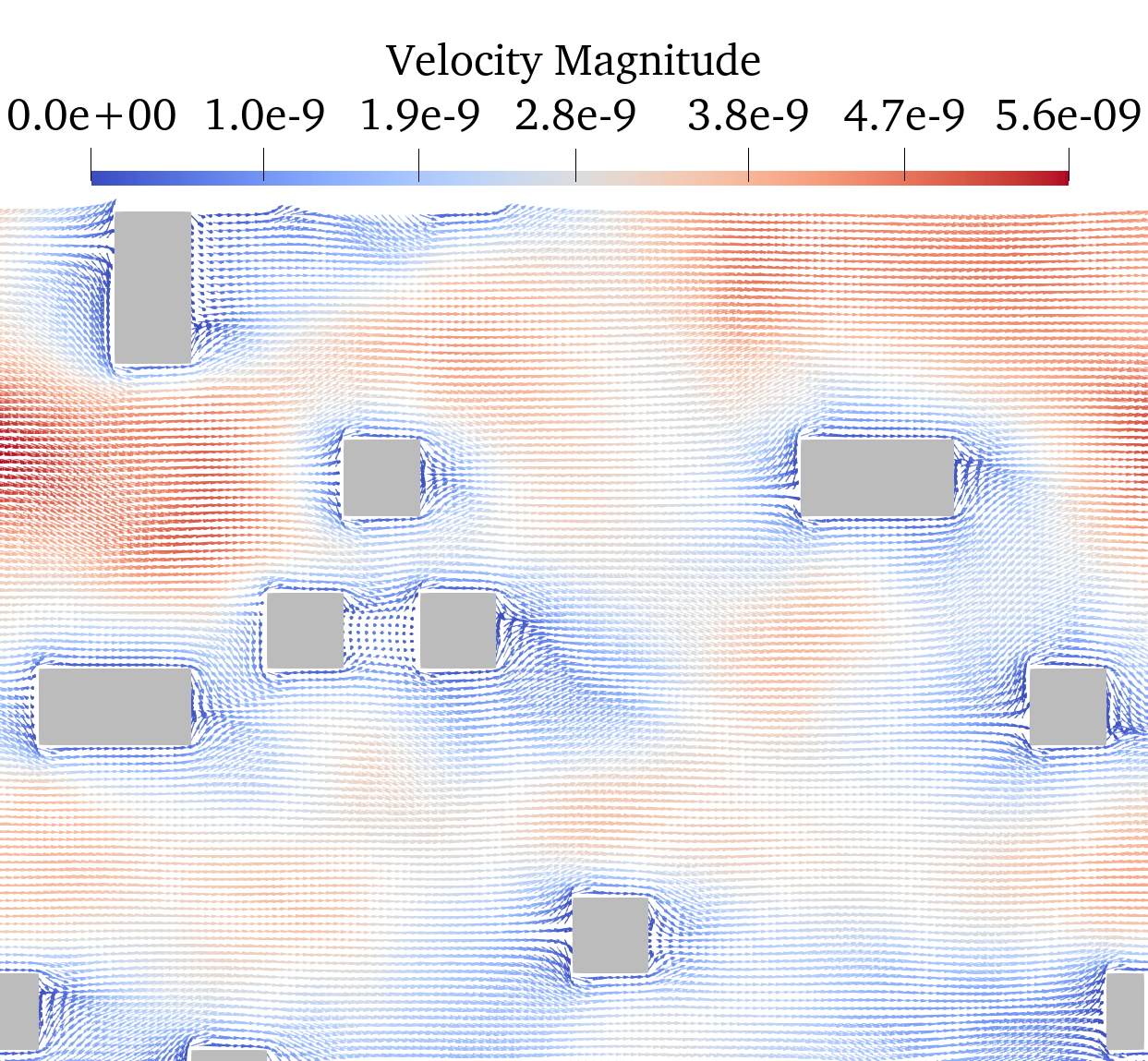}
		\includegraphics[width=.49\linewidth]{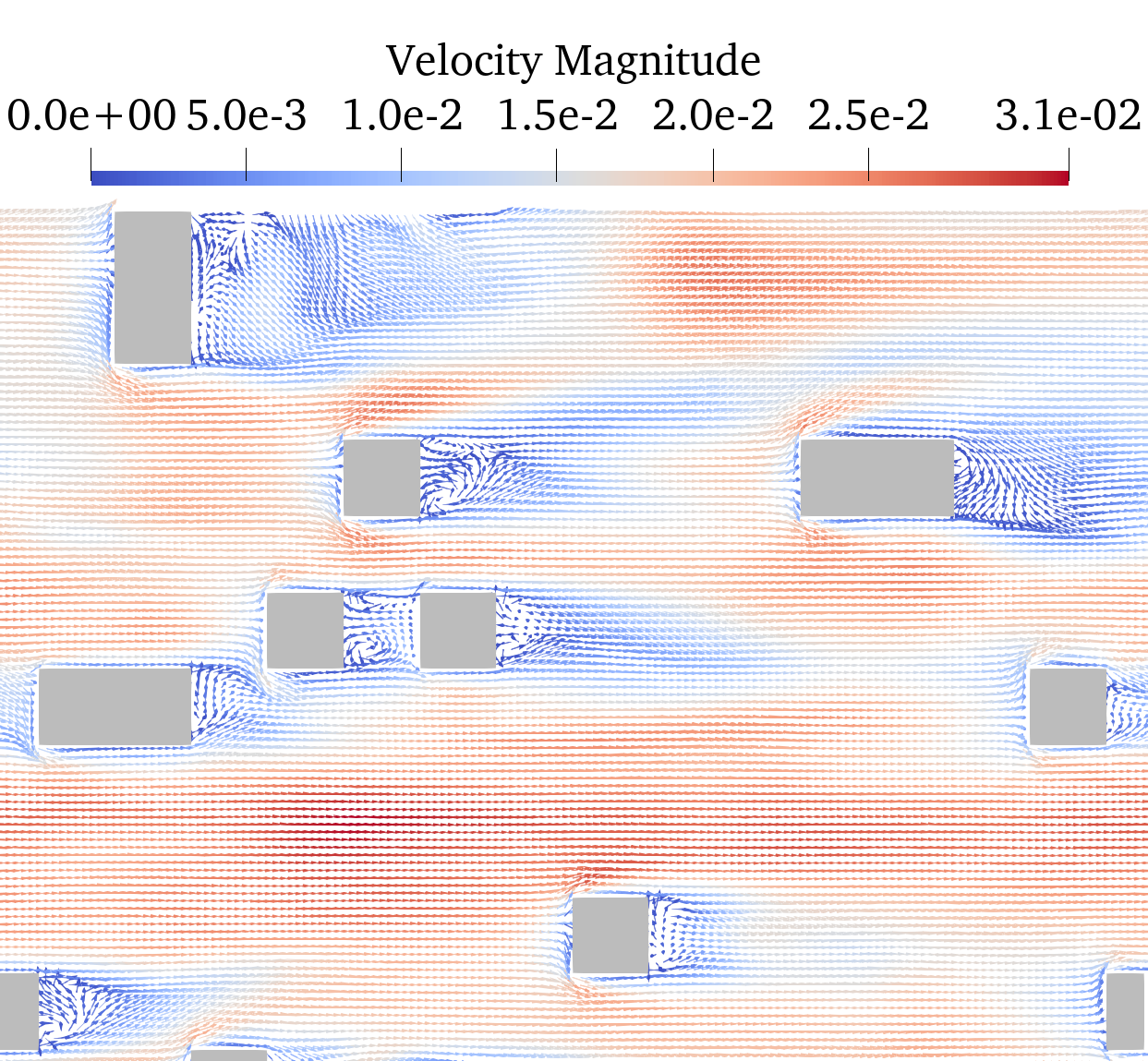}
		
		\caption{Velocity vectors at chosen sections for the stochastic porous media of porosity $\phi=0.7$ (\textit{top row}) and $\phi=0.9$ (\textit{bottom row}) at low (\textit{left column}) and high (\textit{right column}) $Re_k$. The flow is from left to right. For $\phi=0.7$, only every 3rd velocity vector in each direction is shown for clarity.}
		\label{fig:stochastic_PM_viz_section}
	\end{figure}

	We first investigate the transition from the Darcy to inertial regime for two stochastic porous samples of porosities $\phi=0.7$ and $\phi=0.9$. Top subplot of Fig.~\ref{fig:TV-Re_sotchastic_PM} shows the dependence of the tortuosity $T_\Omega$ defined in Eq.~\eqref{eq:TV} on the Reynolds number
	\iftoggle{showdeleted}{
		\begin{equation}\label{eq:Re}
			Re_k = \frac{\cancel{\add{\sqrt{k(Re_k \rightarrow 0)}}} \add{\sqrt{k_0}}\langle V_0 \rangle \phi}{\nu}, \quad \langle V_0 \rangle = \frac{I[V_0]}{I[1]}
		\end{equation}
	}{
		\begin{equation}\label{eq:Re}
			Re_k = \frac{\add{\sqrt{k_0}}\langle V_0 \rangle \phi}{\nu}, \quad \langle V_0 \rangle = \frac{I[V_0]}{I[1]}
		\end{equation}
	}
	where \del{$k(Re_k \rightarrow 0)$} \add{$k_0$} is the \add{intrinsic} permeability \del{at the vanishing Reynolds number}. The same data are shown rescaled in the bottom subplot of Fig.~\ref{fig:TV-Re_sotchastic_PM}, where by $T_\Omega(Re_k \rightarrow 0)$ we mean the value of $T_\Omega$ at the vanishing Reynolds number (denoted by the horizontal dashed line in the top subplot). We note that for $Re_k$ values beyond the investigated range, the solution went unsteady. We observe that in the Stokes regime, the tortuosity for each sample has a constant value, $1.07$ for $\phi = 0.9$ and $1.175$ for $\phi=0.7$. As $Re_k$ increases, the tortuosity first decreases and then increases for both samples. However, the magnitude of the initial decrease and subsequent increase is considerably different for the two porosities. In the case of the higher-porosity sample, the difference between the minimum of $T_\Omega$ and the $T_\Omega(Re_k \rightarrow 0)$ is about $0.02$ ($0.3$ relative to the Darcy-regime value above unity, see the bottom subplot), and the increase is about $0.01$. For the lower-porosity sample, these values are $0.003$ ($0.01$ relative to the Darcy-regime value above unity) and $0.125$ ($0.67$ relative to the Darcy-regime value above unity), respectively. Such differences between the minimal and maximal values achieved by $T_\Omega$ in the steady-state regime were previously observed in stochastic high-porosity media~\cite{Sniezek2024} and in low-porosity geological samples~\cite{Naqvi2025}. The onset of inertial effects is also visible in the visualizations of the velocity fields, as the weaker channeling at the outlets of the samples (Fig.~\ref{fig:stochastic_PM_viz_3d}) and the emergence of the recirculation zones in the wakes of the obstacles (Fig.~\ref{fig:stochastic_PM_viz_section}).

	\bigskip
	
	\iftoggle{showdeleted}{
		\subsection{\del{Splitting the volume-integrated tortuosity $T_\Omega$}}
		
		\del{To explain such a different behavior of $T_\Omega(Re_k)$ in the low- and high-porosity stochastic systems, we first take a closer look at the definition of $T_\Omega$ and observe that both the numerator and denominator can be split into separate contributions coming from the percolating volume $\Omega_p$ and the recirculation volume $\Omega_v$}
		{\color{changecolor}
			\begin{equation}
				\cancel{T_\Omega = \frac{I[V]}{I[V_0]} = 
					\frac{I[V]_p + I[V]_v}{I[V_0]_p + I[V_0]_v}.}
			\end{equation}
		}
		\del{According to the definition of the vortex from Sec.~\ref{ssec:tortuosity}, the mean velocity of the vortex in steady-state flow is zero, i.e., $I[\bsym{V}]_v=\bsym{0}$, so the term $I[V_0]_v$ in the denominator can be omitted. Thus, we can rewrite Eq.~\eqref{eq:TV_split} as}
		{\color{changecolor}
			\begin{equation}
				\cancel{T_\Omega = \frac{I[V]_p}{I[V_0]_p} + \frac{I[V]_v}{I[V_0]_p}.}
			\end{equation}
		}
		\del{Now, using the same procedure as in~\cite{Duda2011}, it is possible to show that the surface integrals over $A \cap \Omega_p$ in Eq.~\eqref{eq:T_sV0_over_percolating_volume} can be transformed into volume integrals over the entire percolating volume $\Omega_p$, i.e., $T_s = I[V]_p/I[V_0]_p$. This allows to rewrite Eq.~\eqref{eq:TV_split_reduced} as}
		{\color{changecolor}
			\begin{equation}
				\cancel{T_\Omega = T_s + \frac{I[V]_v}{I[V_0]_p}.}
			\end{equation}
		}
		\del{Such a form of the volume-integrated tortuosity allows one to clearly distinguish between the contributions from the percolating and the recirculation zones and directly points to the sources of inequality in Eq.~\eqref{eq:Duda_inequality}. The denominator, $I[V_0]_p$, is proportional to the Reynolds number (cf. Eq.~\eqref{eq:Re}) and does not introduce any new unknown variable into our considerations.}
		
		\begin{figure}[!ht]
			\centering
			\includegraphics[width=.75\linewidth]{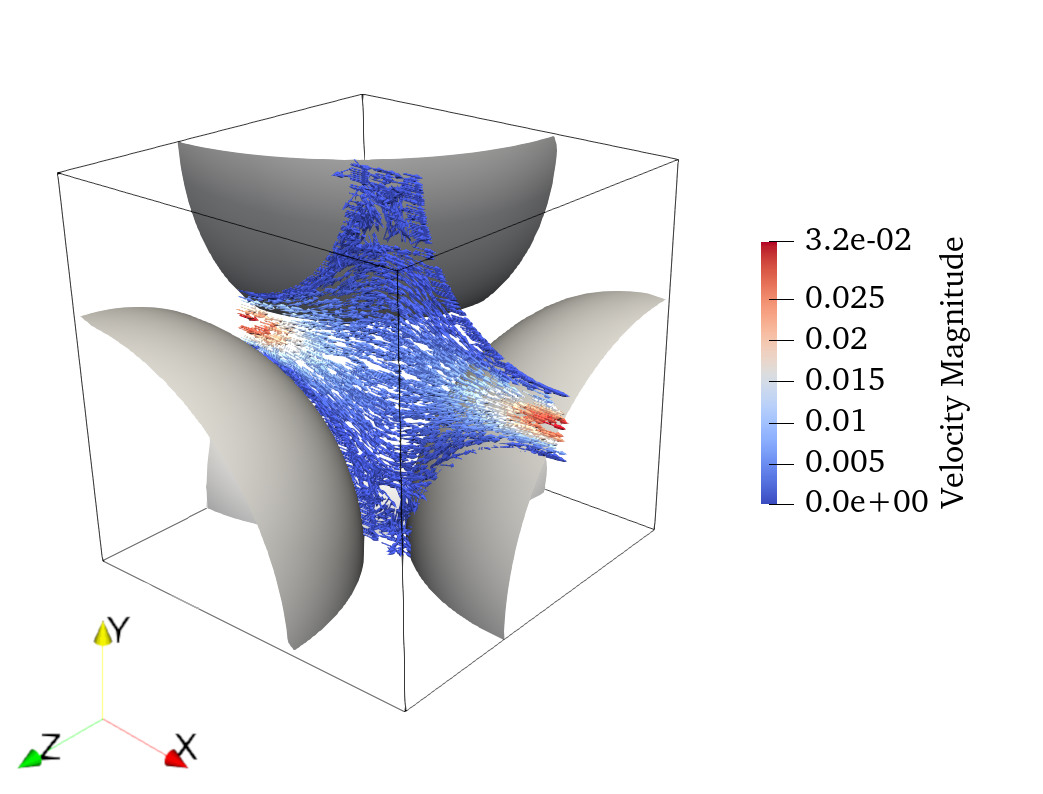}
			\hspace{.25cm}
			\includegraphics[width=.75\linewidth]{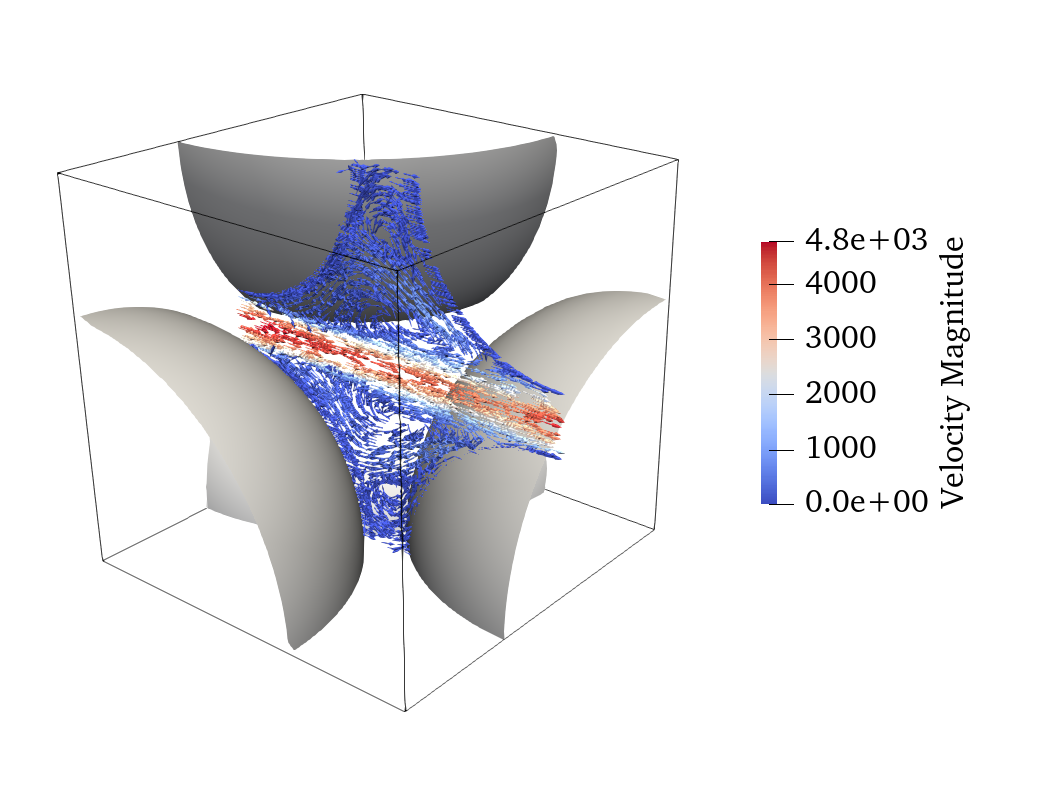}
			\caption*{\del{FIG. 10: Visualizations of the sections of the velocity field in $\phi=0.1$ SC system for $Re_k = 2.61 \cdot 10^{-7}$ (\textit{top}) and $Re_k = 0.049$ (\textit{bottom}). The gray surfaces are the obstacles' walls, and only a few chosen walls are shown for clarity.}}
		\end{figure}
		
		\begin{figure}[!ht]
			\centering
			\includegraphics[height=.45\linewidth]{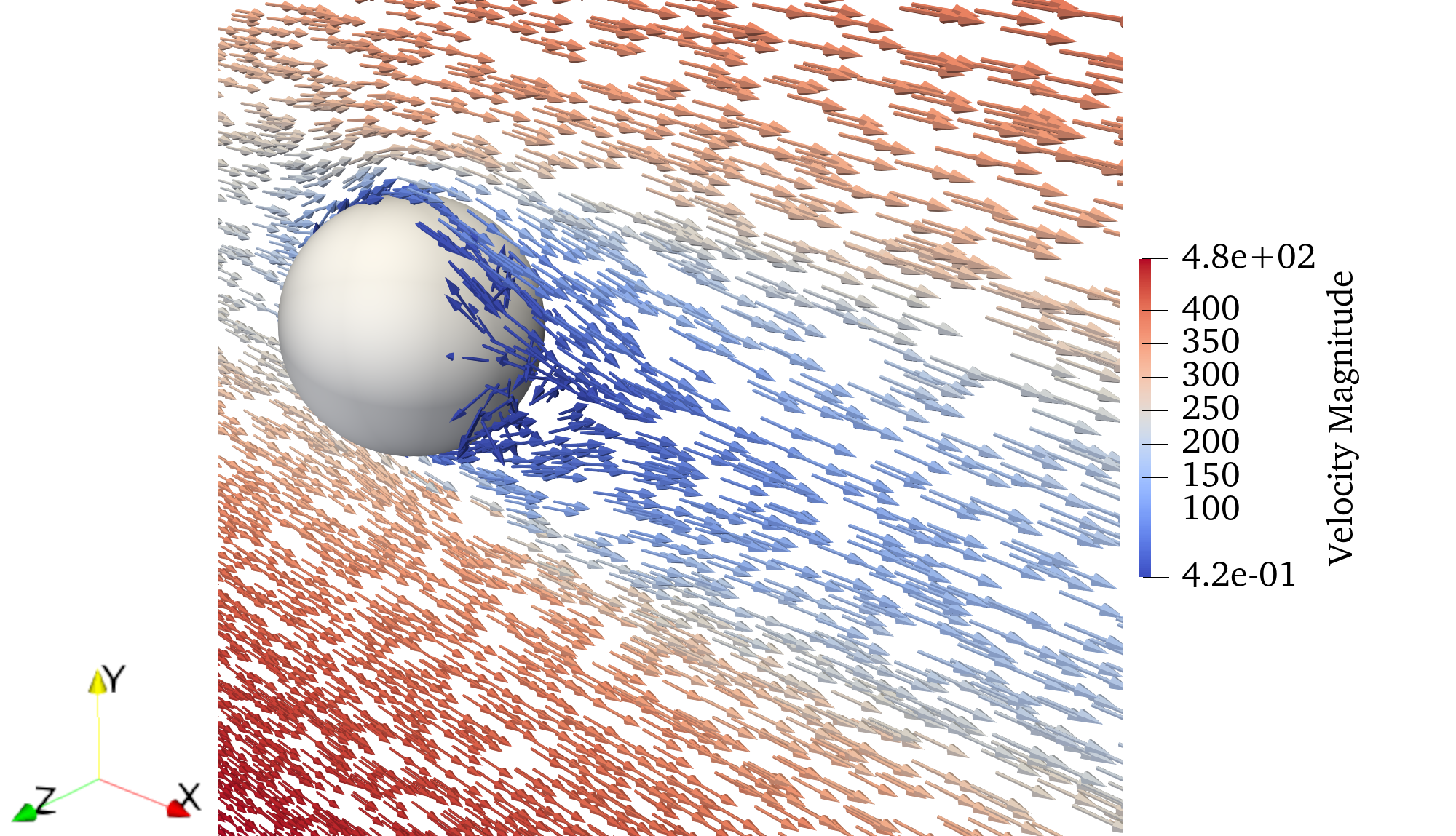}
			
			\vspace{.5cm}
			
			\includegraphics[height=.45\linewidth]{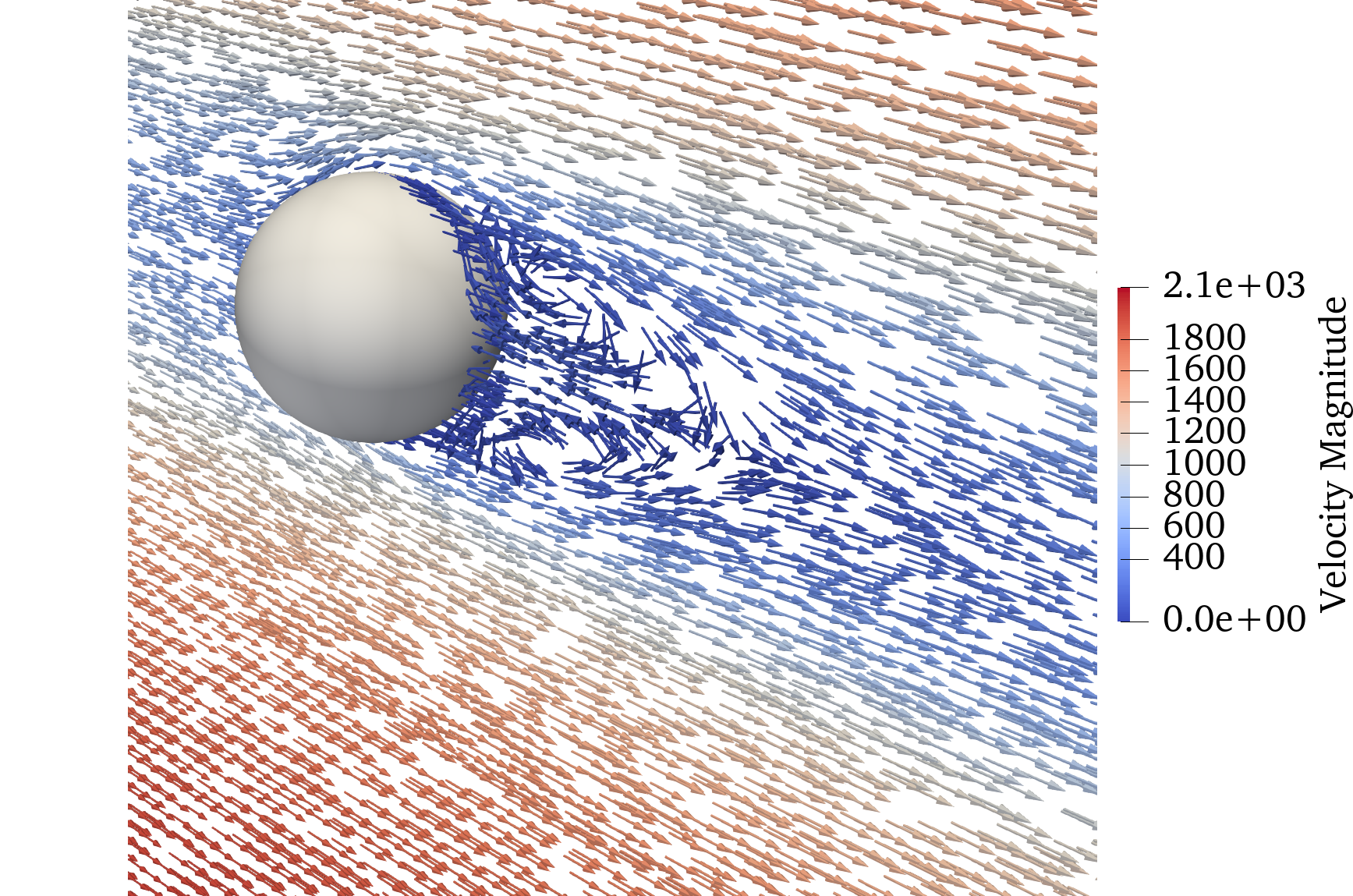}
			\caption*{\del{FIG. 11: Visualizations of the sections of the velocity field in $\phi=0.999$ SC system for $Re_k = 376$ (\textit{top}) and $Re_k = 1557$ (\textit{bottom}). The gray surfaces are the obstacles' walls.}}
		\end{figure}
	}{}
	
	\bigskip
	
	\subsection{Tortuosity as the inertia onset indicator in simple cubic systems}
	
\begin{figure}[!ht]
		\centering
		\includegraphics[width=.75\linewidth]{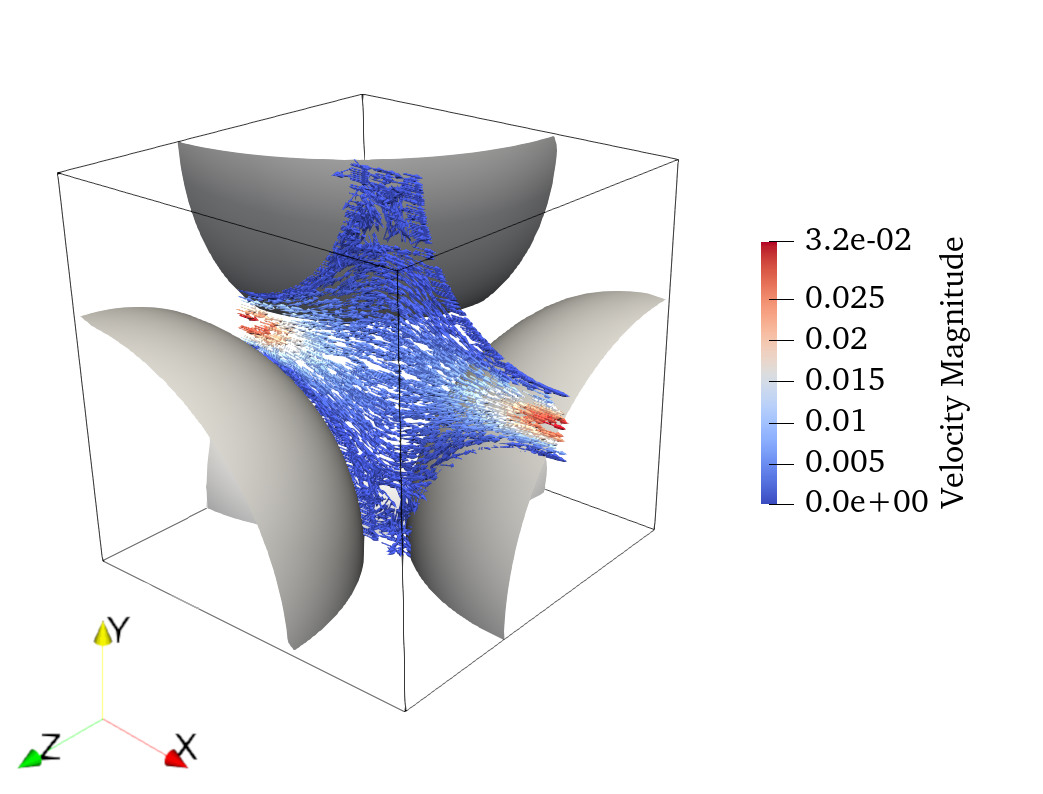}
		\hspace{.25cm}
		\includegraphics[width=.75\linewidth]{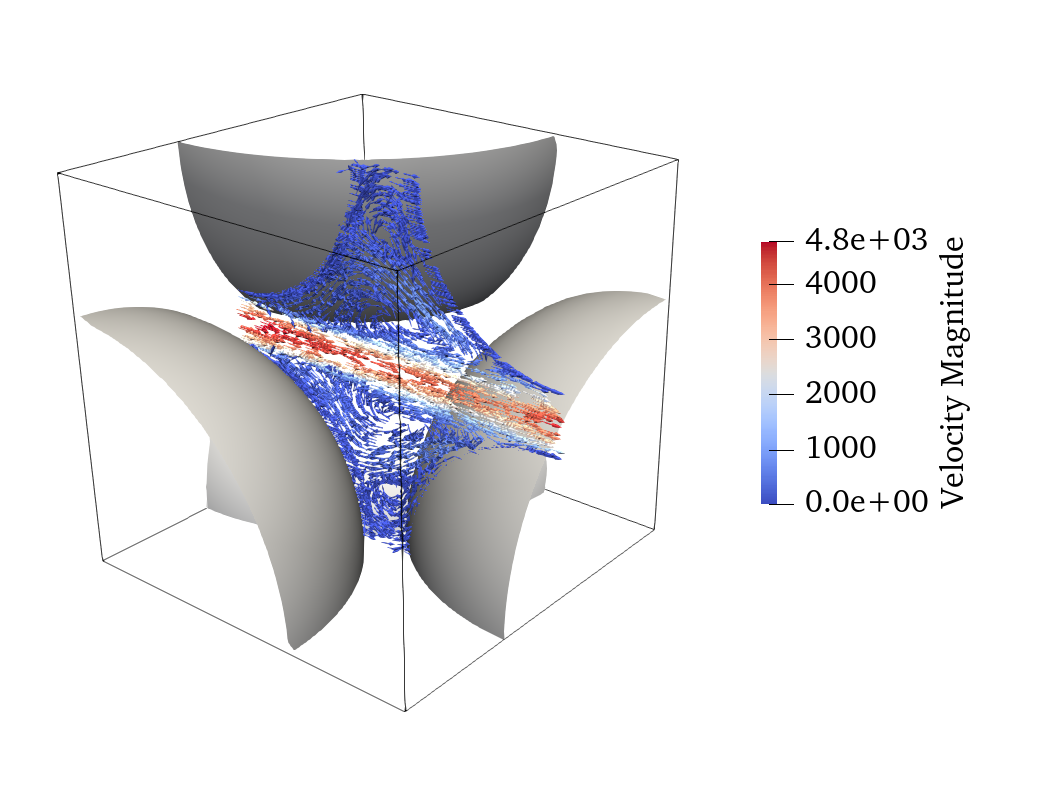}
		\caption{\add{Visualizations of the sections of the velocity field in $\phi=0.1$ SC system for \del{$Re_k = 2.61 \cdot 10^{-7}$} $Re_k = 8.34 \cdot 10^{-7}$ (\textit{top}) and \del{$Re_k = 0.049$} $Re_k = 0.155$ (\textit{bottom}). The gray surfaces are the obstacles' walls, and only a few chosen walls are shown for clarity.}}
		\label{fig:velocity_section_r0-6526}
	\end{figure}

	\add{To deeper investigate the origin of such different behavior of the tortuosities observed in the stochastic samples, we make use of the split of the volume-integrated tortuosity, Eq.~\eqref{eq:TV_split_reduced_using_Ts}, and consider the terms $T_s$ and $I[V]_v/I[V_0]_p$ separately. As noted previously, the proper identification of the percolating and recirculating volumes in the fluid is very sensitive to stray streamlines (i.e., those that, due to numerical errors, cross the boundary between the percolating and recirculating volumes). We observed that even several stray streamlines can significantly alter the value of the streamline-based tortuosity. Because of this, we conduct the numerical investigation of the two contributions to $T_\Omega$ on simple cubic (SC) systems, and use the obtained results to explain the behavior of $T_\Omega$ in the stochastic samples. SC samples can be thought of as models of particular fragments of a stochastic porous medium, at the same time having the geometry simple enough to allow for the accurate calculation of streamlines and identification of $\Omega_p$ and $\Omega_v$. In our analysis, we consider SC systems of porosities $\phi=0.1,0.59,0.93,0.999$, which include the cases with extremely high and low geometrical confinement. Exemplary visualizations of the steady-state velocity fields in the studied SC systems are shown in Figs.~\ref{fig:velocity_section_r0-6526} \del{and} \add{--} \ref{fig:velocity_section_r0-062}.}
	
	\begin{figure}[!ht]
		\centering
		\includegraphics[width=.75\linewidth]{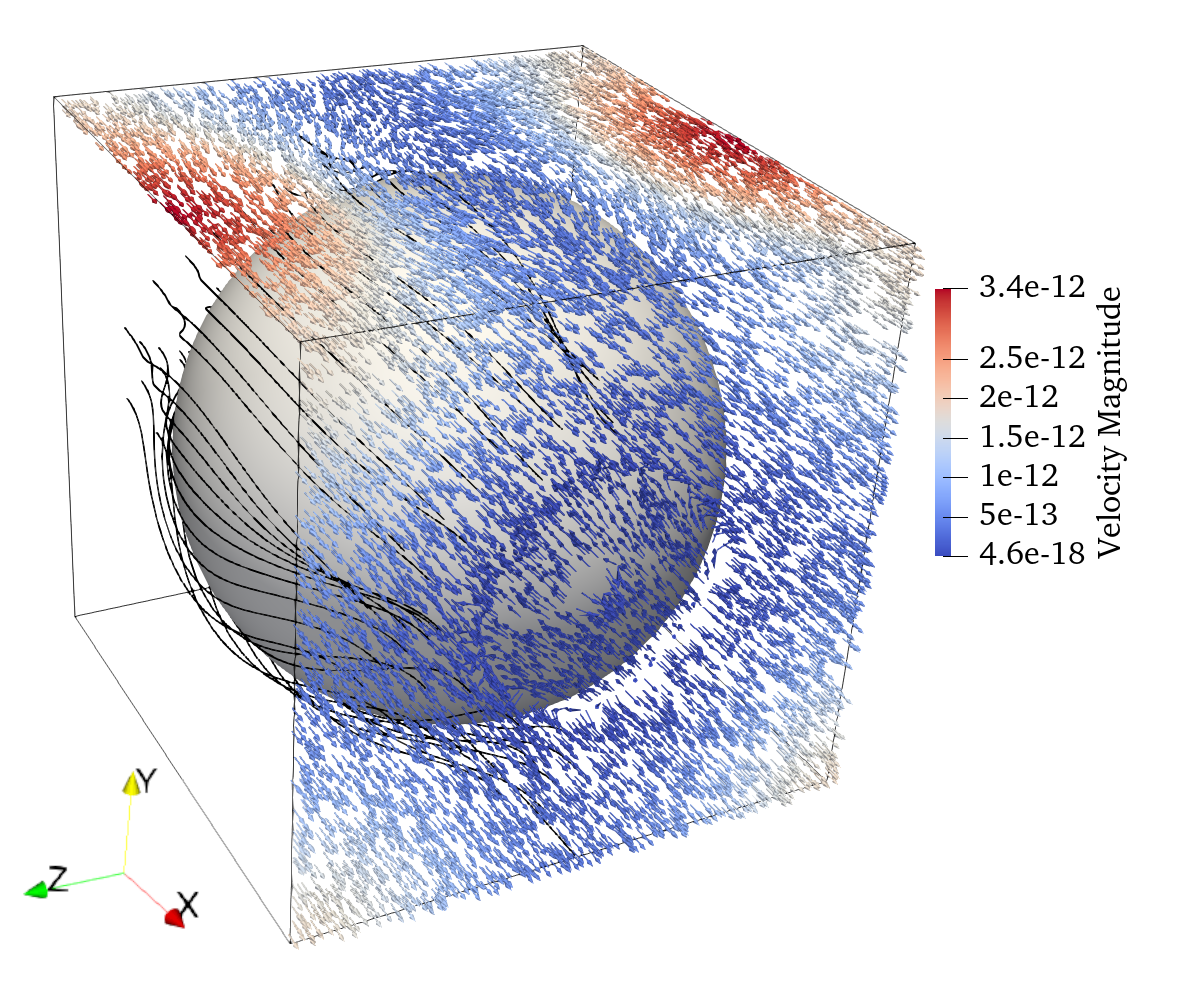}
		\hspace{.25cm}
		\includegraphics[width=.75\linewidth]{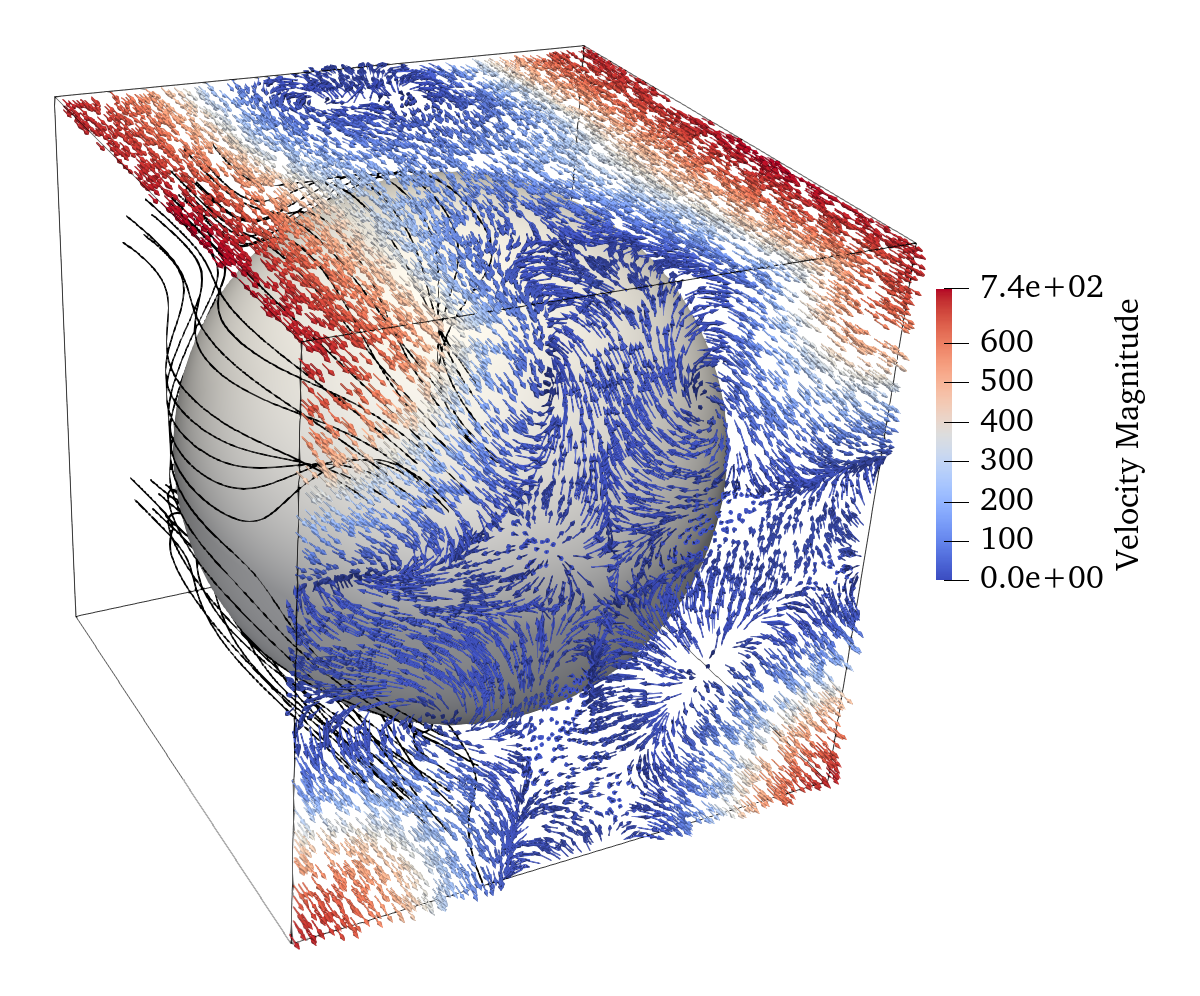}
		\caption{\add{Visualizations of the sections of the velocity field in $\phi=0.59$ SC system for $Re_k = 3.72 \cdot 10^{-14}$ (\textit{top}) and $Re_k = 8.70$ (\textit{bottom}). The gray surfaces are the obstacles' walls and black lines are chosen streamlines close to the obstacle's surface.}}
		\label{fig:velocity_section_r0-46}
	\end{figure}
	
	\begin{figure}[!ht]
		\centering
		\includegraphics[width=.75\linewidth]{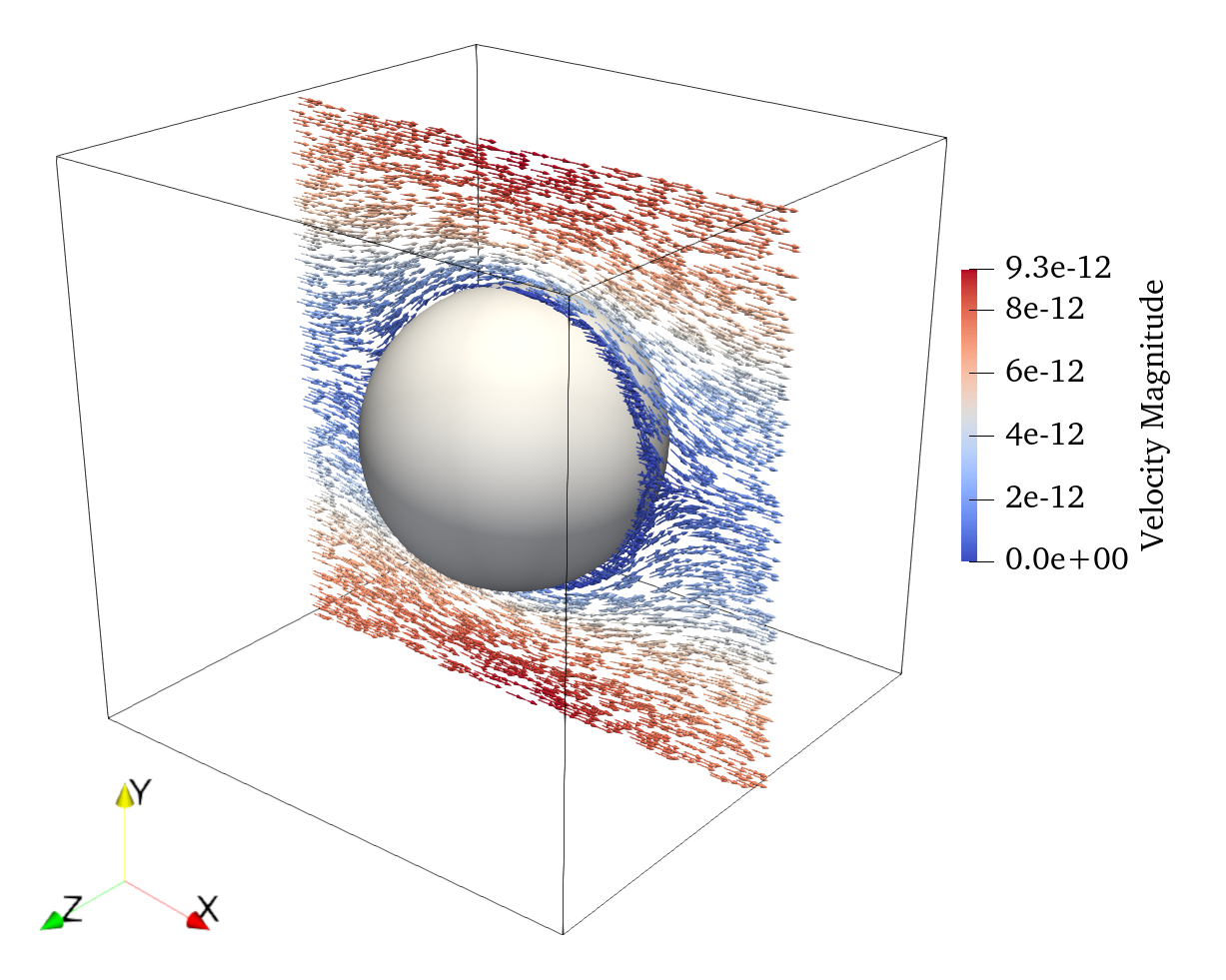}
		\hspace{.25cm}
		\includegraphics[width=.75\linewidth]{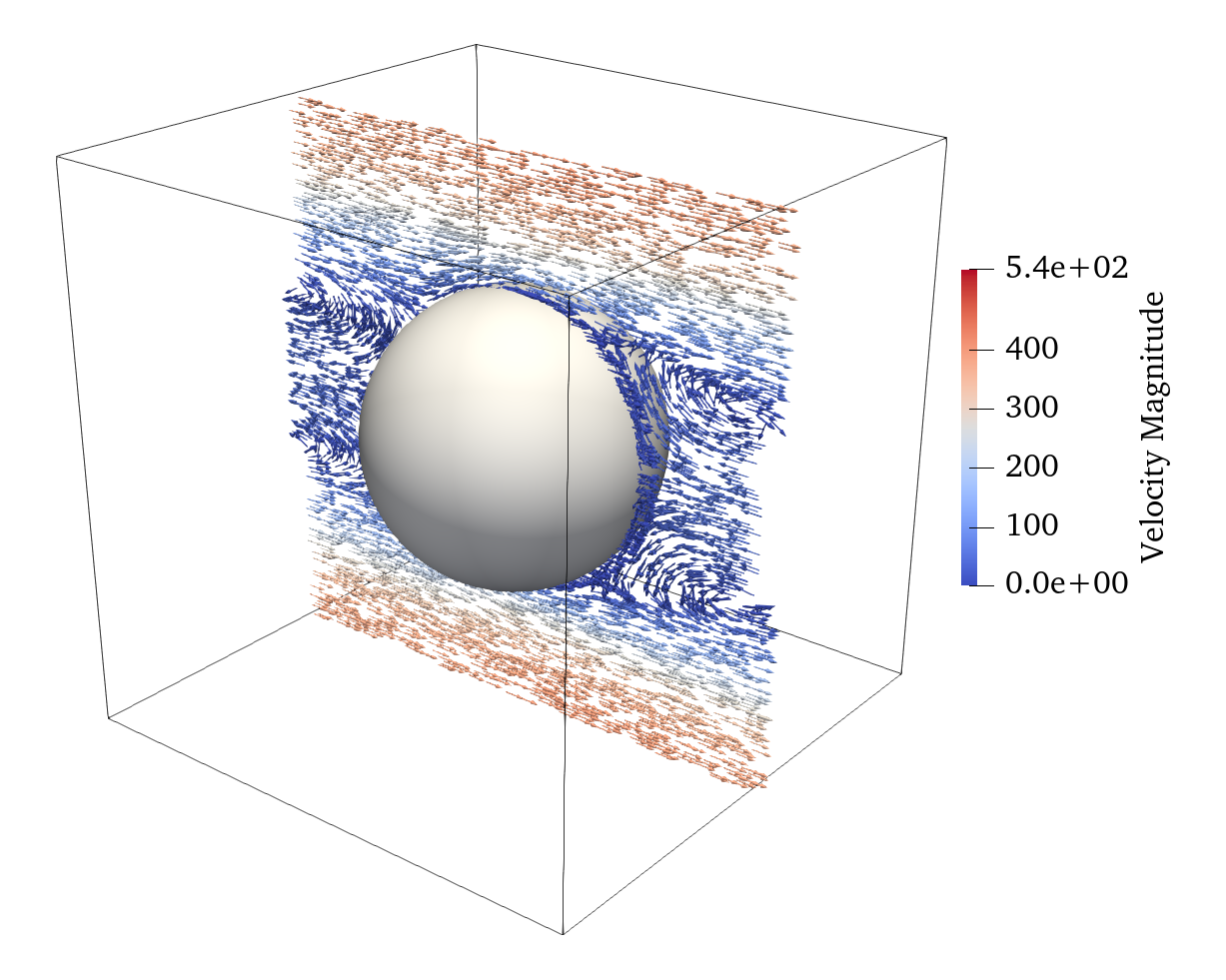}
		\caption{\add{Visualizations of the sections of the velocity field in $\phi=0.93$ SC system for $Re_k = 1.73 \cdot 10^{-12}$ (\textit{top}) and $Re_k = 2.01$ (\textit{bottom}). The gray surfaces are the obstacles' walls.}}
		\label{fig:velocity_section_r0-26}
	\end{figure}
	
	\begin{figure}[!ht]
		\centering
		\includegraphics[height=.45\linewidth]{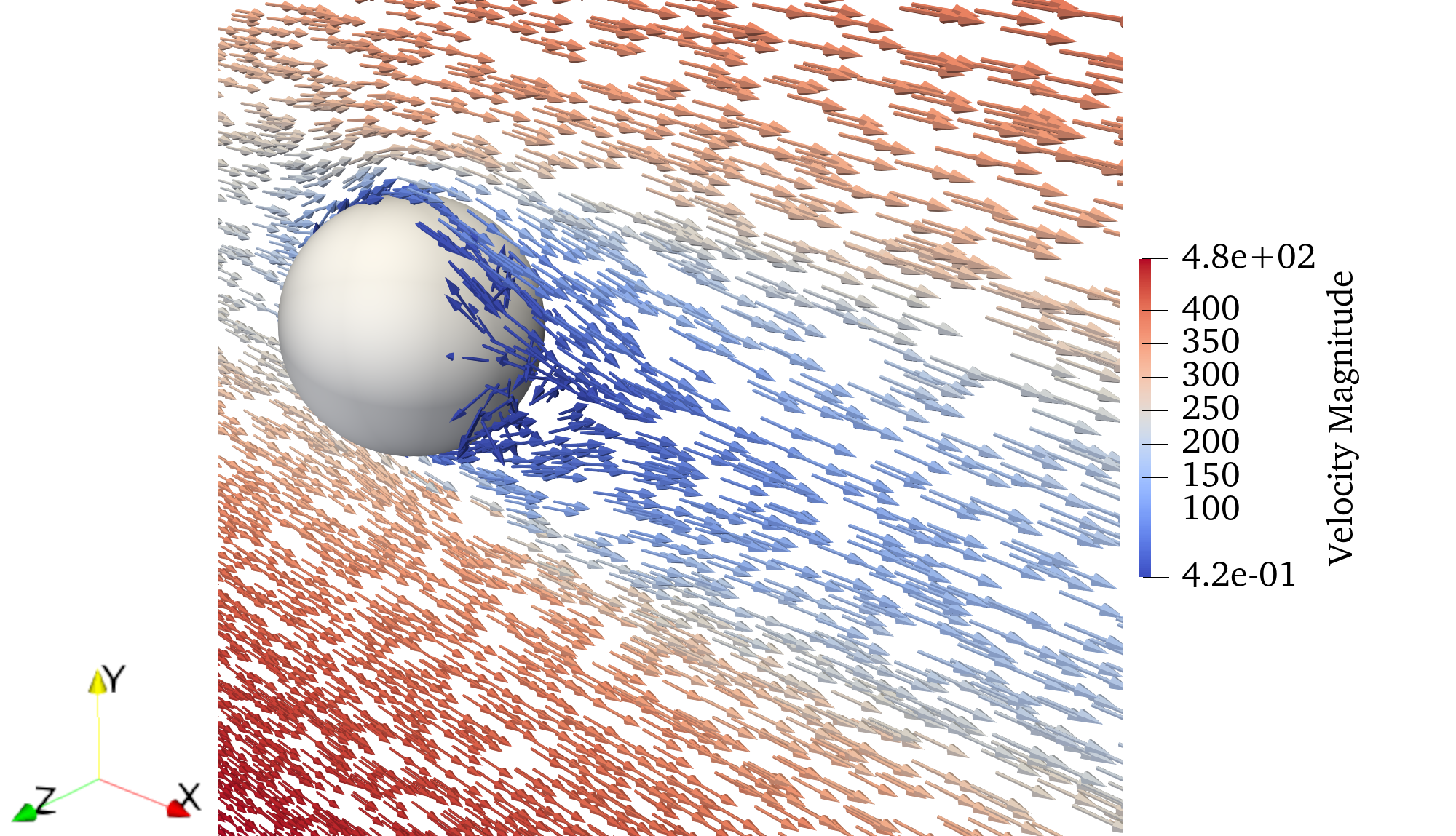}
		
		\vspace{.5cm}
		
		\includegraphics[height=.45\linewidth]{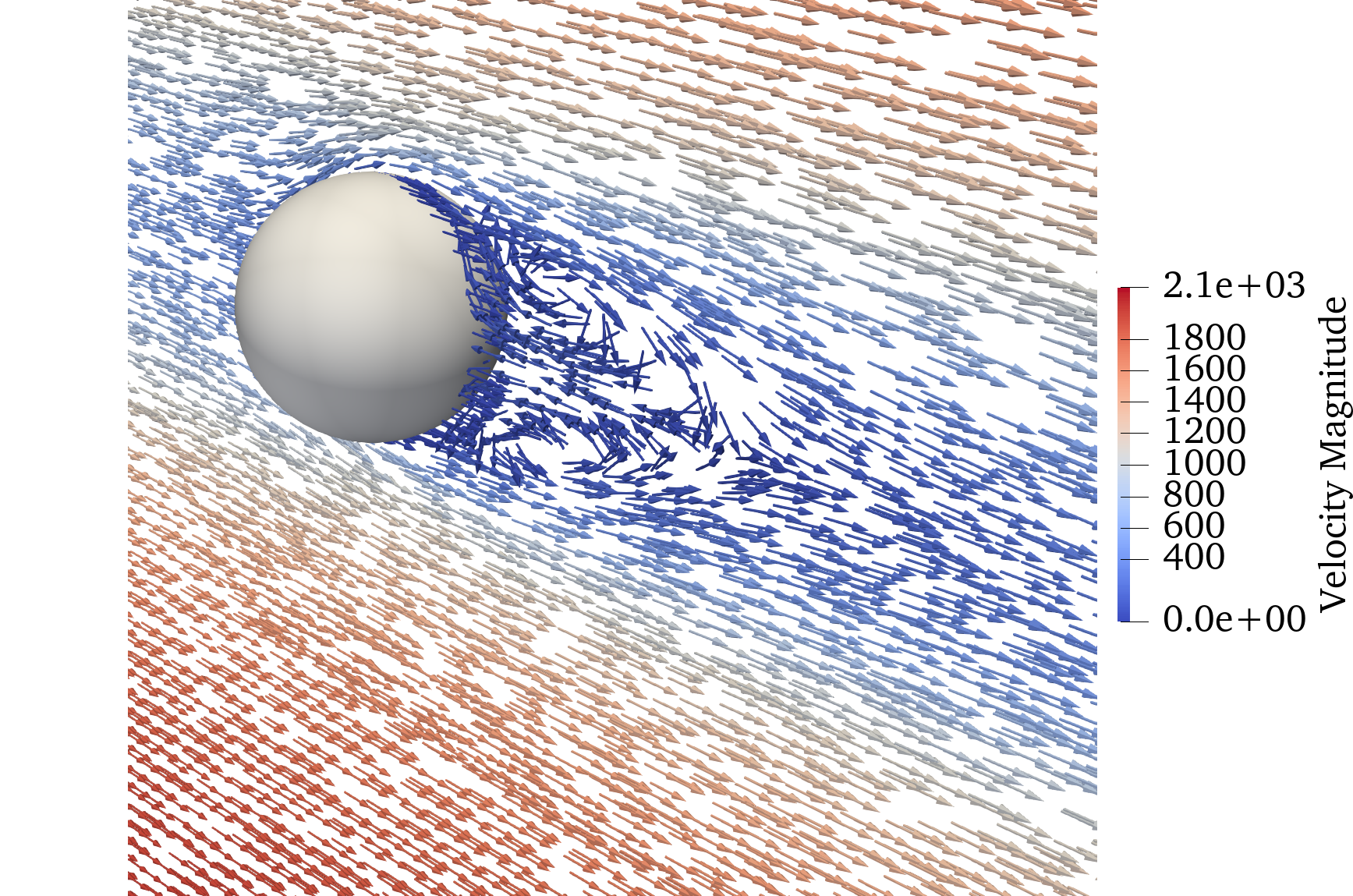}
		\caption{\add{Visualizations of the sections of the velocity field in $\phi=0.999$ SC system for $Re_k = 376$ (\textit{top}) and $Re_k = 1557$ (\textit{bottom}). The gray surfaces are the obstacles' walls.}}
		\label{fig:velocity_section_r0-062}
	\end{figure}
	
\begin{figure}[!ht]
		\centering
		\includegraphics[width=\linewidth]{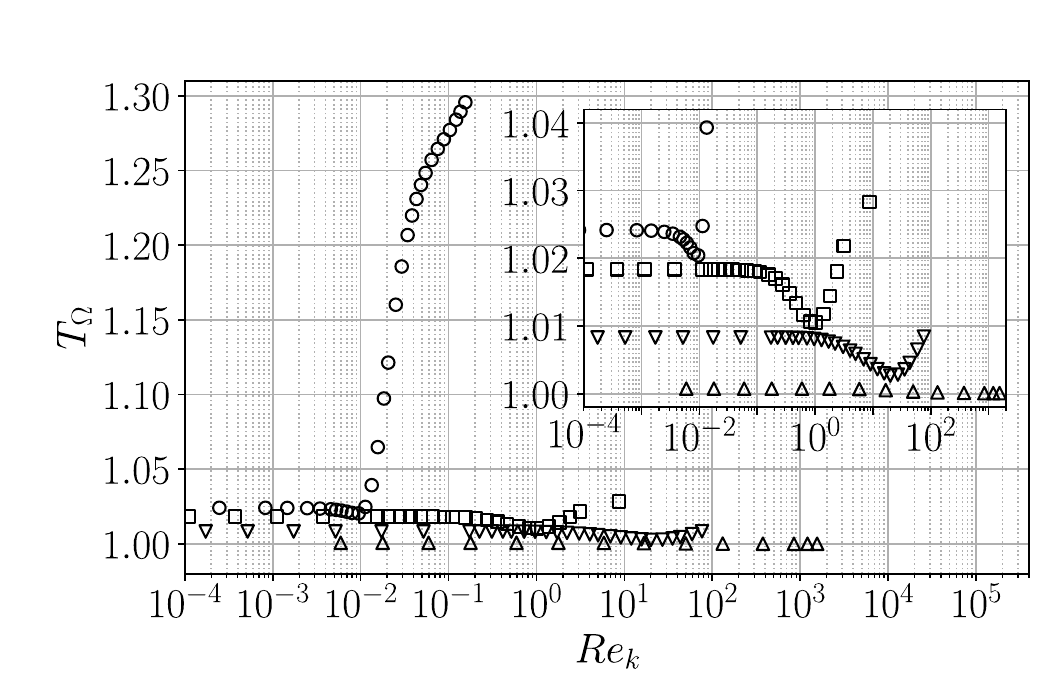}
		\includegraphics[width=\linewidth]{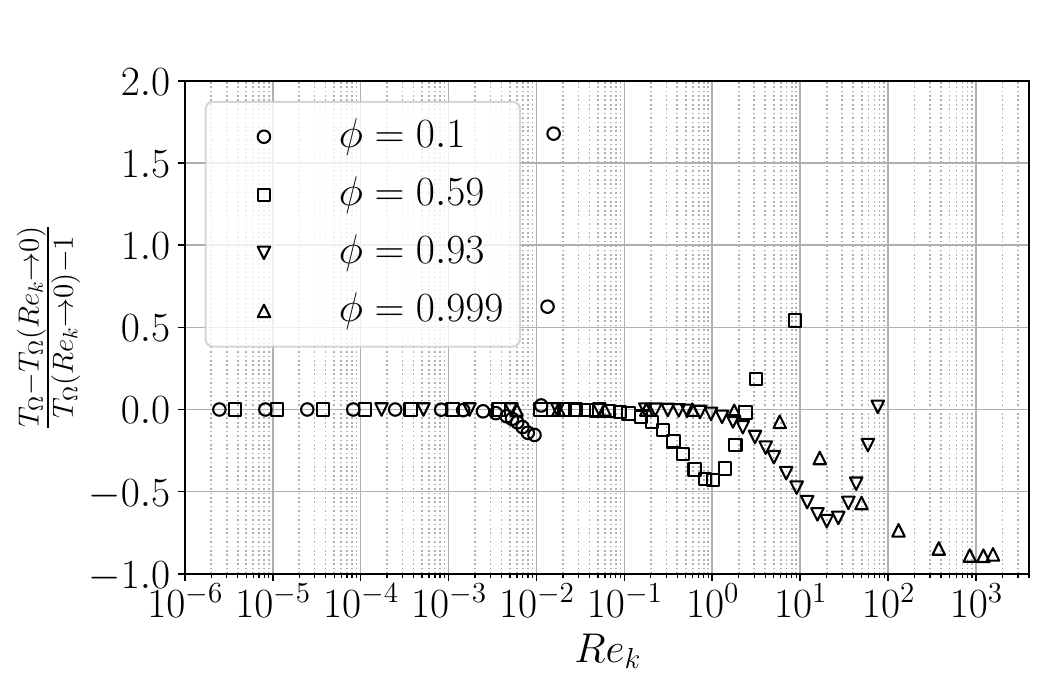}
		\caption{\add{Volume-integrated tortuosity for all SC samples. The top plot shows the raw values of $T_\Omega$ with the inset showing the same data but in different values range. The bottom plot shows the values rescaled by $T_\Omega$ at vanishingly small $Re_k$.}}
		\label{fig:TV_vs_Rek_SC}
	\end{figure}

	\add{We first look at the Reynolds number dependence of the volume-integrated tortuosity $T_\Omega$ for all SC samples in Fig.~\ref{fig:TV_vs_Rek_SC}. As the porosity increases, the value of the volume-integrated tortuosity at vanishingly small Reynolds number, $T_\Omega(Re_k \rightarrow 0)$, decreases. This is an expected result, since in this regime $T_\Omega \equiv T_s$, and a smaller spherical obstacle means fewer streamlines significantly deviate from the straight shape. Secondly, looking at the minima of the functions in the bottom plot, one sees that as the porosity decreases, the value by which $T_\Omega$ falls before reaching the minimum decreases as well, relative to $T_\Omega-1$ at $Re_k \rightarrow 0$. At the same time, when this value of the volume-integrated tortuosity drop is not normalized by $T_\Omega(Re_k \rightarrow 0)-1$, it does not behave monotonically with porosity, i.e., the largest value of the drop is observed for the $\phi=0.59$ sample (see the inset of the top plot). Smaller values of the normalized $T_\Omega$ drop in the bottom subplot suggest the sooner dominance of the recirculation zones-related phenomena over the ones related to the percolating volume. The positions of the $T_\Omega$ minima increase with porosity, which is due to the reference length $\sqrt{k_0}$ in the Reynolds number changing by orders of magnitude between the samples (cf. Eq.~\eqref{eq:Re}). When the diameter of the spherical obstacle is taken as the reference length (cf. Table~\ref{tab:re_conversion}), the profiles of $T_\Omega(Re_k)$ fall more into a single curve, see Fig.~\ref{fig:TV_vs_Rek_SC_Re_d}. On the other hand, what the choice of different reference lengths in the Reynolds number cannot change is the range of $Re$ over which $T_\Omega$ is decreasing (in terms of decades, i.e., the width of the decreasing part of $T_\Omega(Re_k)$ plot on logarithmic $Re_k$-axis). This range increases with increasing porosity. This suggests that as the porosity decreases and the geometrical confinement increases, the inertial effects appear in the system less abruptly.}
	
\begin{figure}[!ht]
		\centering
		\includegraphics[width=.75\linewidth]{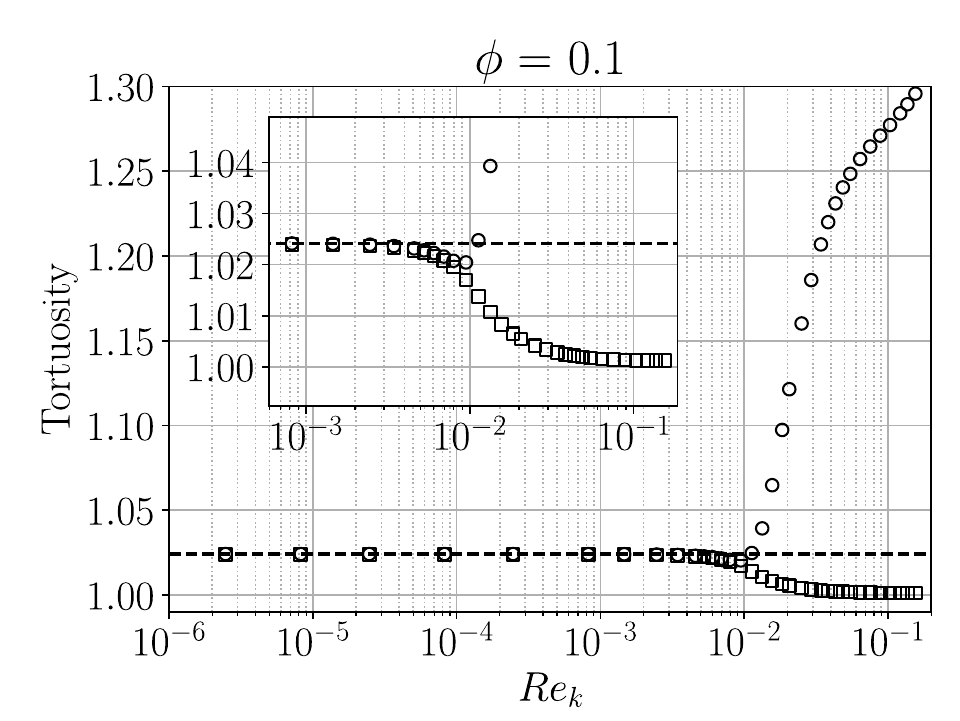}
		\includegraphics[width=.75\linewidth]{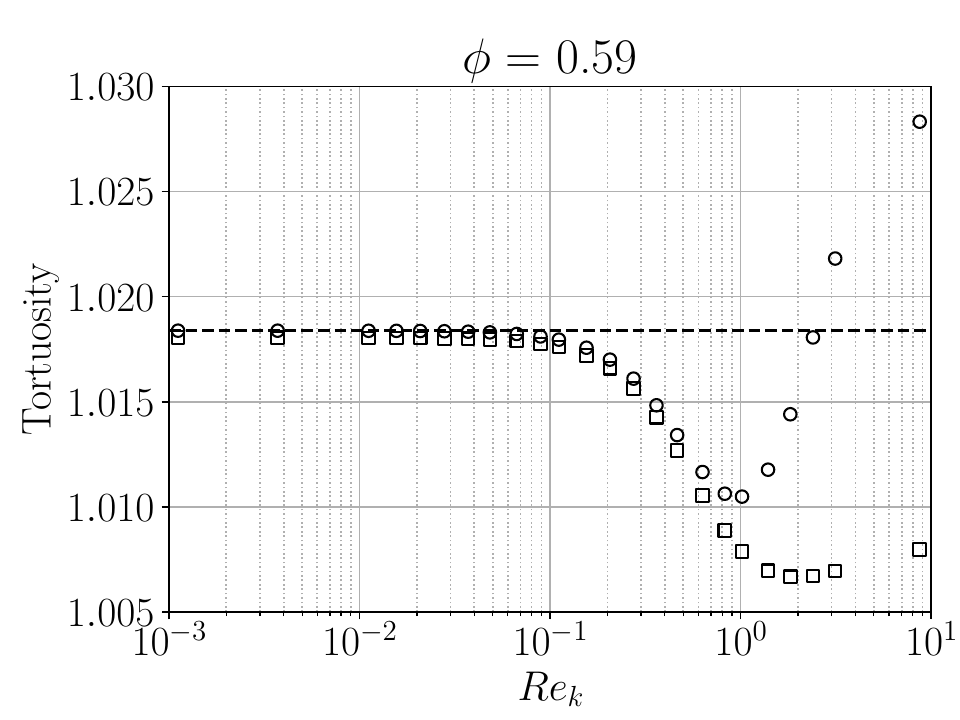}
		\includegraphics[width=.75\linewidth]{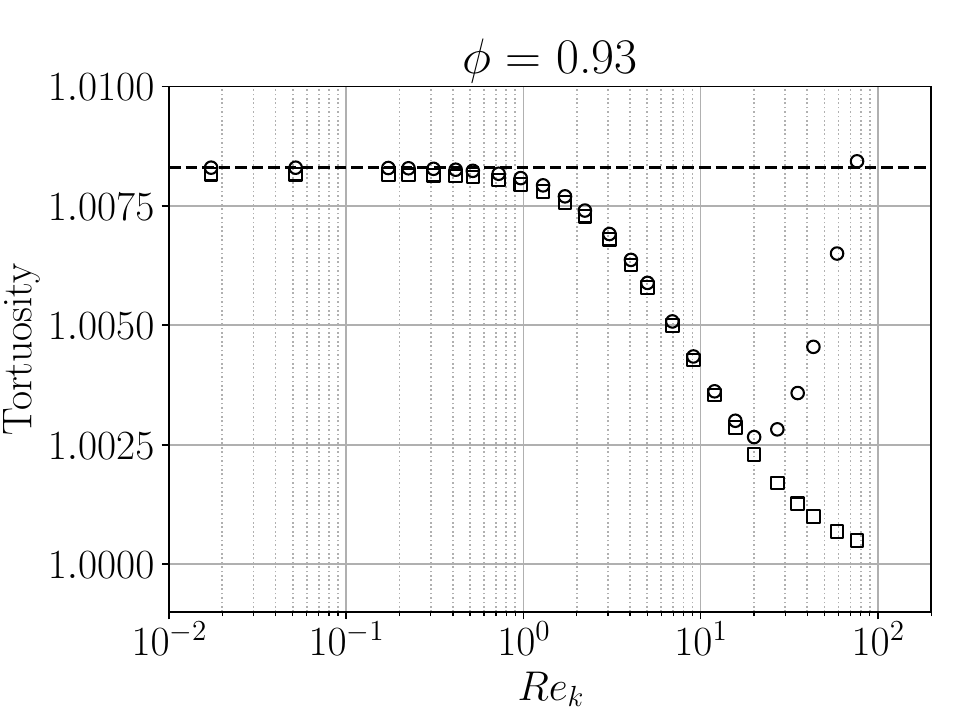}
		\includegraphics[width=.75\linewidth]{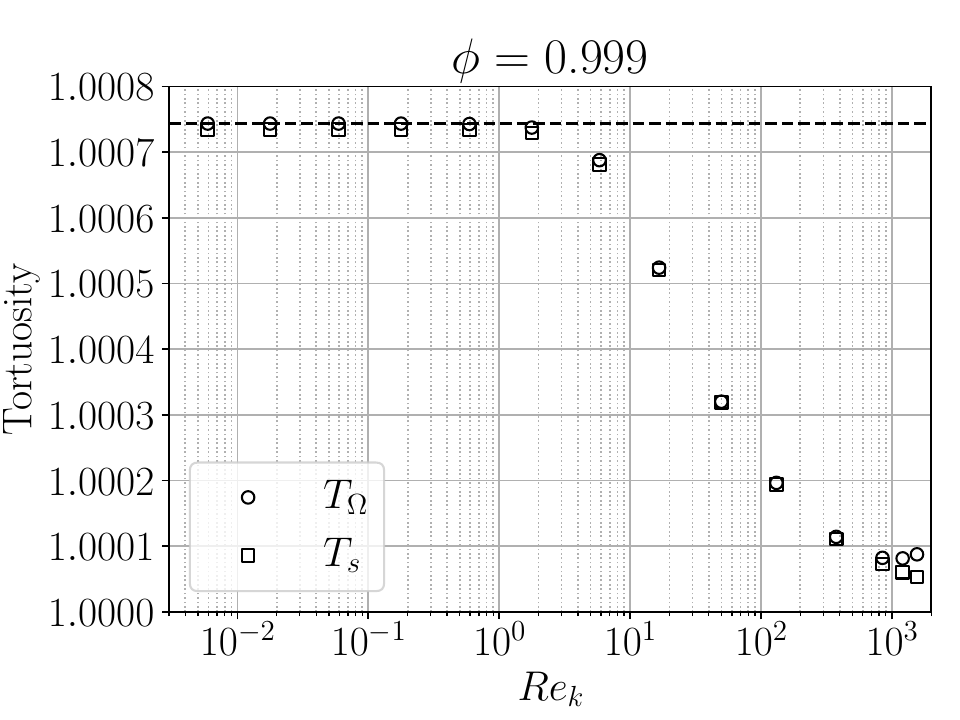}
		\caption{\add{$T_\Omega$ ({\itshape circles}) and $T_s$ ({\itshape squares}) for SC samples. The inset in the $\phi=0.1$ plot shows the magnification of the data to the range of $T_s$ values. The horizontal dashed lines represent the values of $T_\Omega$ calculated for vanishingly small $Re_k$.}}
		\label{fig:tortuosities_vs_Rek}
	\end{figure}

	\del{We continue our analysis by looking} \add{We next look} at the Reynolds number dependence of $T_\Omega$ and $T_s$ for SC samples, shown in Fig.~\ref{fig:tortuosities_vs_Rek}. In the Darcy regime ($Re_k \rightarrow 0$), the values of both tortuosities stay constant and are compliant with each other, as shown in~\cite{Duda2011}. We attribute small visible discrepancies between the two tortuosities to the errors introduced during the generation of streamlines and the identification of percolating/recirculation volumes. When the Reynolds number increases, the streamline-based tortuosity, $T_s$, generally decreases for all systems, which was observed in stochastic and periodic porous media by other authors~\cite{Sivanesapillai2014,Agnaou2017}. \add{The exception from this behavior is visible in the last decade of $Re_k$ in the case of $\phi=0.59$ sample, however even in this case the eventual increase of $T_s$ is small compared to the increase of $T_\Omega$ in the same range of Reynolds number.} On the other hand, when the volume-based tortuosity, $T_\Omega$, is concerned, the inertial effects manifest themselves differently depending on the porosity, similarly to the stochastic porous media (cf. Fig.~\ref{fig:TV-Re_sotchastic_PM}). \add{The overall tendency is that $T_\Omega$ starts to deviate significantly from $T_s$ earlier when the porosity is decreased. For instance, \del{In the lower-porosity} in the lowest-porosity} case, its value decreases slightly in the range \del{$Re_k \in [5 \cdot 10^{-4};3 \cdot 10^{-3}]$} \add{$Re_k \in [3 \cdot 10^{-3};10^{-2}]$} to later increase significantly until the steady-state regime ends. Such behavior resembles the one observed for the lower-porosity stochastic porous media in Fig.~\ref{fig:TV-Re_sotchastic_PM} and also for a real-life porous sample~\cite{Naqvi2025}. \add{At the same time,} for the \del{higher} \add{highest}-porosity sample, the volume-based tortuosity follows the streamline-based one closely and diverges from it only at $Re_k \approx 10^3$. Similarly to the \del{lower-porosity} \del{lowest-porosity} system, the initial decline of $T_s$ is followed by its rise, \add{however} to a much smaller extent, \del{however}. This is in line with the behavior of $T_\Omega$ for the higher-porosity stochastic porous media shown in Fig.~\ref{fig:TV-Re_sotchastic_PM}, and in~\cite{Sniezek2024}.
	
	The \del{strictly} decreasing behavior of the streamline-based tortuosity can be attributed to the straightening of the percolating streamlines, which follows the growth of the recirculation zones volumes in the cavities and obstacles wakes. \add{The rise of $T_s$ for $Re_k>2$ in the case of $\phi=0.59$ sample indicates that perhaps a non-trivial reorganization of the percolating streamlines shapes takes place there, reminiscent of those observed in porous media of more complex geometries~\cite{Naqvi2025,Arbabi2024}.} The volume-integrated tortuosity exhibits a more complex behavior, dependent on the sample's porosity. Eq.~\eqref{eq:TV_split_reduced_using_Ts} shows that the difference between $T_\Omega$ and $T_s$ is related to the integral of velocity magnitude inside the recirculation zones. Thus, we will further investigate the fall and the rise of the volume-based tortuosity in terms of the volume of the recirculation zones and the field of the velocity magnitude therein.
	
	\bigskip
	
	\subsection{Origin of the \del{$T_\Omega$} decrease and rise \add{of the volume-integrated tortuoisty} -- volume of the recirculation zones}
	
\begin{figure}[!ht]
		\centering
		\includegraphics[height=.45\linewidth]{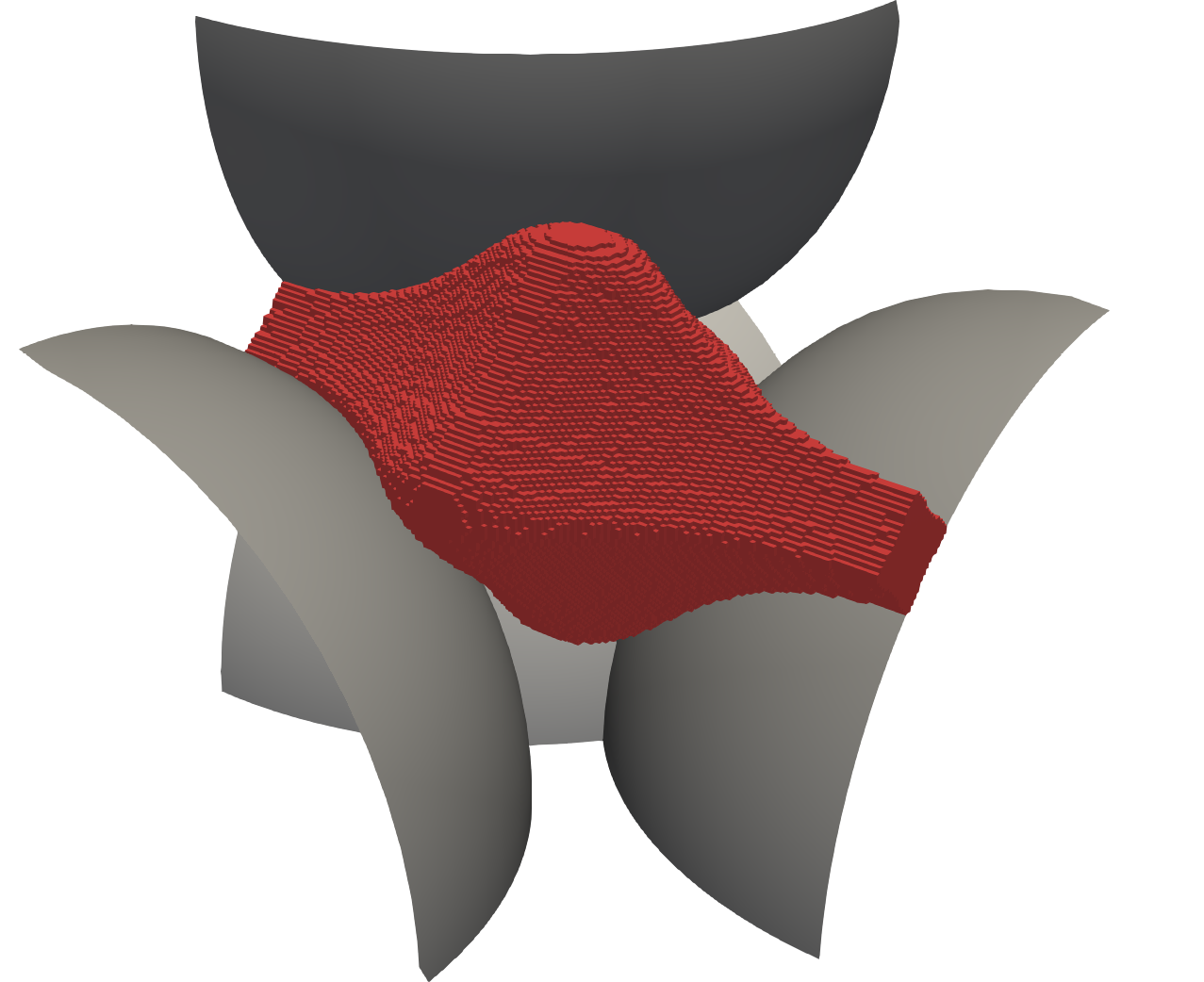}
		
		\vspace{.5cm}
		
		\includegraphics[height=.45\linewidth]{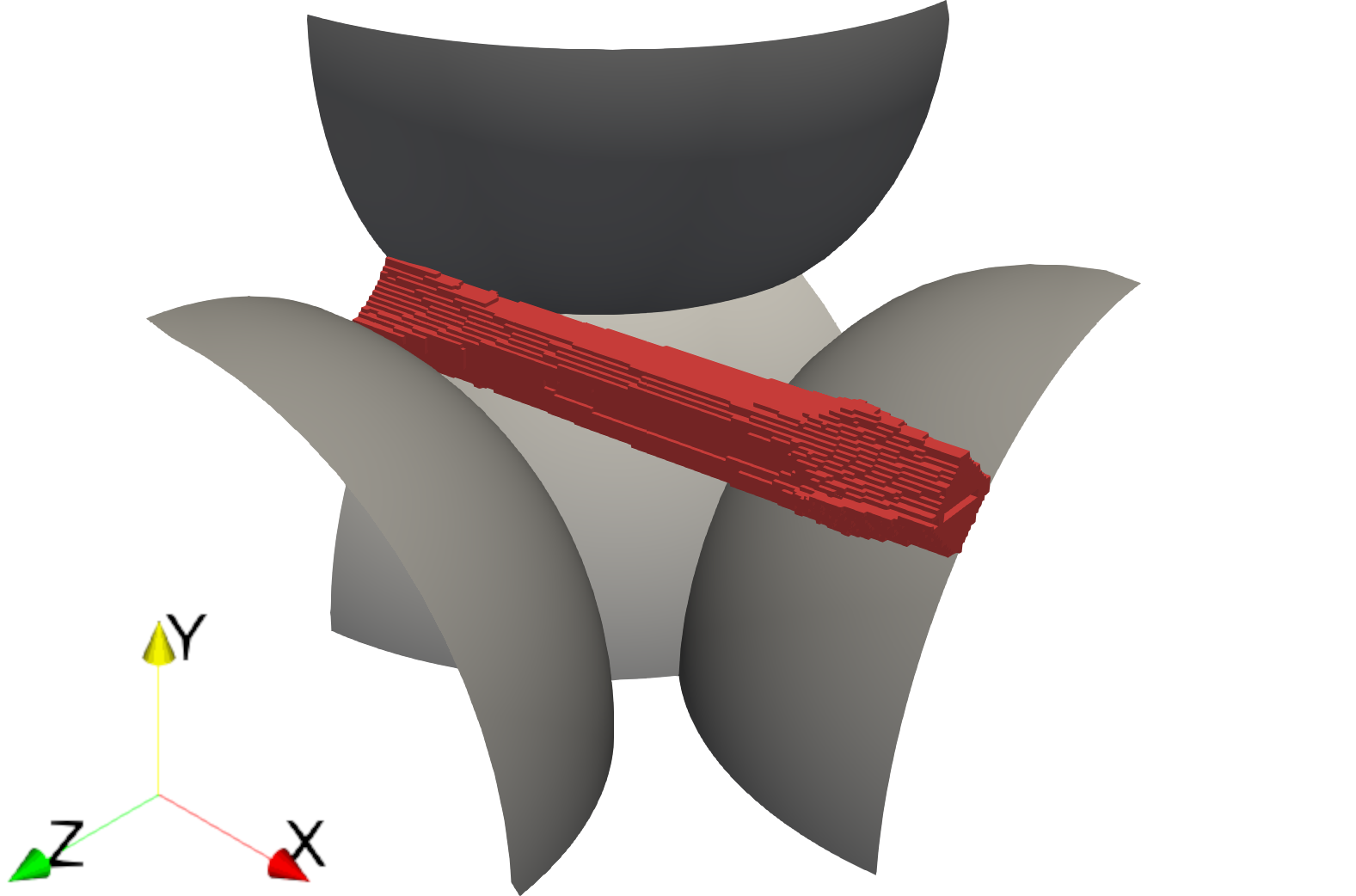}
		\caption{Visualizations of the percolating volumes in $\phi=0.1$ SC system for the same values of $Re_k$ as in Fig.~\ref{fig:velocity_section_r0-6526}. The gray surfaces are the walls of the obstacles, and only a few chosen walls are shown for clarity. The visualizations use the histograms from which the percolating volumes were found. \del{We show only the percolating volumes for clarity}.}
		\label{fig:perVol_viz_r0-6526}
	\end{figure}

\begin{figure}[!ht]
		\centering
		\includegraphics[height=.5\linewidth]{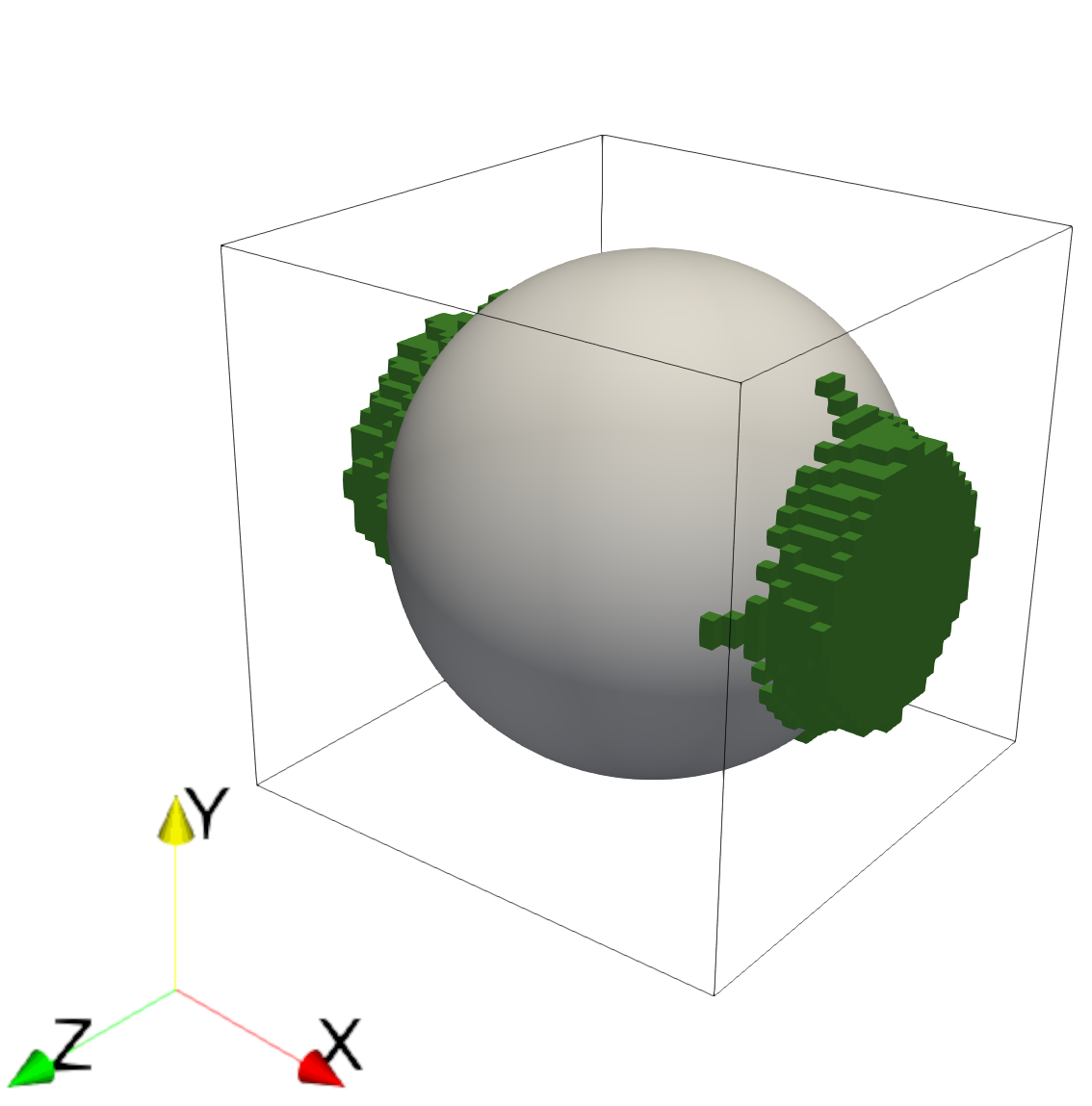}
		\includegraphics[height=.5\linewidth]{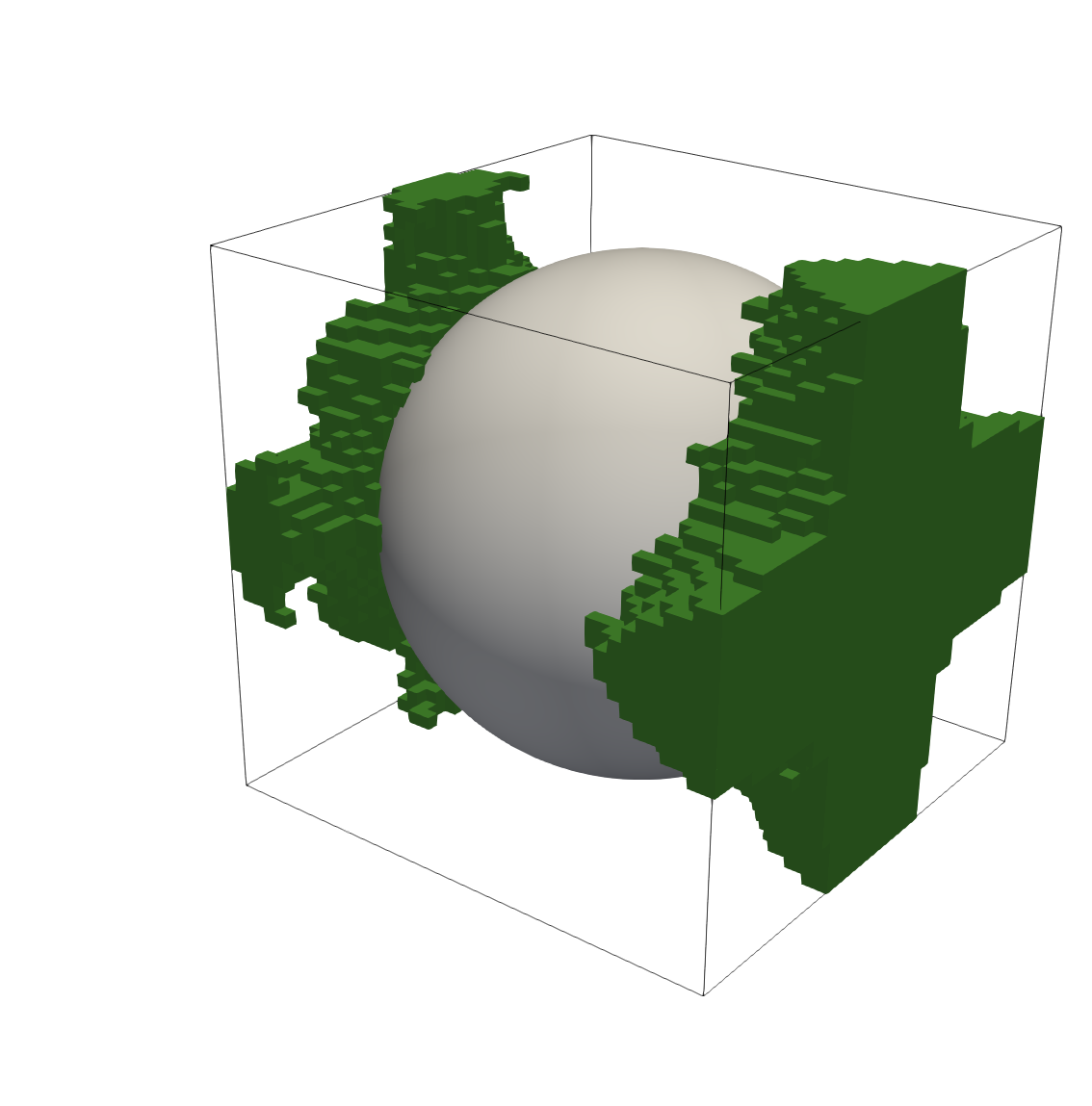}
		\caption{\add{Visualizations of the recirculating volumes for $\phi=0.59$ SC system for the same values of $Re_k$ as in Fig.~\ref{fig:velocity_section_r0-46}. The gray surfaces are the walls of the obstacles. The visualizations use the histograms from which the recirculation volumes were found.}}
		\label{fig:perVol_viz_r0-46}
	\end{figure}

\begin{figure}[!ht]
		\centering
		\includegraphics[height=.5\linewidth]{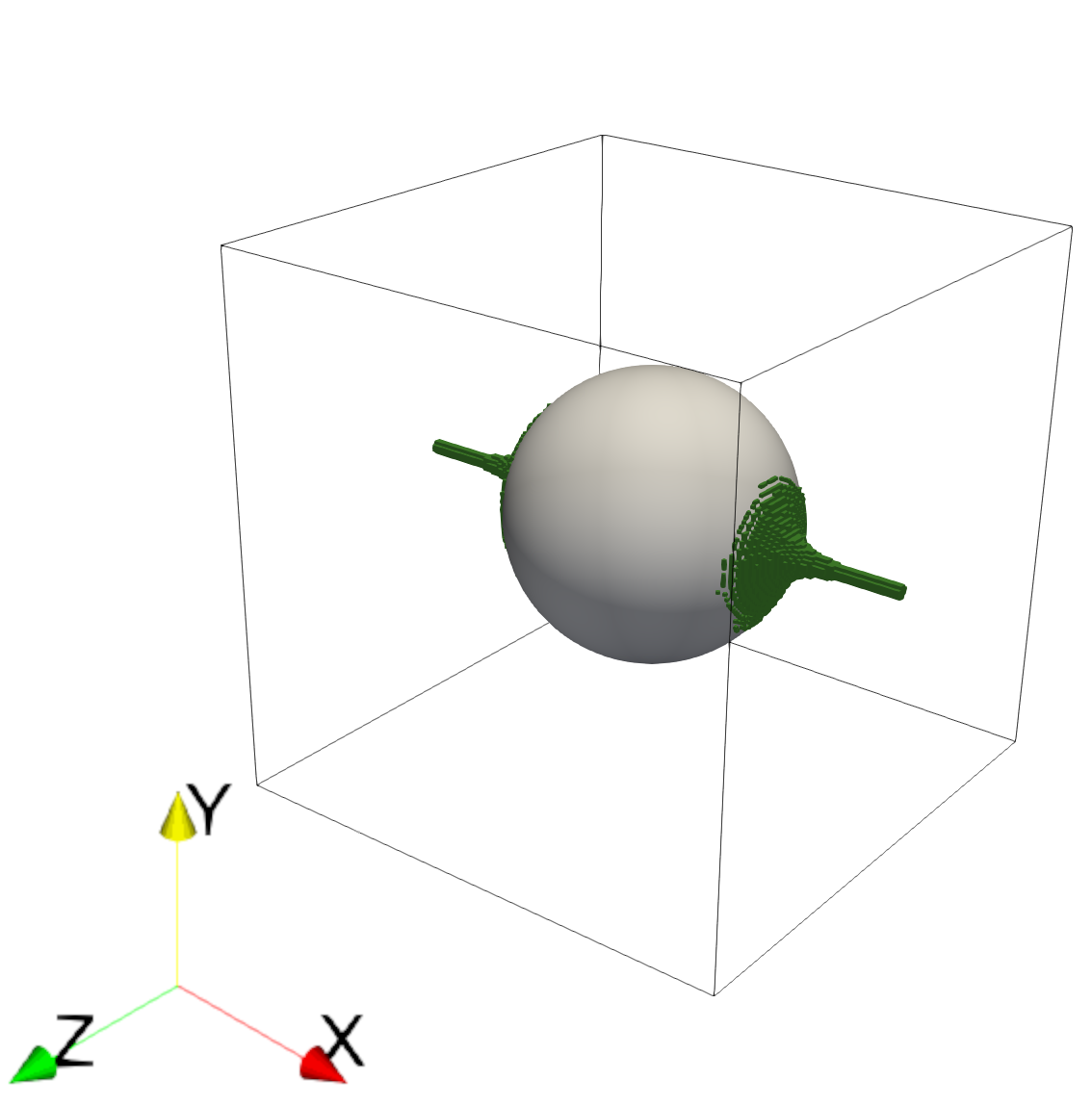}
		\includegraphics[height=.5\linewidth]{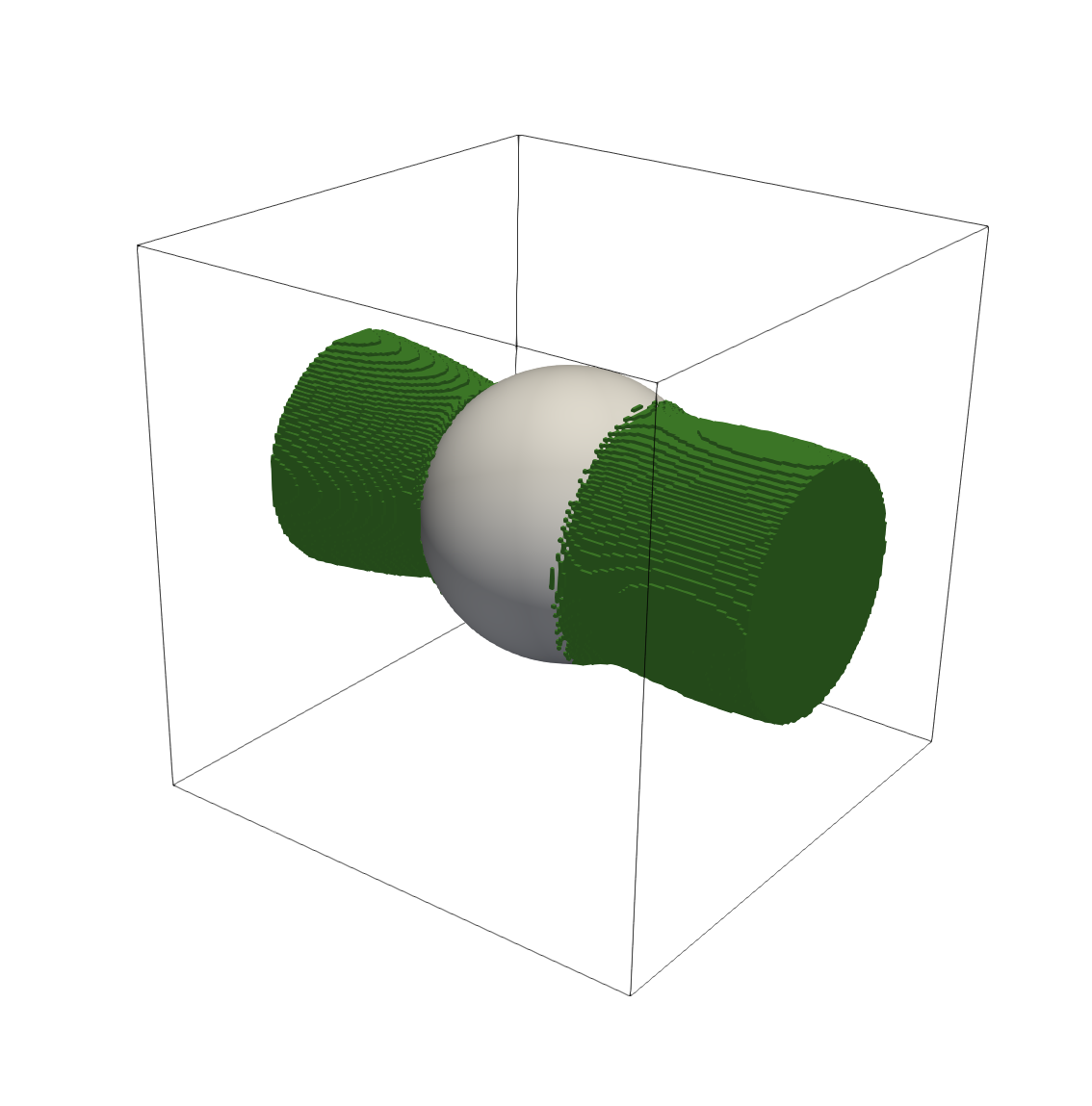}
		\caption{\add{Visualizations of the recirculating volumes for $\phi=0.93$ SC system for the same values of $Re_k$ as in Fig.~\ref{fig:velocity_section_r0-26}. The gray surfaces are the walls of the obstacles. The visualizations use the histograms from which the recirculation volumes were found.}}
		\label{fig:perVol_viz_r0-26}
	\end{figure}

\begin{figure}[!ht]
		\centering
		\includegraphics[height=.5\linewidth]{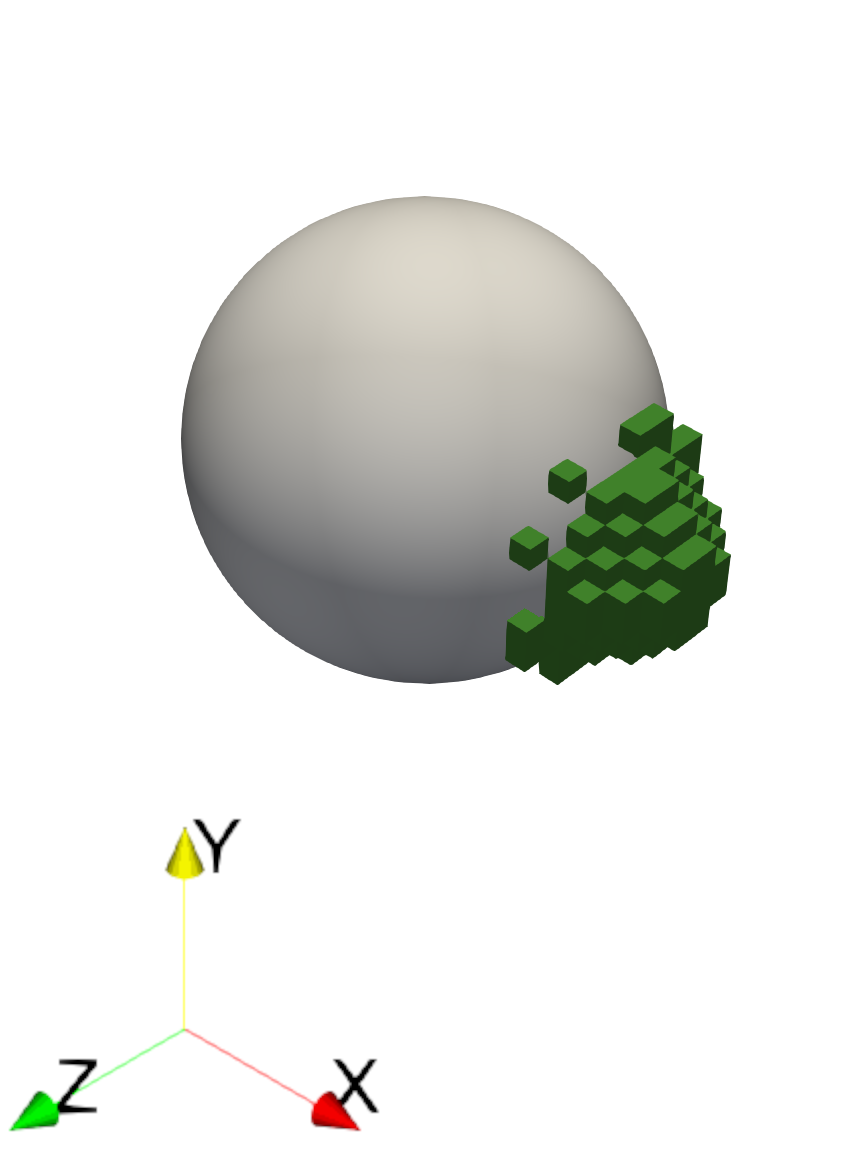}
		\includegraphics[height=.5\linewidth]{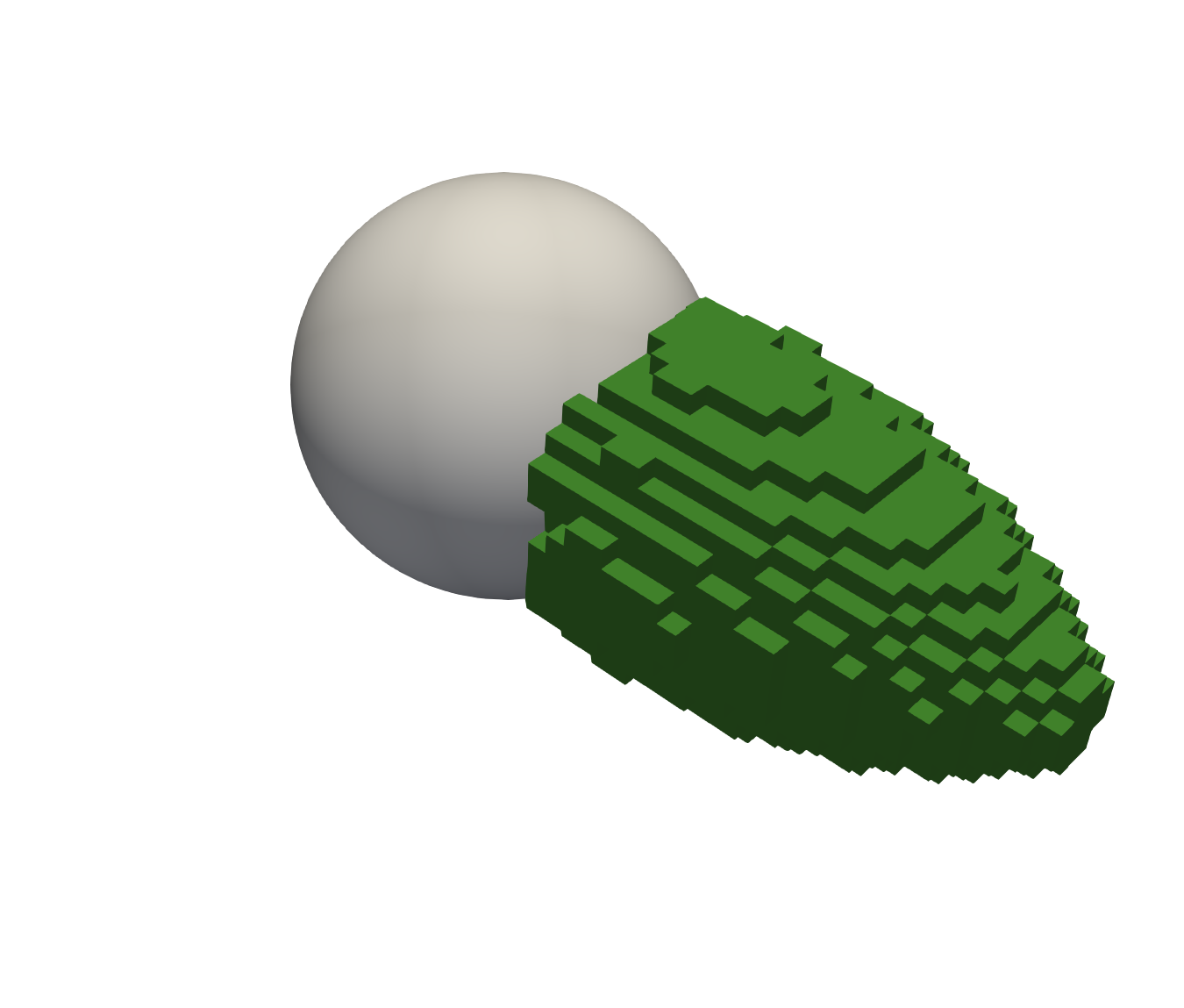}
		\caption{Visualizations of the recirculating volumes for $\phi=0.999$ SC system for the same values of $Re_k$ as in Fig.~\ref{fig:velocity_section_r0-062}. The gray surfaces are the walls of the obstacles. The visualizations use the histograms from which the recirculation volumes were found. \del{We show only the recirculation volumes for clarity}.}
		\label{fig:perVol_viz_r0-062}
	\end{figure}

	Let us \del{first} focus on the numerator of the recirculation volume-related term from Eq.~\eqref{eq:TV_split_reduced_using_Ts}, $I[V]_v / I[V_0]_p$. For convenience, we divide it by the total fluid volume, $I[1]$, i.e.
	\begin{equation}\label{eq:TV_vortex_contributions}
		\frac{I[V]_v}{I[1]} = \frac{I[1]_v}{I[1]} \cdot \frac{I[V]_v}{I[1]_v}
	\end{equation}
	where the first term on the right-hand side represents the vortex volume normalized by the total volume of fluid, and the other one -- the total velocity magnitude per unit recirculation zone volume. We consider each of those two components separately. Figs.~\ref{fig:perVol_viz_r0-6526} \del{and} \add{--} \ref{fig:perVol_viz_r0-062} show the visualization of the percolating volumes \del{and} \add{or} recirculation zones for \del{low- and high-porosity} SC samples, \del{respectively,} for two different Reynolds numbers each. In \del{both} \add{all} systems, as $Re_k$ rises, the recirculation zones grow in volume. \del{which in} \add{In} the case of \del{the low-porosity} \add{$\phi=0.1$} system \add{(Fig.~\ref{fig:perVol_viz_r0-6526})}, \add{it} is visible through the percolating volume shrinking to a straight jet connecting the inlet and the outlet. \add{$\phi=0.59$ system behaves similarly (Fig.~\ref{fig:perVol_viz_r0-46}), although because there is more space between the obstacles in the neighboring periodic cells, the high-$Re_k$ streamlines still feature a slight curvature (cf. Fig.~\ref{fig:velocity_section_r0-46}). In the two higher-porosity systems, there is no recirculation in the obstacles' wakes for low $Re_k$, but with the increasing velocity they emerge, and in the case of $\phi=0.93$ sample (Fig.~\ref{fig:perVol_viz_r0-26}) they fill the whole fluid volume between the spheres in the streamwise direction at higher $Re_k$. We note that for low $Re_k$ in $\phi=0.26$ sample, a thin recirculation zone around the $y=z=0.5$ line is visible. However, the later examination of the velocity fields (cf. Fig.~\ref{fig:velocity_section_r0-26}) confirmed that the recirculation zone is not present there, and its appearance in the histograms is most likely due to the numerical errors introduced in the streamlines generation.}
	
\begin{figure}[!ht]
		\centering
		\includegraphics[height=.55\linewidth]{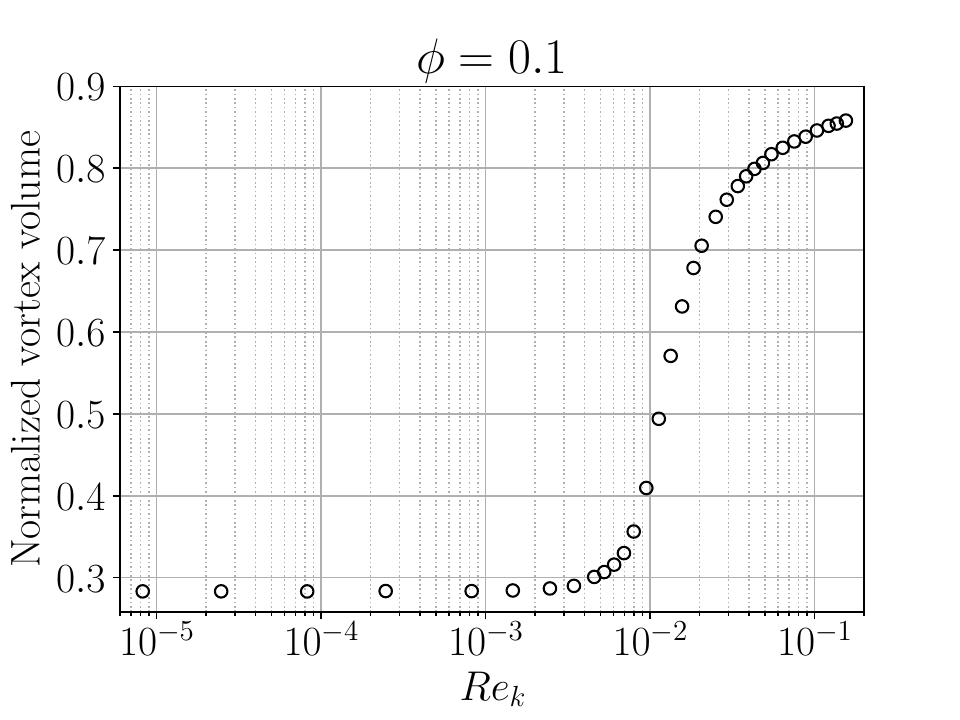}
		\includegraphics[height=.55\linewidth]{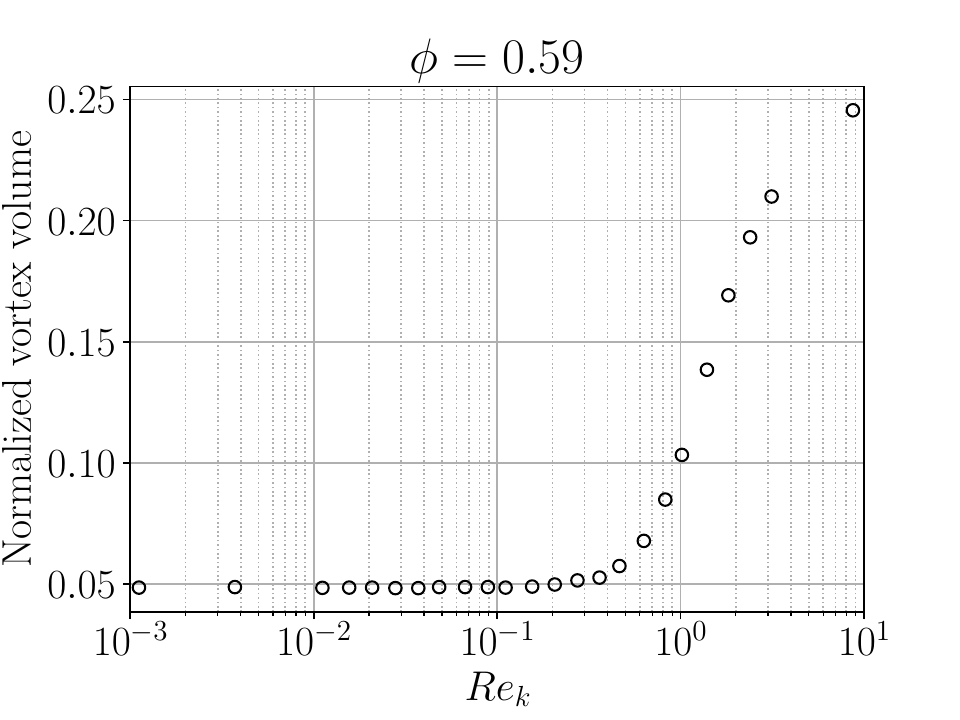}
		\includegraphics[height=.55\linewidth]{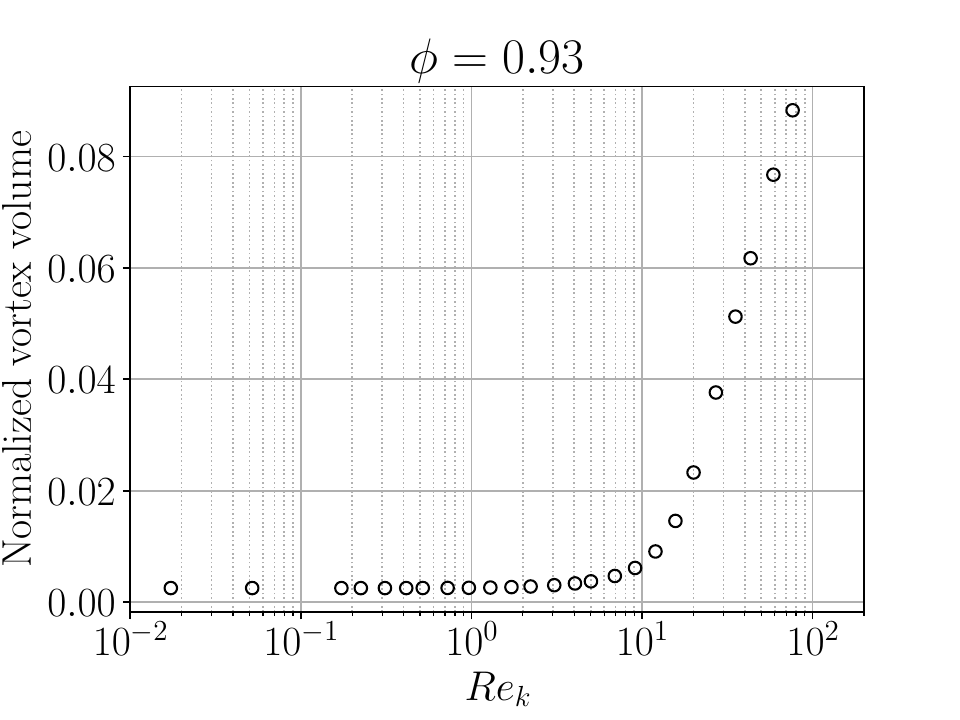}
		\includegraphics[height=.55\linewidth]{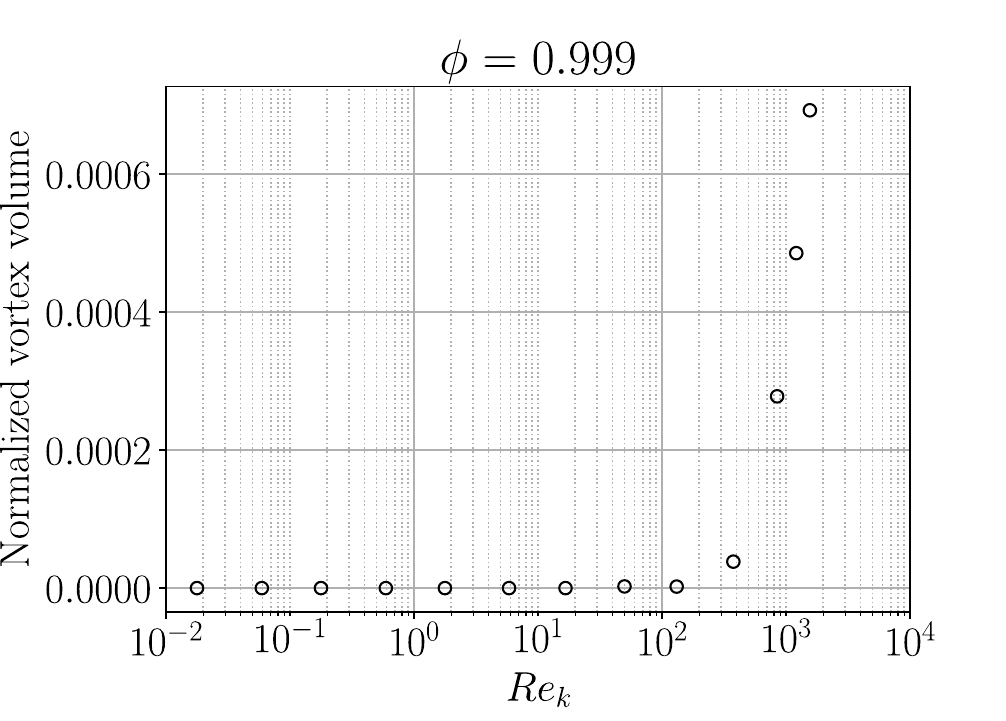}
		\caption{\add{Relative vortex volumes, $I[1]_v / I[1]$, for the \del{two} studied SC systems.}}
		\label{fig1:vortex_volume_relative}
	\end{figure}

	To quantify the changes in the size of the recirculation zones, we plot their volume, $I[1]_v$, normalized by the total volume of fluid, $I[1]$, in Fig.~\ref{fig1:vortex_volume_relative}. In \del{the low-porosity} \add{$\phi=0.1$} system, there already exist recirculation zones in the Darcy regime (in the channels perpendicular to the main flow, cf. top visualization in Fig.~\ref{fig:velocity_section_r0-6526}). The increase in their volume starts at \del{$Re_k \approx 5 \cdot 10^{-4}$} \add{$Re_k \approx 2 \cdot 10^{-3}$}, which corresponds to the start of $T_\Omega$ decrease in Fig.~\ref{fig:tortuosities_vs_Rek}. The vortex volume then continues to grow until the end of the steady state regime, with significant slowdown for \del{$Re_k > 10^{-2}$} \add{$Re_k > 4 \cdot 10^{-2}$}, the value at which the slope of the $T_\Omega$ rise in Fig.~\ref{fig:tortuosities_vs_Rek} decreases. This reduction in the recirculation volume growth pace indicates the strong geometrical confinement for the vortex development. We highlight that at the highest $Re_k$, there already exist secondary vortices in the perpendicular channels (cf. bottom visualization in Fig.~\ref{fig:velocity_section_r0-6526}). \del{The total increase in the volume of the recirculation zone is about 56\% of the fluid volume, or -- equivalently -- $0.56 \cdot 0.1 / 0.9 \approx 0.062$ of the volume of the solid obstacles.} \add{The behavior of the vortex volume and $T_\Omega$ in the $\phi=0.59$ sample is similar, with the start of rise of the vortices volume and fall of the tortuosity at $Re_k=10^{-1}$, and a visible decrease of the slope in both quantities visible for the highest-$Re_k$ case.}
	
	On the other hand, in the \add{two } \del{high-porosity} \add{higher-porosity} system\add{s}, there is no recirculation in the Darcy regime. \add{Contrary to the two lower-porosity samples, there is no visible decrease in the slope of the recirculation zone volume growth in the highest Reynolds numbers. This might be caused by a lower geometrical confinement of the recirculation volumes when the porosity is high}. \add{For $\phi=0.93$ sample, the growth of the recirculation zones volume starts at about $Re_k=2$, slightly after the start of $T_\Omega$ decrease. We note once again that for this sample, the non-zero value of the vortex volume for $Re_k \rightarrow 0$ in Fig.~\ref{fig1:vortex_volume_relative} is caused by numerical errors. For the highest-porosity sample, the} increase of the recirculation volume size starts at $Re_k \approx 2 \cdot 10^2$ (cf. top visualization in Fig.~\ref{fig:velocity_section_r0-062}), which is about two decades later than the visible start of $T_\Omega$ decrease for this system in Fig.~\ref{fig:tortuosities_vs_Rek}. This discrepancy can be related to the `formation of inertial cores'~\cite{Dybbs1984} when the inertia already starts to change the velocity field, although no separation yet occurs. As the recirculation volume starts to grow, it does so with an increasing slope until the end of the steady-state regime (cf. bottom visualization in Fig.~\ref{fig:velocity_section_r0-062}). \del{The increase in the volume of the recirculation zone in this case is about $0.63$ of the solid obstacle volume.} The growth of the recirculation zones in the case of \del{high-porosity} \add{the highest-porosity} SC systems was highlighted by other authors~\cite{Hill2001} as the reason for lower drag compared to the flow over a single obstacle.
	
	\add{Table.~\ref{tab:recirculation_volume_growth} shows the difference between the volume of the recirculation zone in the highest and lowest investigated $Re_k$ for each SC sample. While the difference relative to the fluid volume decreases with increasing porosity, it reaches a maximum in the intermediate porosity when calculated relative to the volume of the obstacle.}
	
	Finally, we note that as the considered SC systems have a regular and simple geometry, phenomena like the non-strictly-monotonic growth of the volume of the recirculation zones~\cite{Arbabi2024} related to the instantaneous reorganization of the percolating flow paths are not visible here. Nevertheless, the start of the \add{growth of the } recirculation zones volume \del{growth} can still be related to the deviation from Darcy's law for \add{the lowest-porosity system} (compare the data for $\phi=0.1$ in Figs.~\ref{fig1:vortex_volume_relative} and \ref{fig:darcy}). \del{Interestingly, no high-wavenumber oscillations can be seen in the values of $T_\Omega(Re_k)$ in the case of the stochastic samples (cf. Fig.~\ref{fig:TV-Re_sotchastic_PM}), which suggests that the reorganization of the percolating flow paths might not impact the velocity integrals of the volume-integrated tortuosity.} \add{Interestingly, no oscillations can be seen as well in the values of $\rho^-$~\cite{Naqvi2025,Sniezek2024} -- the fraction of fluid volume with negative $V_0$ -- in the case of the stochastic samples (results not shown, but are available in the repository, see Sec.~\ref{sec:data_repo}), which suggests that their geometry might still not be complex enough for the spontaneous reorganization of the percolating flow paths to occur.}
	
	\begin{table}
		{\color{changecolor}
			\begin{tabular}{lcc}
				\multicolumn{3}{c}{$I[1]_v(Re_{k,\text{max}}) - I[1]_v(Re_k \rightarrow 0)$} \\
				\hline
				\hline
				$\phi$ & \begin{tabular}{c}Rel. to fluid \\ volume $[\%]$\end{tabular} & \begin{tabular}{c}Rel. to obstacle's \\ volume $[\%]$\end{tabular} \\
				\hline
				$0.1$ & 56 & 6.2\\
				$0.59$ & 20 & 28\\
				$0.93$ & 8.8 & 117\\
				$0.999$ & 0.07 & 63\\
			\end{tabular}
			\caption{\add{The difference between the volume of the recirculation zone at the minimal and maximal investigated Reynolds number for SC samples. Data are given relative to the volume of fluid and the solid obstacle.}}
			\label{tab:recirculation_volume_growth}
		}
	\end{table}
	
	\subsection{Origin of the \del{$T_\Omega$} decrease and rise \add{of the volume-integrated tortuoisty} -- energy confinement in the recirculation zones}
	
\begin{figure}[!ht]
		\centering
		\includegraphics[height=.55\linewidth]{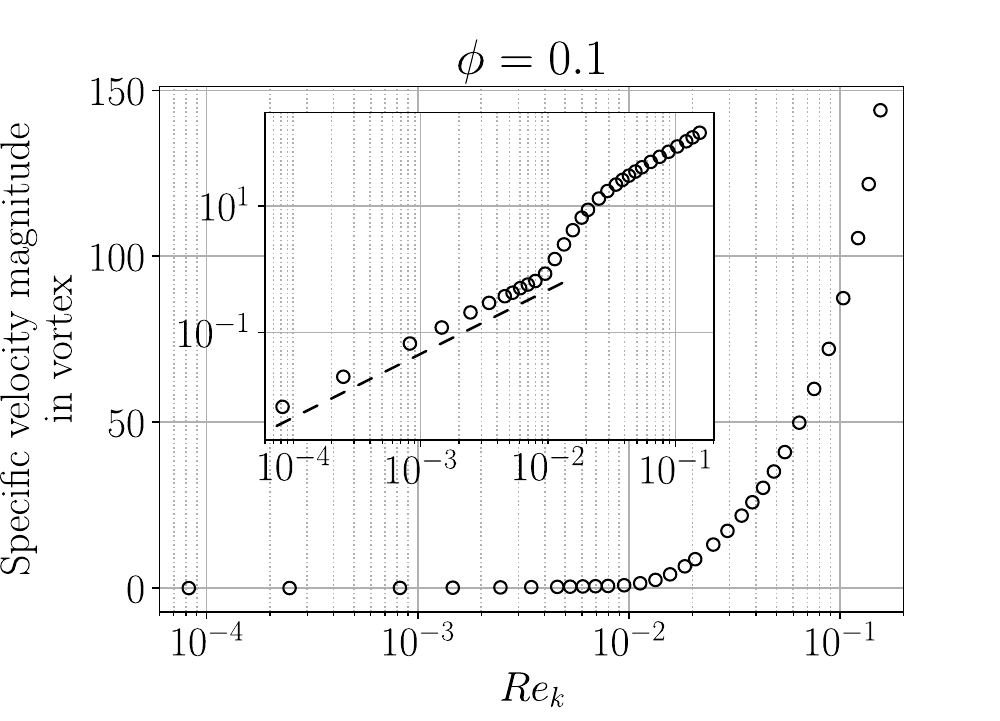}
		\includegraphics[height=.55\linewidth]{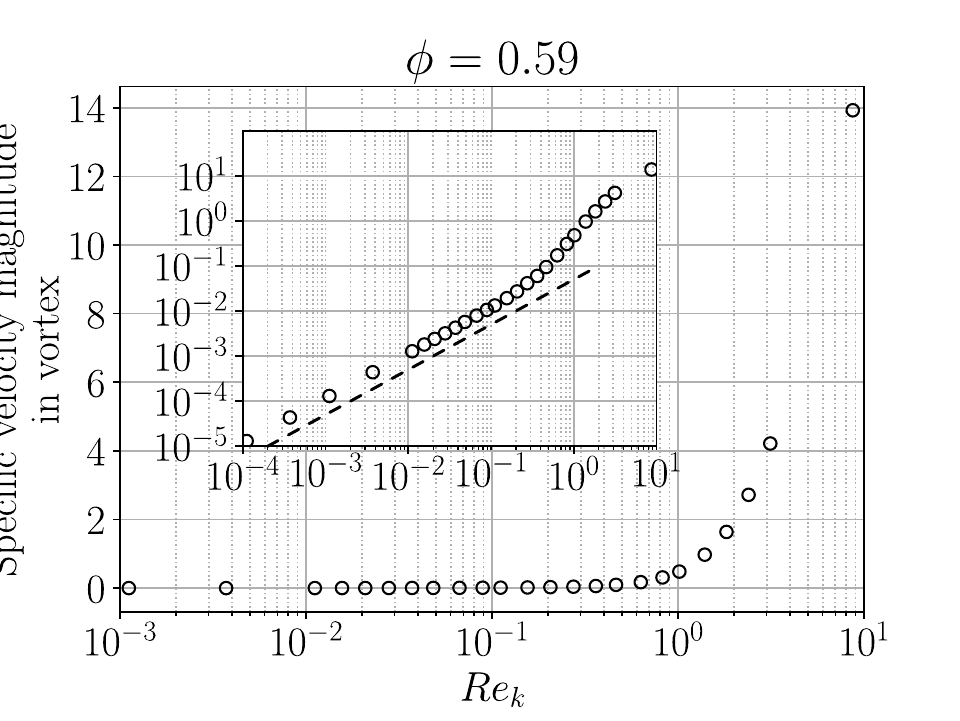}
		\includegraphics[height=.55\linewidth]{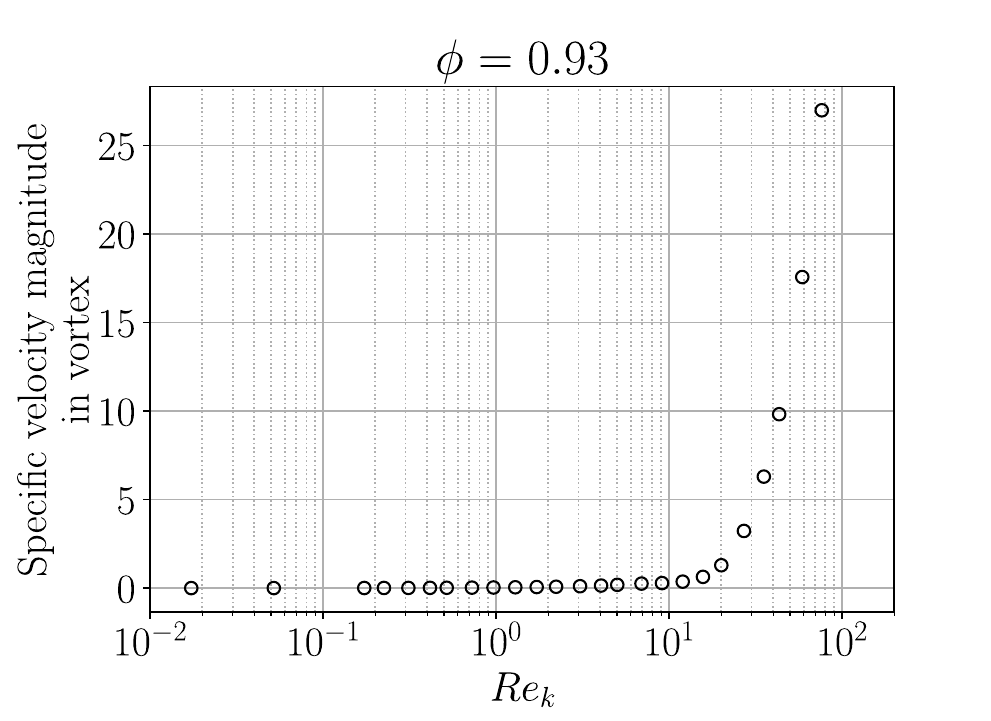}
		\includegraphics[height=.55\linewidth]{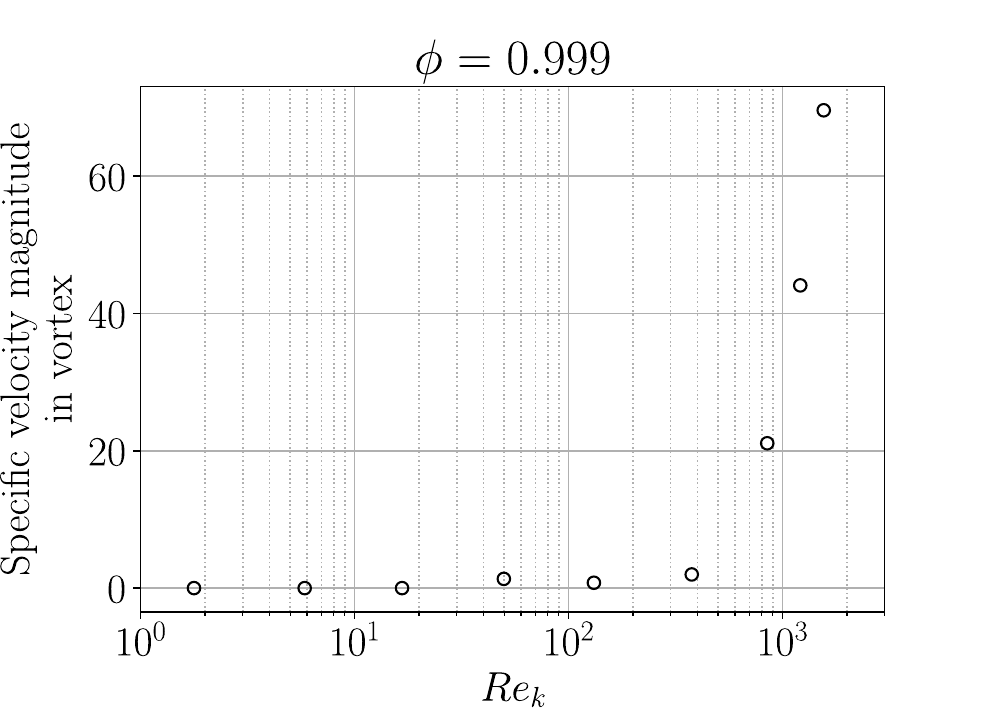}
		\caption{\add{Integral of the absolute value of velocity in the vortex per vortex volume for SC systems (for brevity called the \textit{specific velocity magnitude in vortex} in the text, $I[V]_v / I[1]_v$). The insets in the two topmost subplots show the same data as the main subplot but on a log-log scale.}}
		\label{fig:vmag_per_vortexVol}
	\end{figure}

	The increase of the recirculation zone volume can explain the decrease of $T_\Omega$ because it directly implies the straightening of all the percolating streamlines and thus the decrease of $T_s$ in Eq.~\eqref{eq:TV_split_reduced_using_Ts}. However, it fails to address the rise of the volume-based tortuosity at the highest Reynolds numbers. This is why now we turn our attention to the \add{the other term in Eq.~\eqref{eq:TV_vortex_contributions}, the} integral of the velocity magnitude inside the vortex per vortex volume, $I[V]_v/I[1]_v$ (further in the text, we shall refer to this quantity as the \textit{specific velocity magnitude in vortex}, for brevity). Its dependence on the Reynolds number is shown in Fig.~\ref{fig:vmag_per_vortexVol} for \del{both} \add{all} SC systems. We note that since there is no recirculation zone \del{present} \add{found} in the \del{higher-porosity} \add{highest-porosity} SC system in the Darcy regime, we assume that the value of the specific velocity magnitude in the vortex is zero in those cases.
	
	For \del{both} \add{all} SC systems, the specific velocity magnitude in vortices increases with Reynolds number. In the \del{low-porosity sample} \add{two lower-porosity samples}, the rise proceeds from the lowest $Re_k$ with the first-order slope \del{($p=1$)} and experiences a jump in the slope around \del{$Re_k = 3 \cdot 10^{-3}$} \add{$Re_k = 10^{-2}$ for $\phi=0.1$ and $Re_k = 4 \cdot 10^{-1}$ for $\phi=0.59$}. It later attains the slope of the order approximately one again, in the range of $Re_k$ corresponding to those where the growth of $T_\Omega$ slows down (cf. Fig.~\ref{fig:tortuosities_vs_Rek}). The initial first-order dependence is explained by the linearity of the Stokes equation and proportionality between the body force magnitude $g$ and the mean flow velocity $\langle \bsym{V} \rangle$ in the Darcy regime (i.e., when $g$ is increased, the velocity field inside the vortex is only scaled proportionally by a constant). The subsequently higher slope indicates that a non-trivial mechanism involving the momentum transfer to the recirculation zone and the growth of its volume comes into play in the inertial regime. At the same time, an analogy between the Darcy regime and the \add{part of the }inertial regime \del{above $Re_k = 10^{-2}$} \add{where the specific velocity magnitude in vortex decreases its slope again} can be drawn in the sense that both are characterized by the recirculation zone growing more in kinetic energy than volume. This further highlights the importance of the geometrical confinement in the emergence of inertial effects in low-porosity samples~\cite{Naqvi2025}.
	
	In the \del{high-porosity system} \add{two higher-porosity systems}, the start of the rapid growth of the specific velocity magnitude in the vortex happens roughly at $Re_k$ when the recirculation zone volume starts to grow (cf. Fig.~\ref{fig1:vortex_volume_relative}). It is interesting that even though the integral of the velocity magnitude over the whole fluid volume, $I[V]$, achieved at the highest Reynolds number is about $40$ times larger in the \del{high} \add{highest}-porosity system (\del{results not shown in the plots} \add{1855 vs. 47, see Table~\ref{tab:specific_momentum_growth}}), the specific velocity magnitude in vortex is about 2 times higher in \del{low} \add{the lowest}-porosity system ($\approx 140$ vs. $\approx 70$\add{, see also Table~\ref{tab:specific_momentum_growth}}). This suggests that strong geometrical confinement leads to the accumulation of relatively large amounts of the kinetic energy in the recirculation zones, quantified also by other authors using the variance of the kinetic energy~\cite{Andrade1999,Agnaou2017,Sniezek2024,Naqvi2025}.

	\begin{table}
		{\color{changecolor}
			\begin{tabular}{lcc}
				\multicolumn{3}{c}{For maximal $Re_k$} \\
				\hline
				\hline
				$\phi$ & $I[V]$ & $I[V]_v / I[1]_v$ \\
				\hline
				$0.1$ & 47 & 142 \\
				$0.59$ & 138 & 14 \\
				$0.93$ & 298 & 27 \\
				$0.999$ & 1855 & 70 \\
				
			\end{tabular}
			\caption{\add{The integral of the velocity magnitude in the whole fluid, $I[V]$ and the specific velocity magnitude in vortex, $I[V]_v/I[1]_v$, for the studied SC systems at the highest Reynolds number investigated for each porosity.}}
			\label{tab:specific_momentum_growth}
		}
	\end{table}
	
	In the end, \del{in the low-porosity stochastic and SC systems, the greater kinetic energy confinement in the recirculation zones and the cessation of their volume growth at high Reynolds numbers lead to the more pronounced increase of $T_\Omega$, compared to the higher-porosity counterparts.}
	\add{at high Reynolds numbers, as the porosity of either stochastic or SC system decreases, more kinetic energy stays confined in the recirculation zones and the recirculation zones occupy a larger fraction of the fluid volume. Those two phenomena, reflected in the growth of the terms $I[V]_v/I[1]_v$ and $I[1]_v/I[1]$, respectively, lead to a greater increase of $T_\Omega$ close to the steady-state regime limit, compared to the higher-porosity samples. The dynamics of the growth of the recirculation zones also determine the amount of the fall of $T_\Omega$ between its $Re_k \rightarrow 0$ value and its minimum. A more pronounced growth of the volume and kinetic energy inside them in the lower-porosity samples leads to a sooner domination of the $I[V]_v / I[V_0]_p$ term over the $T_s$ term in Eq.~\eqref{eq:TV_split_reduced_using_Ts}. This, in turn, causes the divergence between the two tortuosities to begin at their values closer to $T_\Omega(Re_k \rightarrow 0) \equiv T_s(Re_k \rightarrow 0)$, rather than the minimum of $T_s$, achieved at high Reynolds numbers, which explains the tendencies visible in the bottom subplot of Fig.~\ref{fig:TV_vs_Rek_SC} and in Fig.~\ref{fig:tortuosities_vs_Rek}.}
	
	\bigskip
	
	\section{Conclusions}
	
	
	The presented findings on the differences in behavior of the streamline-based and volume-integrated tortuosity during the Darcy-inertial transition in simple cubic porous media advance the understanding of the mechanisms behind the inertia onset in the studied porous systems. We show that porosity and the associated geometrical confinement are two of the key factors determining the behavior of the system in the inertial regime. The growth of the volume of the recirculation zones and the corresponding straightening of the percolating streamlines are captured by both types of tortuosity via their decrease in the initial stage of inertia onset. The later dynamics of the emergence of the inertial effects are strongly dependent on the system's geometry. In low-porosity media, the geometrical confinement prevents the recirculation zones from growing freely in volume, thus mainly the kinetic energy confined within them increases. This is represented by the saturation of the streamline-based tortuosity at its minimal value, while the volume-integrated tortuosity grows rapidly in this regime. This is because the volume-integrated tortuosity contains information about the whole fluid volume, while the streamline-based one -- only about the percolating volume. On the other hand, in high-porosity systems, the growth of the volume of the recirculation zones is not limited by the solid walls. Due to this, the increase of the volume-integrated tortuosity is less pronounced compared to the low-porosity samples, and the range of Reynolds numbers over which the two definitions of tortuosity are compliant with one another can extend almost to the end of the steady-state flow regime.
	
	The investigated physical phenomena are closely related to the tools used to quantify them. The presented results show the utility of the two discussed measures of the inertial effects in porous media -- streamline-based and volume-integrated tortuosity. Each of them separately carries information about various forms of the inertial effects in the flow -- $T_s$ is more sensitive to the changes occurring solely in the percolating volume, while in $T_\Omega$ the contribution of the recirculation zones is also visible. The difference between the values of the two, as well as the magnitude of the \add{fall and the} growth of $T_\Omega$, can be treated as an indicator of the confinement of the recirculation zones.
	
	The presented results also highlight which aspects of the inertia onset in porous media are better captured by which definition of tortuosity and why, allowing for a more sensitive choice between the two in more application-oriented problems. One may stipulate, for instance, that the information about the recirculation zones carried by the volume-integrated tortuosity might prove useful when the mixing and reaction phenomena in the porous sample are of interest. On the other hand, the exclusive contribution of the percolating volume to the streamline-based tortuosity may carry more valuable information for studies concerned with the mass transport through the sample, e.g., during gas extraction, where the gaseous phase is often transported through preferential channels and the residuals of the fluid phase occupy the dead-end pores and cavities. We refer the reader to Appendix~\ref{app:other_measures} for an additional discussion on the relation between the volume-integrated tortuosity and other, commonly used, measures of inertia in porous media flows.
	
	\bigskip
	
	\section{Acknowledgments}
	Funded by National Science Centre, Poland under the OPUS call in the Weave programme 2021/43/I/ST3/00228.
	This research was funded in whole or in part by National Science Centre (2021/43/I/ST3/00228). For the purpose of Open Access, the author has applied a CC-BY public copyright licence to any Author Accepted Manuscript (AAM) version arising from this submission. D. Strzelczyk and M. Matyka acknowledge the financial support from the Slovenian Research And Innovation Agency (ARIS) research core funding No. P2-0095.
	
	\section{Author contributions}
	\noindent \textbf{D.S.}: Conceptualization, Methodology, Software, Validation, Formal analysis, Investigation, Writing - Original Draft, Visualization
	
	\noindent \textbf{G.K.}: Resources, Writing - Review \& Editing, Project administration, Funding acquisition
	
	\noindent \textbf{M.M.}: Conceptualization, Methodology, Resources, Writing - Review \& Editing, Supervision, Project administration, Funding acquisition
	
	\section{Data availability}\label{sec:data_repo}
	\add{The velocity and pressure (density) fields generated in the present study are available at a public repository under the URL: \href{https://doi.org/10.5281/zenodo.16794913}{https://doi.org/10.5281/zenodo.16794913}. The repository also contains chosen scripts, source and configuration files used in the present study.}
	
	\bigskip
	
	\appendix
	
	\section{Relation to other measures of the onset of inertia}\label{app:other_measures}
	
\begin{figure}[!ht]
		\centering
		\includegraphics[width=.75\linewidth]{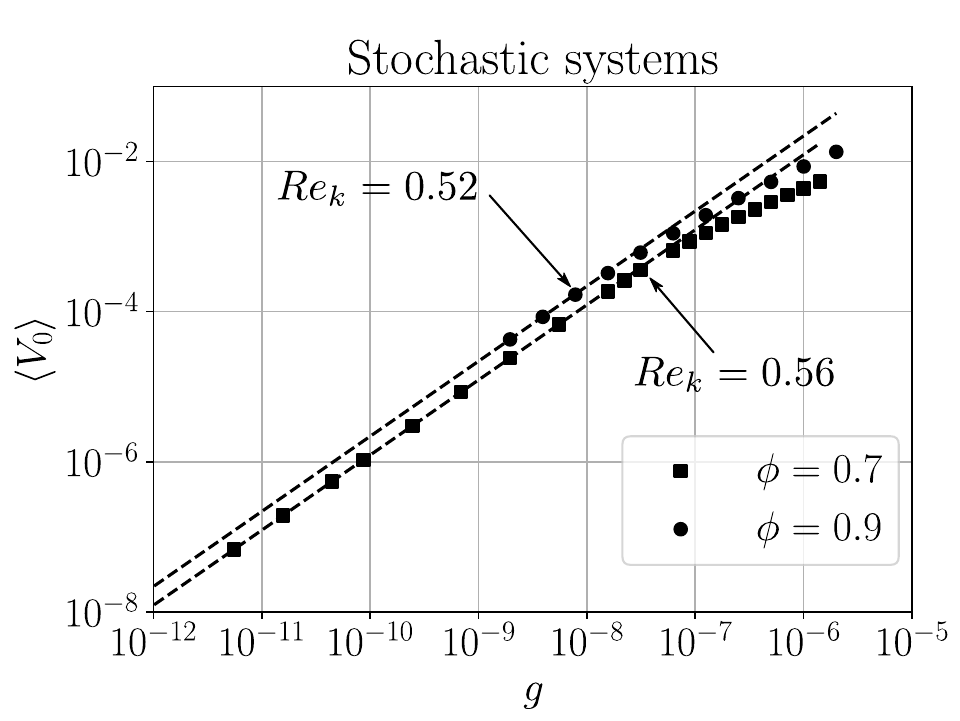}
		\includegraphics[width=.75\linewidth]{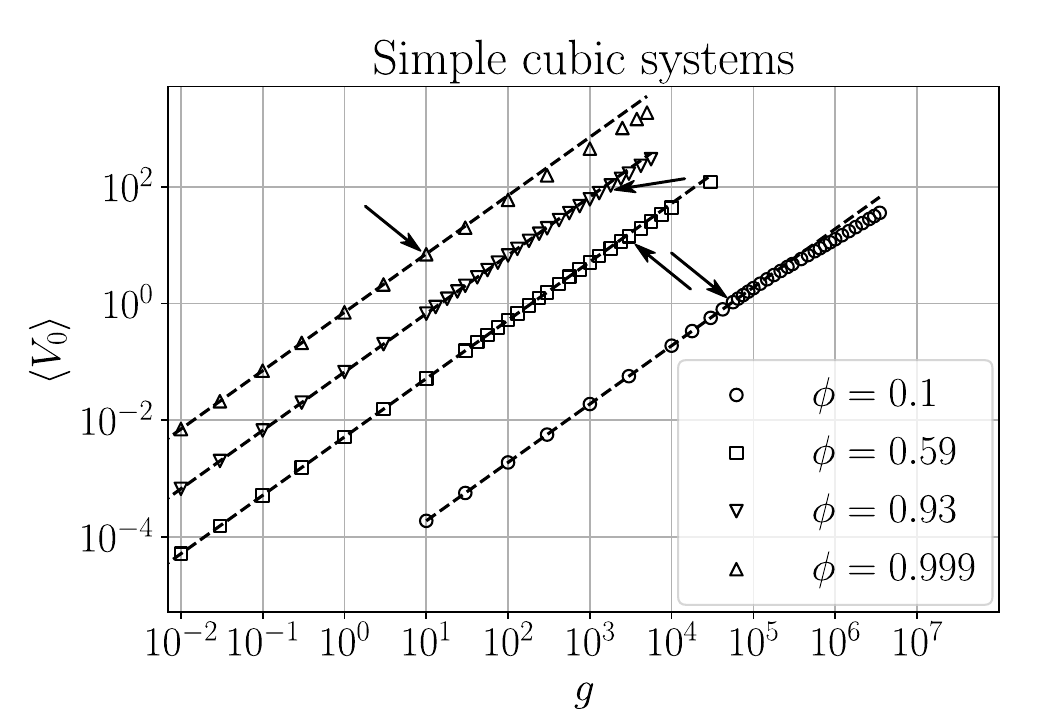}
		\caption{\add{Mean flow velocity in the $x_0$-direction versus the magnitude of the forcing term for the stochastic systems (\textit{top}) and simple cubic systems (\textit{bottom}). The dashed lines present the first-order scaling slopes anchored at the data points with the lowest value of $g$ (for some systems, those points are not visible at the presented axes ranges). The values at the arrows show the accelerations around which, according to this plot, the relation becomes non-linear, corresponding to $Re_k=4.58 \cdot 10^{-3}, 1.02, 27.1, 5.84$, in the order of the increasing porosity. The acceleration and velocity for stochastic systems are given in lattice units.}}
		\label{fig:darcy}
	\end{figure}

	To give a better perspective on our present findings, for the discussed stochastic and simple cubic systems, we additionally calculate other measures of the inertia onset. Fig.~\ref{fig:darcy} shows the relation between the mean flow velocity in the $x_0$-direction versus the magnitude of the forcing term (the Darcy's law, Eq.~\eqref{eq:darcy}). A comparison with Figs.~\ref{fig:TV-Re_sotchastic_PM} and ~\ref{fig:tortuosities_vs_Rek} shows that for the \del{higher}\add{highest}-porosity samples of each type the Reynolds number at which the mean velocity--forcing term relation becomes nonlinear corresponds roughly to the beginning of the fall of the tortuosities (\del{$Re_k \approx 0.49$ and $5.75$}\add{$Re_k \approx 0.52$ and $5.84$} for the stochastic and simple cubic system, respectively). For the \del{lower}\add{lowest}-porosity samples, the transition to the nonlinear dependence occurs later, around the minimum of $T_\Omega$ (\del{$Re_k \approx 0.48$ and $1.45 \cdot 10^{-3}$} \add{$Re_k \approx 0.56$ and $4.58 \cdot 10^{-3}$}, for the stochastic and simple cubic system, respectively). Note also that for the simple cubic samples, the divergence from linearity in $\langle V_0 \rangle$--$g$ relation is weaker than in the stochastic systems. This is especially in contrast to the behavior of $T_\Omega$ in the \del{low}\add{lowest}-porosity SC sample, which experiences the biggest jump in value between the Darcy regime and the highest $Re_k$ from among all the studied systems. This suggests that the volume-integrated tortuosity might be a good indicator for the onset of inertia for extremely low-porosity systems.
	
\begin{figure}[!ht]
		\centering
		\includegraphics[height=.55\linewidth]{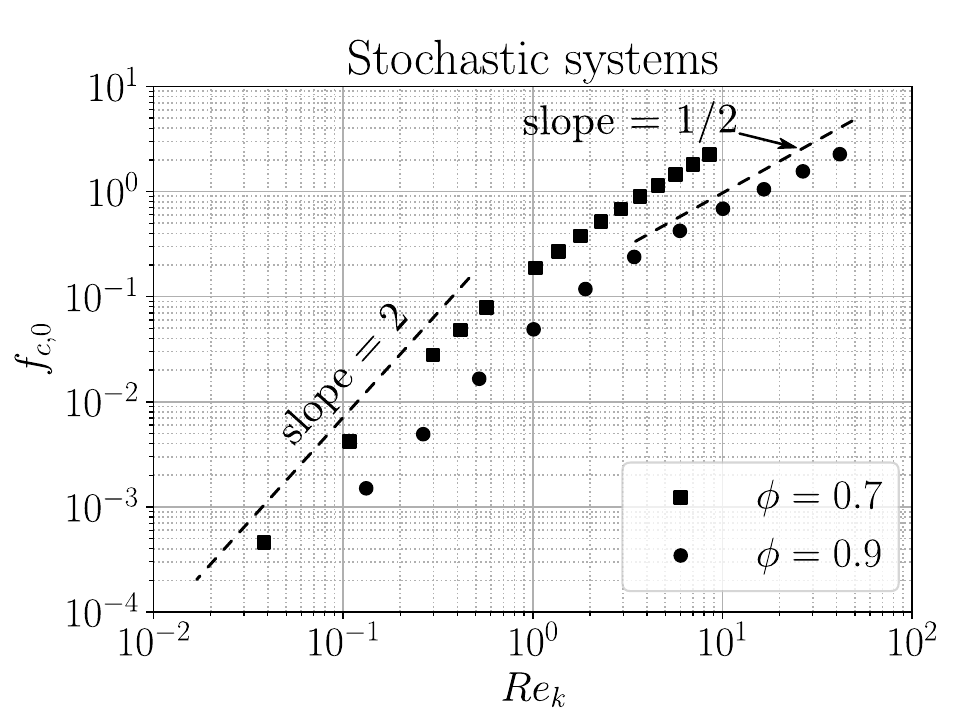}
		\includegraphics[height=.55\linewidth]{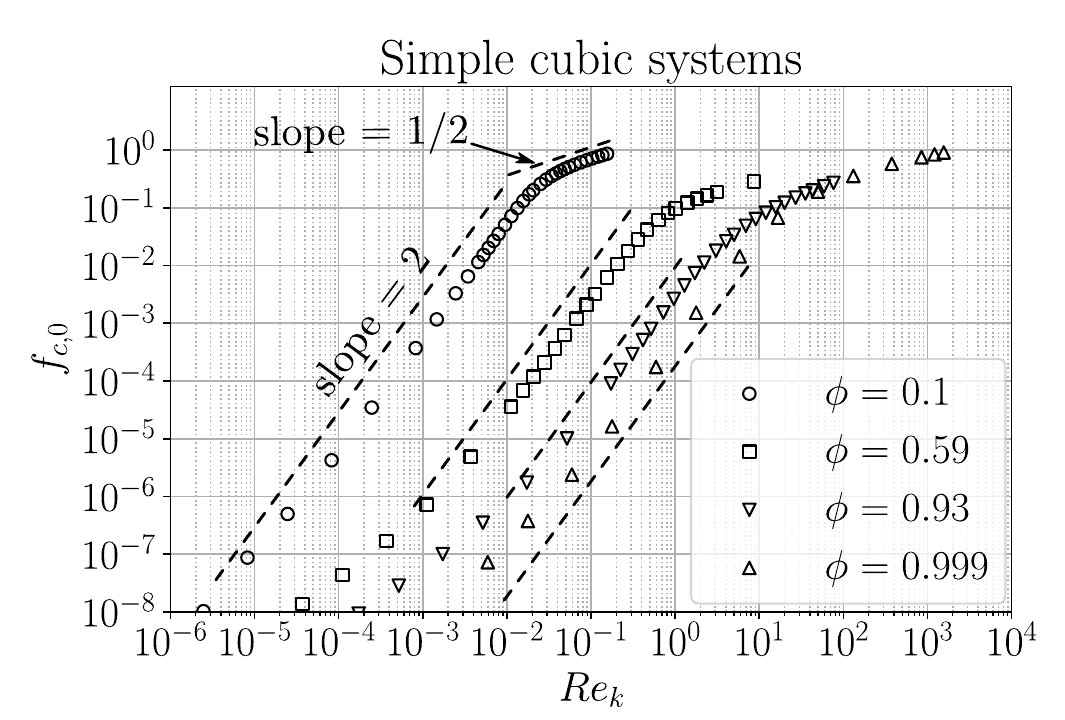}
		\caption{\add{The inertial correction $f_{c,0}$ (Eq.~\eqref{eq:inertial_correction}) versus the Reynolds number for the stochastic systems (\textit{top}) and simple cubic systems (\textit{bottom}). The dashed lines present the first- and second-order scaling slopes for the stochastic samples and $1/2$-- and second-order scaling slopes for the simple cubic samples.}}
		\label{fig:inertial_correction}
	\end{figure}

	Fig~\ref{fig:inertial_correction} shows the dependence of the dimensionless inertial correction $f_{c,0}$ defined as~\cite{Lasseux2011}
	\iftoggle{showdeleted}{
		\begin{equation}\label{eq:inertial_correction}
			f_{c,0}
			=
			-\left(
			\frac{\langle \bsym{V} \rangle}{| \langle \bsym{V} \rangle | \phi} - \frac{k_0\bsym{g}}{{\color{changecolor}\cancel{\mu} \nu} | \langle \bsym{V} \rangle |}
			\right)
			\cdot
			\hat{\bsym{x}}_0
			=
			-1 + \frac{k_0g}{{\color{changecolor}\cancel{\mu} \nu} \langle V_0\rangle}
		\end{equation}
	}{
		\begin{equation}\label{eq:inertial_correction}
			f_{c,0}
			=
			-\left(
			\frac{\langle \bsym{V} \rangle}{| \langle \bsym{V} \rangle | \phi} - \frac{k_0\bsym{g}}{\nu | \langle \bsym{V} \rangle |}
			\right)
			\cdot
			\hat{\bsym{x}}_0
			=
			-1 + \frac{k_0g}{\nu \langle V_0\rangle}
		\end{equation}
	}
	on the Reynolds number $Re_k$ for the studied stochastic and simple cubic systems. In Eq.~\eqref{eq:inertial_correction}, $\hat{\bsym{x}}_0$ is the unit vector aligned with the $x_0$-axis, and the last equality holds under the assumption of the mean flow aligned with the $x_0$ direction, namely $\langle \bsym{V} \rangle = \langle V_0 \rangle\hat{\bsym{x}}_0$. The weak inertia regime is characterized by the $f_{c,0} \propto Re_k^2$ scaling, while the strong inertia regime -- by $f_{c,0} \propto Re_k$ scaling. In the case of stochastic samples, the latter regime is vaguely visible, which suggests that the transition between the two regimes takes place over the majority of the considered Reynolds numbers. In the case of simple cubic systems, the weak inertia regime is clearly visible for both porosities. In the several lowest $Re_k$ cases \add{for each porosity}, the scaling becomes below-second order, which was also observed by other authors~\cite{Lasseux2011}. In the \del{lower}\add{lowest}-porosity sample, the deviation from the weak inertia regime appears approximately at \del{$Re_k=4 \cdot 10^{-3}$}\add{$Re_k=10^{-2}$}, for which the $T_\Omega$ reaches its minimum (cf. Fig~\ref{fig:tortuosities_vs_Rek}). For the highest Reynolds numbers, the relation changes into sublinear; however, it endures over a short range of $Re_k$ before the steady-state regime ends. It is possible that one observes the regime above strong inertia where $f_{c,0} \propto Re_k^2 + Re_k$. One may relate this regime to the reduction \del{in the increase} of the slope of the volume-based tortuosity at \del{$Re_k > 2 \cdot 10^{-2}$}\add{$Re_k > 6 \cdot 10^{-2}$}. \del{For the  higher-porosity simple cubic sample, the relation becomes sub-quadratic around $Re_k = 5$, the value at which the tortuosities in Fig.~\ref{fig:tortuosities_vs_Rek} begin to rapidly decrease.}\add{As the porosity increases, the deviation from the quadratic scaling of $f_{c,0}$ shifts towards the lower Reynolds numbers, closer to the start of the $T_\Omega$ decrease -- $Re_k=0.3,\>2,\>5$ for porosities $\phi=0.59,\>0.93,\>0.999$, respectively. This, along with the similar observations made for the deviation from Darcy's law, supports the idea that in low-porosity samples, the onset of inertia manifests itself rather through the reorganization of the flow paths (e.g., straightening of the streamlines) than significant changes in the averaged transport coefficients (like the apparent permeability or the magnitude of the inertial correction.)} Also, in \del{this case}\add{all SC samples}, for the highest Reynolds numbers, the $f_{c,0}(Re_k)$ relation is clearly non-linear. \add{We note that if one considered the {\itshape dimensional} inertial correction, $f_{i,0} \propto \langle V_0 \rangle f_{c,0}$, the second order scaling $f_{c,0} \propto Re_k^2$ would turn into the cubic scaling $f_{i,0} \propto Re_k^3$ which would directly highlight the dominance of the $\langle V_0 \rangle^3$ term in the corrections to the Darcy's law, Eq.~\eqref{eq:darcy_forchhiemer}.}
	
\begin{figure}[!ht]
		\centering
		\includegraphics[width=\linewidth]{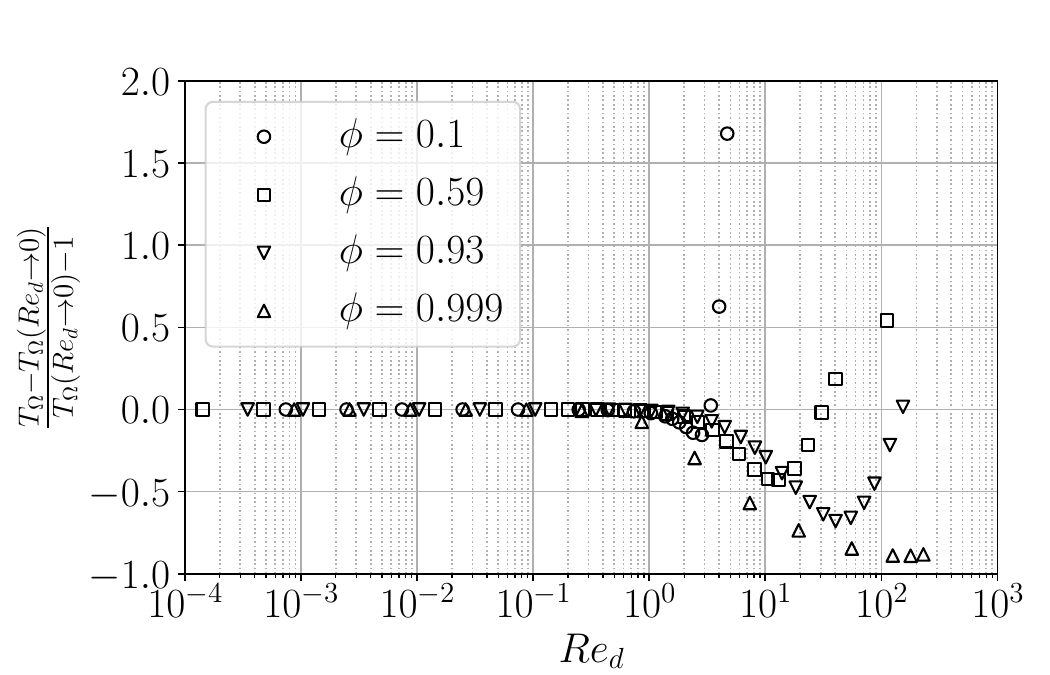}
		\caption{\add{Volume-integrated tortuosity for all SC samples rescaled by $T_\Omega$ at vanishingly small Reynolds number. Note that the horizontal axis shows $Re_d$, not $Re_k$.}}
		\label{fig:TV_vs_Rek_SC_Re_d}
	\end{figure}

	\add{Finally, to allow for easy comparison with other works, in Table~\ref{tab:re_conversion} we provide the conversion factors from $Re_k$ used in the present study, to the Reynolds number $Re_d$, which uses the grain size (diameter in case of spheres and side length in case of cubes) as the reference length. In Fig.~\ref{fig:TV_vs_Rek_SC_Re_d} we show the same data as in the bottom subplot of Fig.~\ref{fig:TV_vs_Rek_SC} but plotted against $Re_d$.}
	
	\begin{table}[]
		\centering
		{\color{changecolor}
			\begin{tabular}{cccc}
				System type & $\phi$ & $L_\text{ref}$ & $Re_d/Re_k$ \\
				\hline
				\hline
				\multirow{2}{*}{Stochastic} & $0.7$ & $19$ & $5.99$\\
				& $0.9$ & $10$ & $ 1.82$\\
				\hline
				\multirow{4}{*}{SC} & $0.1$ & $1.31$ & $300$\\
				& $0.59$ & $0.92$ & $12.8$\\
				& $0.93$ & $0.52$ & $2.01$\\
				& $0.999$ & $0.124$ & $0.148$\\
			\end{tabular}
		}
		\caption{\add{Conversion factors from the Reynolds number used in the present study, $Re_k$, to the Reynolds number $Re_d$ using the size of the grain as the reference length.}}
		\label{tab:re_conversion}
	\end{table}
	
	\section{Discretization independence study}\label{app:convergence}
	
	To assess the discretization independence of $T_\Omega$ for the SC systems, we calculate its values from the solution of the governing equations obtained at different space discretization parameters ($h$ for uniform meshes and $h_\text{min}$ for refined meshes, cf. Section~\ref{sssec:setup_sc_systems}). For each porosity, the highest $Re_k$ is considered. In the case of the \del{higher} \add{highest}-porosity sample, we use the refined discretization and solve the transient problem. The results are presented in Table~\ref{tab:grid_indenpendece_TOmega}. The change of $T_\Omega$ between the subsequent discretizations is calculated relative to the value at the finer discretization. \add{The results show that increasing the discretization density by a factor of $1.3$ from the one used to produce the data presented in the Results section is not expected to alter the value of $T_\Omega-1$ by more than $3\%$.}
	
	\begin{table}[!ht]
		\centering
		\begin{tabular}{l}
			\begin{tabular}{ccc}
				\multicolumn{3}{c}{$\phi=0.1$, $g=3.5 \cdot 10^6$}\\
				\hline \hline
				$h$ & $T_\Omega-1$ & $\%$ change\\
				\hline
				$1/115$ & $0.279$ & --\\
				$1/150$ & $0.255$ & $9.444$ \\
				$\mathbf{1/195}$ & $\mathbf{0.296}$ & $\mathbf{13.697}$ \\
				$1/253$ & $0.299$ & $0.996$ \\
				$1/330$ & $0.301$ & $0.822$
			\end{tabular}
			\\ \vspace{.1cm} \\
			{\color{changecolor}
				\begin{tabular}{ccc}
					\multicolumn{3}{c}{$\phi=0.59$, $g=3 \cdot 10^4$}\\
					\hline \hline
					$h$ & $T_\Omega-1$ & $\%$ change \\
					\hline
					$\mathbf{1/150}^*$ & $\mathbf{2.832 \cdot 10^{-2}}$ & \textbf{--} \\
					$1/195$ & $2.848 \cdot 10^{-2}$ & $0.565$ \\
					$1/253$ & $2.853 \cdot 10^{-2}$ & $0.193$ \\
				\end{tabular}
			}
			\\ \vspace{.1cm} \\
			{\color{changecolor}
				\begin{tabular}{ccc}
					\multicolumn{3}{c}{$\phi=0.93$, $g=5.6 \cdot 10^3$}\\
					\hline \hline
					$h$ & $T_\Omega-1$ & $\%$ change \\
					\hline
					$1/115$ & $ 8.451 \cdot 10^{-3}$ & -- \\
					$\mathbf{1/150}$ & $\mathbf{8.432 \cdot 10^{-3}}$ & \textbf{0.232} \\
					$1/195$ & $8.398 \cdot 10^{-3}$ & $0.405$ \\
				\end{tabular}
			}
			\\ \vspace{.1cm} \\
			\begin{tabular}{cccc}
				\multicolumn{4}{c}{$\phi=0.999$, $g=5 \cdot 10^3$}\\
				\hline \hline
				$h_\text{min}$ & $\Delta t$ & $T_\Omega - 1$ & $\%$ change \\
				\hline
				$1/116$ & $2.25 \cdot 10^{-6}$ & $8.193 \cdot 10^{-5}$ & --\\
				$\mathbf{1/150}$ & $\mathbf{2.25 \cdot 10^{-6}}$ & $\mathbf{8.778 \cdot 10^{-5}}$ & $\mathbf{6.665}$\\
				$1/196$ & $2.25 \cdot 10^{-6}$ & $9.035 \cdot 10^{-5}$ & $2.851$\\
				$1/254$ & $1.5 \cdot 10^{-6}$ & $ 9.137 \cdot 10^{-5}$ & $1.112$\\
				&&&
			\end{tabular}
		\end{tabular}
		\caption{Values above unity of the volume-integrated tortuosity, $T_\Omega-1$, obtained for the highest forcing term $g$ for SC systems at various space discretization parameters used for the solution of the governing equations. The relative change in each row (rightmost column) is calculated with respect to the value of $T_\Omega-1$ from the row above (one coarser discretization). The bold case row indicates the values discussed in the work. The asterisk denotes the data obtained from \texttt{pimpleFoam} simulation with $\Delta t=3.5 \cdot 10^{-6}$.}
		\label{tab:grid_indenpendece_TOmega}
	\end{table}
	
	We additionally perform the study of the independence of the $T_s$ value on the spacing $h_s$ and timestep value \del{$\Delta t$} \add{$\Delta t_s$} of the streamline generation procedure in SC samples. In this case, we take a look at the samples of the lowest $Re_k$, as these are the ones where the streamlines are expected to be the most tortuous, thus introducing larger errors at the segments of high curvature. We use the velocity fields obtained on the uniform $h=1/150$ or refined $h_\text{min}=1/150$ discretizations described in Sec.~\ref{sssec:setup_sc_systems}. We scale streamline seeding distance $h_s$ and the timestep for the integration along characteristics $\Delta t_s$ by a factor of $2$ between subsequent setups. The results are presented in Table~\ref{tab:grid_indenpendece_Ts}. \add{Also here, doubling the discretizations density is not expected to change $T_s-1$ by more than $3\%$.}
	
	\begin{table}[!ht]
		\centering
		\begin{tabular}{l}
			\begin{tabular}{rrcc}
				\multicolumn{4}{c}{$\phi=0.1$, $Re_k=8.24 \cdot 10^{-7}$}\\
				\hline \hline
				\multicolumn{1}{c}{$h_s$} &  \multicolumn{1}{c}{$\Delta t_s$} & $T_s-1$ & $\%$ change\\
				\hline\
				$4 \cdot 10^{-3}$ & $4 \cdot 10^{-3}$ & $0.0255$ & --\\
				$2 \cdot 10^{-3}$ & $2 \cdot 10^{-3}$ & $0.0244$ & $4.598$\\
				$\mathbf{10^{-3}}$ & $\mathbf{10^{-3}}$ & $\mathbf{0.0240}$ & $\mathbf{1.676}$ \\
				$5 \cdot 10^{-4}$ & $5 \cdot 10^{-4}$ & $0.0238$ & $0.861$ \\
			\end{tabular}
			\\ \vspace{.1cm} \\
			{\color{changecolor}
				\begin{tabular}{rrcc}
					\multicolumn{4}{c}{$\phi=0.59$, $Re_k=3.72 \cdot 10^{-14}$}\\
					\hline \hline
					\multicolumn{1}{c}{$h_s$} &  \multicolumn{1}{c}{$\Delta t_s$} & $T_s-1$ & $\%$ change\\
					\hline\
					$1 \cdot 10^{-2}$ & $1 \cdot 10^{-2}$ & $1.766 \cdot 10^{-2}$ & --\\
					$5 \cdot 10^{-3}$ & $5 \cdot 10^{-3}$ & $1.792 \cdot 10^{-2}$ & $1.454$ \\
					$\mathbf{2.5 \cdot 10^{-3}}$ & $\mathbf{2.5 \cdot 10^{-3}}$ & $\mathbf{1.805  \cdot 10^{-2}}$ & $\mathbf{0.724}$\\
					$1.25 \cdot10^{-3}$ & $1.25 \cdot 10^{-3}$ & $1.819 \cdot 10^{-2}$ & $0.785$ \\
				\end{tabular}
			}
			\\ \vspace{.1cm} \\
			{\color{changecolor}
				\begin{tabular}{rrcc}
					\multicolumn{4}{c}{$\phi=0.93$, $Re_k=1.73 \cdot 10^{-12}$}\\
					\hline \hline
					\multicolumn{1}{c}{$h_s$} &  \multicolumn{1}{c}{$\Delta t_s$} & $T_s-1$ & $\%$ change\\
					\hline\
					$10^{-2}$ & $10^{-2}$ & $8.076 \cdot 10^{-3}$ & --\\
					$\mathbf{5 \cdot 10^{-3}}$ & $\mathbf{5 \cdot 10^{-3}}$ & $\mathbf{8.156 \cdot 10^{-3}}$ & $\mathbf{0.992}$\\
					$2.5 \cdot 10^{-3}$ & $2.5 \cdot 10^{-3}$ & $8.196 \cdot 10^{-3}$ & $0.495$\\
					$1.25 \cdot10^{-3}$ & $1.25 \cdot 10^{-3}$ & $8.240 \cdot 10^{-3}$ & $0.530$ \\
				\end{tabular}
			}
			\\ \vspace{.1cm} \\
			\begin{tabular}{rrcc}
				\multicolumn{4}{c}{$\phi=0.999$, $Re_k=5.93 \cdot 10^{-11}$}\\
				\hline \hline
				\multicolumn{1}{c}{$h_s$} & \multicolumn{1}{c}{$\Delta t_s$} & $T_s - 1$ & $\%$ change \\
				\hline
				$2 \cdot 10^{-2}$ & $2 \cdot 10^{-2}$ & $7.190 \cdot 10^{-4}$ & --\\
				$\mathbf{10^{-2}}$ & $\mathbf{10^{-2}}$ & $\mathbf{7.343 \cdot 10^{-4}}$ & $\mathbf{2.085}$ \\
				$5 \cdot 10^{-3}$ & $5 \cdot 10^{-3}$ & $7.385 \cdot 10^{-4}$ & $0.561$ \\
				$2.5 \cdot 10^{-3}$ & $2.5 \cdot 10^{-3}$ & $7.404 \cdot 10^{-4}$ & $0.254$ \\
			\end{tabular}
		\end{tabular}
		\caption{Values above unity of the streamline-based tortuosity, $T_s-1$, obtained for the lowest $Re_k$ for SC systems at various discretization parameters used for the streamline generation. The relative change in each row (rightmost column) is calculated with respect to the value of $T_\Omega-1$ from the row above (one coarser discretization). The bold case row indicates the values discussed in the work.}
		\label{tab:grid_indenpendece_Ts}
	\end{table}
	
	\section{Details of the numerical setup and results analysis}\label{app:numerical_details}
	
	\subsection{{\color{changecolor} Dimensionalization of the governing equations}}\label{subapp:dimensions}
	
	{\color{changecolor}
		As noted in Section~\ref{ssec:setup}, in the numerical simulations we consider the non-dimensional form of the Navier-Stokes equations. To relate them and the obtained results to real-life porous samples let us use the reference (physical) length of the samples' side $L_\text{ph} = 10^{-3} \> \text{m} = LC_x$ and the kinematic viscosity of the medium equal to the one of water in $20^\circ \text{C}$ -- $\nu_\text{ph} = 10^{-6} \> \text{m}^2/\text{s} = \nu C_x^2/C_t$. $C_x$ and $C_t$ denote the conversion factors form the non-dimensional to the dimensional form for the length and time, respectively. As the non-dimensional values of the reference length and kinematic viscosity used in the present work are
		\begin{itemize}
			\item $L=160$ and $\nu=c_s^2(\tau^+-1/2)=5/3 \cdot 10^{-3}$ for $\phi=0.9$ stochastic sample,
			\item $L=304$ and $\nu=5/3 \cdot 10^{-3}$ for $\phi=0.7$ stochastic sample,
			\item $L=1$ and $\nu=1$ for SC samples,
		\end{itemize}
		this sets the conversion factor for the velocity, $C_V = C_x/C_t$ at
		\begin{itemize}
			\item $9.6 \cdot 10^1$ for $\phi=0.9$ stochastic sample,
			\item $1.83 \cdot 10^2$ for $\phi=0.7$ stochastic sample,
			\item $10^{-3}$ for SC samples,
		\end{itemize}
		all in $[\text{m}/\text{s}]$. Further, assuming the reference density $\rho_\text{ph} = 10^3 \> \text{kg} / \text{m}^3$ corresponding to $\rho=1$ in the simulations, one obtains the conversion factor for the pressure gradient or, equivalently, the body force $C_P/C_x = C_gC_\rho = C_xC_\rho/C_t^2$ equal to
		\begin{itemize}
			\item $2.94 \cdot 10^6$ for $\phi=0.9$ stochastic sample,
			\item $1.44 \cdot 10^7$ for $\phi=0.7$ stochastic sample,
			\item $1$ for SC samples,
		\end{itemize}
		all in $[\text{Pa}/\text{m}]$. The maximal velocities and the pressure gradients in dimensional form in each of the studied systems are summarized in Table~\ref{tab:dimensional_qunatities}.
		\begin{table}[]
			\centering
			{\color{changecolor}
				\begin{tabular}{lcc}
					& \multicolumn{2}{c}{\textbf{Quantity}}  \\
					\textbf{System} & $|\mathbf{V}| \> [\text{m}/\text{s}] \quad$ & $\mathbf{\Delta P/L} \> [\text{Pa}/\text{m}]$ \\
					\hline
					Stochastic, $\phi=0.7 \quad$ & $4.4$ & $1.4 \cdot 10^7$ \\
					Stochastic, $\phi=0.9$ & $3.6$ & $2.9 \cdot 10^6$ \\
					SC, $\phi=0.1$ & $4.8$ & $3.5 \cdot 10^6$ \\
					SC, $\phi=0.59$ & $0.74$ & $3 \cdot 10^4$ \\
					SC, $\phi=0.93$ & $0.54$ & $5.6 \cdot 10^3$ \\
					SC, $\phi=0.999$ & $2.1$ & $5 \cdot 10^3$
				\end{tabular}
			}
			\caption{\add{Maximal velocities and the pressure gradients in each of the studied systems in dimensional forms.}}
			\label{tab:dimensional_qunatities}
		\end{table}
		
	}
	
	\subsection{\add{The Lattice Boltzmann Method for stochastic porous media simulation}}\label{subapp:stochastic}
	
	{\color{changecolor}
		The two-relaxation time Lattice Boltzmann Method used for the simulations of the stochastic samples solves the Lattice Boltzmann Equation for $q$ distribution functions $f_k$
		\begin{equation}\label{eq:LBM_streaming}
			f_k\left(t+1,\bsym{x}\right) = f^\text{post}_k\left(t,\bsym{x}+\boldsymbol{e}_{k'}\right), \quad k=0,1,\dots,q-1
		\end{equation}
		where $\boldsymbol{e}_k=-\boldsymbol{e}_{k'}$ is the $k$-th discrete microscopic velocity and its inverse. In this study we use three dimensional models with $q=19$ and $q=27$ discrete velocities (denoted D3Q19 and D3Q27, respectively). The discrete velocities used in D3Q19 model are
		\begin{equation}
			\begin{array}{c}
				\begin{array}[t]{cc}
					k&\bsym{e}_k\\
					\hline
					0 & (0,0,0)\\
					1 & (1,0,0)\\
					3 & (0,1,0)\\
					5 & (0,0,1)\\
				\end{array}\quad
				\begin{array}[t]{cc}
					k&\bsym{e}_k\\
					\hline
					7 & (1,1,0)\\
					9 & (-1,1,0)\\
					11 & (1,0,1)\\
					13 & (1,0,-1)\\
					15 & (0,1,1)\\
					17 & (0,1,-1)\\
				\end{array}\quad
				\\
				\vspace{-.15cm}
				\\
				\bsym{e}_k = - \bsym{e}_{k-1} \text{ for } k=2,4,\dots,18.
			\end{array}
		\end{equation}
		For the D3Q27 model, the discrete velocity set is the tensor product of the one-dimensional velocity set $\{-1,0,1\}$, i.e. $\bsym{e}_k \in \{-1,0,1\}^{\otimes 3}$. The post-collisional distribution function with two-relaxation-time collision~\cite{Ginzburg2008,Krueger2016} is defined as
		\begin{equation}\label{eq:LBM_collision}
			\begin{aligned}
				f^\text{post}_k(t,\bsym{x})=f_k(t,\bsym{x}) - \frac{1}{\tau^+}\left(f_k^+(t,\bsym{x})-f_k^{\text{eq},+}(t,\bsym{x})\right) -\\ +\frac{1}{\tau^-}\left(f_k^-(t,\bsym{x})-f_k^{\text{eq},-}(t,\bsym{x})\right) + F_k
			\end{aligned}
		\end{equation}
		where $\tau^+$ and $\tau^-$ are the symmetric and the anti-symmetric relaxation times, respectively, related by
		\begin{equation}
			\tau^- = \frac{1}{2}+\frac{\Lambda}{\tau^+-0.5}.
		\end{equation}
		We use $\Lambda=1/4$ and $\tau^+=0.505$ in the present study. The symmetric and the anti-symmetric velocity distributions are defined as
		\begin{equation}
			\begin{aligned}
				f_k^+ = \frac{1}{2}(f_k+f_{k'})&\\[.5ex]
				f_k^- = \frac{1}{2}(f_k-f_{k'})&,\\[.5ex]
			\end{aligned}
		\end{equation}
		and similarly for the symmetric and the anti-symmetric equilibrium distributions
		\begin{equation}
			\begin{aligned}
				f_k^{\text{eq},+} = \frac{1}{2}(f_k^\text{eq}+f_{k'}^\text{eq})&\\
				f_k^{\text{eq},-} = \frac{1}{2}(f_k^\text{eq}-f_{k'}^\text{eq})&.\\
			\end{aligned} 
		\end{equation}
		The $k$-th equilibrium distribution is discretized with the second-order accuracy in the discrete velocities and is obtained from the macroscopic density and velocity via
		\begin{equation}\label{eq:feq}
			f^\text{eq}_k = \rho_{lb}\omega_k\left[1+\frac{\boldsymbol{e}_k\cdot \boldsymbol{V}_{lb}}{c_s^2} + \frac{(\boldsymbol{e}_k\cdot \boldsymbol{V}_{lb})^2}{2c_s^4} - \frac{\boldsymbol{V}_{lb}^2}{2c_s^2}\right],
		\end{equation}
		which themselves are related to the distribution functions through the quadratures in the velocity space
		\begin{equation}\label{eq:macro_var}
			\renewcommand{\arraystretch}{2.5}
			\begin{aligned}
				\rho_{lb} = &\>\sum\limits_{k=0}^{q-1} f_k \\
				\bsym{V}_{lb} = &\>\frac{1}{\rho_{lb}}\sum\limits_{k=0}^{q-1} f_k \bsym{e}_k.
			\end{aligned}
		\end{equation}
		The lattice speed of sound is $c_s = 1/\sqrt{3}$. We use the subscript `$lb$' to distinguish the quantities expressed in the LBM system of units from the quantities stated in the physical units or non-dimensionalized form. The sets of weights $\omega_k$ is specific for each LBM model. For D3Q19 we have
		\begin{equation}
			\omega_k=
			\begin{cases}
				12/36 \text{ for } \bsym{e}_k^2=0, \\
				2/36 \text{ for } \bsym{e}_k^2=1, \\
				1/36 \text{ for } \bsym{e}_k^2=2,
			\end{cases}
		\end{equation}
		while for D3Q27 the following holds
		\begin{equation}
			\omega_k=
			\begin{cases}
				8/27 \text{ for } \bsym{e}_k^2=0, \\
				2/27 \text{ for } \bsym{e}_k^2=1, \\
				1/54 \text{ for } \bsym{e}_k^2=2, \\
				1/216 \text{ for } \bsym{e}_k^2=3. \\
			\end{cases}
		\end{equation}
		We implement acceleration using first-order discretization in velocity space:
		\begin{equation}\label{eq:force_lbm}
			F_k = \omega_k \frac{\bsym{e}_k \cdot \bsym{F}_{lb}}{c_s^2}
		\end{equation}
		where $\bsym{F}_{lb}=\rho_{lb} \bsym{g_{lb}}$ is the body force defined in terms of the acceleration $\bsym{g_{lb}}$. Finally, the fluid kinematic viscosity is defined by the symmetric relaxation time
		\begin{equation}\label{eq:lbm_viscosity}
			\nu_{lb} = c_s^2\left(\tau^+ - \frac{1}{2}\right)
		\end{equation}
		and the pressure is derived from the density
		\begin{equation}
			p_{lb}=\rho_{lb}c_s^2.
		\end{equation}
	}
	
	\subsection{Setup of simple cubic porous media simulation}\label{subapp:sc}
	
	For the SIMPLE simulations in $\phi=0.1$ system, the convergence criterion based on residuals of the velocity and pressure is set to the values of $10^{-5}$ and $10^{-4}$, respectively. All considered systems reach this convergence criterion, and the verification of the steady-state solution for the system of the highest flow rate is performed with \texttt{pimpleFOAM}. The space discretization is generated by first creating the regular cubic grid with the discretization parameter $h=1/195$ using \texttt{blockMesh} and then adjusting it to the actual fluid volume geometry with \texttt{snappyHexMesh} (see Fig.~\ref{fig:sc_mesh}, top row).
	
	\add{A similar convergence criterion was used for the $\phi = 0.59$ sample. However, the steady state at $g=3 \cdot 10^4$, which did not meet the convergence criteria, was validated using \texttt{pimpleFoam}, using the \texttt{simpleFoam} solution converged to the residuals of $10^{-4}$ as the initial condition. The determination of the steady state in this and other transient cases considered in the present work is based on the relative difference between the current and the estimated final value of the $x_0$-component of velocity in the bulk. In the final iteration, this difference does not exceed $10^{-3}$. The final values of the velocity components at the considered point in the domain are estimated from the fit of the exponential function to the values of those velocities in time.}
	
	\add{For $\phi=0.93$ sample, several \texttt{simpleFoam} simulations at $g > 5.6 \cdot 10^3$ did converge based on the above-mentioned criteria, but the velocity field was not symmetric with respect to the planes $x_1=0.5$ and $x_2=0.5$. Due to this, we limited our analysis only to the accelerations $g \le 5.6 \cdot 10^3$. We did not perform the steady-state confirmation with \texttt{pimpleFoam} for the $g = 5.6 \cdot 10^3$ system.}
	
	For $\phi=0.999$, the convergence criterion based on residuals of the velocity and pressure is set to the values of $10^{-5}$ and $10^{-4}$, respectively. Three systems of the highest flow rate did not reach this criterion; however, the transient simulations performed with \texttt{pimpleFOAM} allow them to reach the steady state, and those data are used in the subsequent analysis. \del{The determination of the steady state in those cases is based on the relative difference between the current and the estimated final value of the $x_0$-component of velocity in the bulk and in the wake of the obstacle. In the final iteration, this difference does not exceed $10^{-3}$. The final values of the velocities were estimated from the fit of the exponential function to the values of those velocities in time.} \add{We used the same steady state-criterion for the transient simulations for this system, as in the $\phi=0.59$ case, with one velocity probe placed in the bulk, and the other -- in the wake of the obstacle.} We note that the $\phi=0.999$ system becomes unsteady in the range of $g \in \> (5 \cdot 10^3; 10^4]$. The discretization for $\phi=0.999$ systems solved using \texttt{simpleFOAM} is created the same way as for the \del{$\phi=0.1$} \add{other SC} systems (see Fig.~\ref{fig:sc_mesh}, bottom-left). For the three highest flow rates solved with \texttt{pimpleFOAM}, we first create a regular cubic grid with the discretization parameter $h_\text{bulk}=1/75$ using \texttt{blockMesh} and then adjust it to the actual fluid volume geometry with \texttt{snappyHexMesh} with local refinement ($h_\text{min}=1/150$) near the sphere and in its wake (see Fig.~\ref{fig:sc_mesh}, bottom-right).
	
	\subsection{Generation of streamlines}\label{subapp:streamlines}
	
\begin{figure}[!ht]
		\centering
		\includegraphics[width=.7\linewidth]{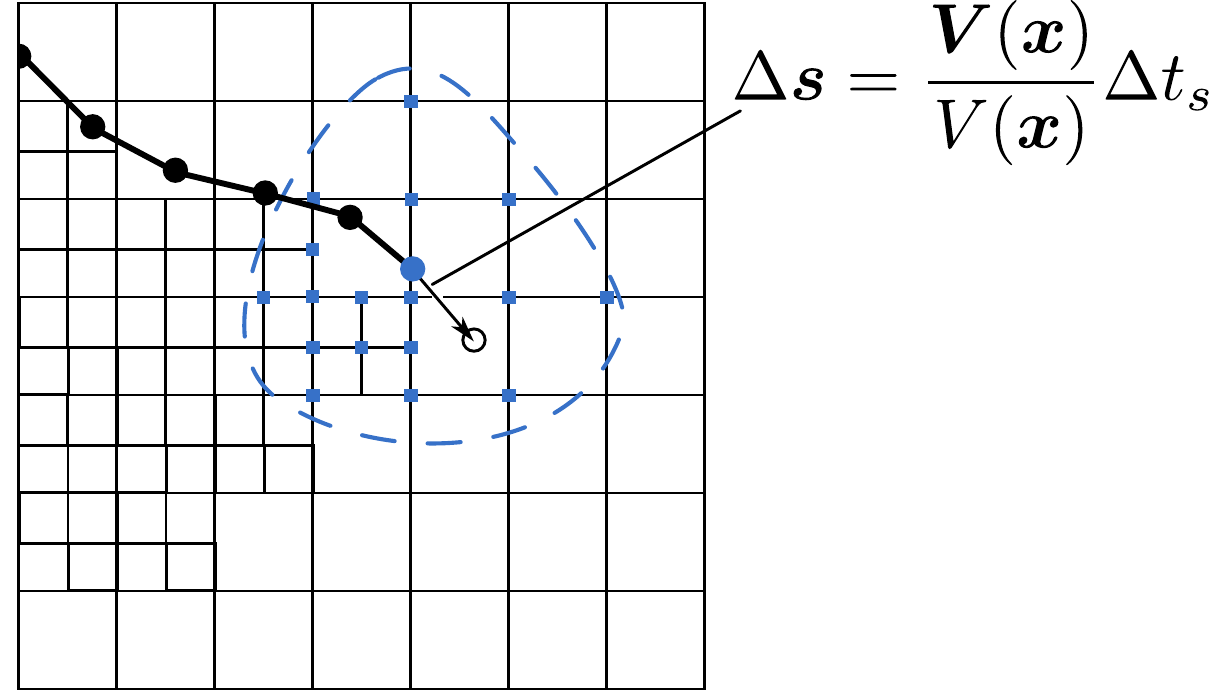}
		\caption{Schematic representation of the streamline generation procedure. The current increment of the streamline, $\Delta \bsym{s}$ (denoted with an arrow), starts with the interpolation of the velocity to the current streamline end (blue filled circle). The interpolation uses a meshless scheme (and is thus naturally capable of operating on non-regular grids) and considers a number of the mesh points closest to the current streamline tip (filled squares, encircled in a dashed loop). The next streamline point (open circle) is calculated using\add{, e.g.,} the \del{first-order Euler} \add{second-order explicit} time integration.}
		\label{fig:streamline_generation}
	\end{figure}

	The meshelss interpolation scheme used for the generation of streamlines uses cubic radial functions augmented with second-order monomials as the interpolation basis, along with 25 closest neighbors in the approximation stencil. During the preprocessing, we normalize the velocity field, $\bsym{V}/V$, i.e., we set the magnitude of the velocity vectors to unity and preserve only their direction (see Fig.~\ref{fig:streamline_generation}). \add{Unless otherwise stated, we use explicit second-order scheme for time integration.}
	
\begin{figure}[!ht]
		\raggedright
		\includegraphics[width=.45\linewidth]{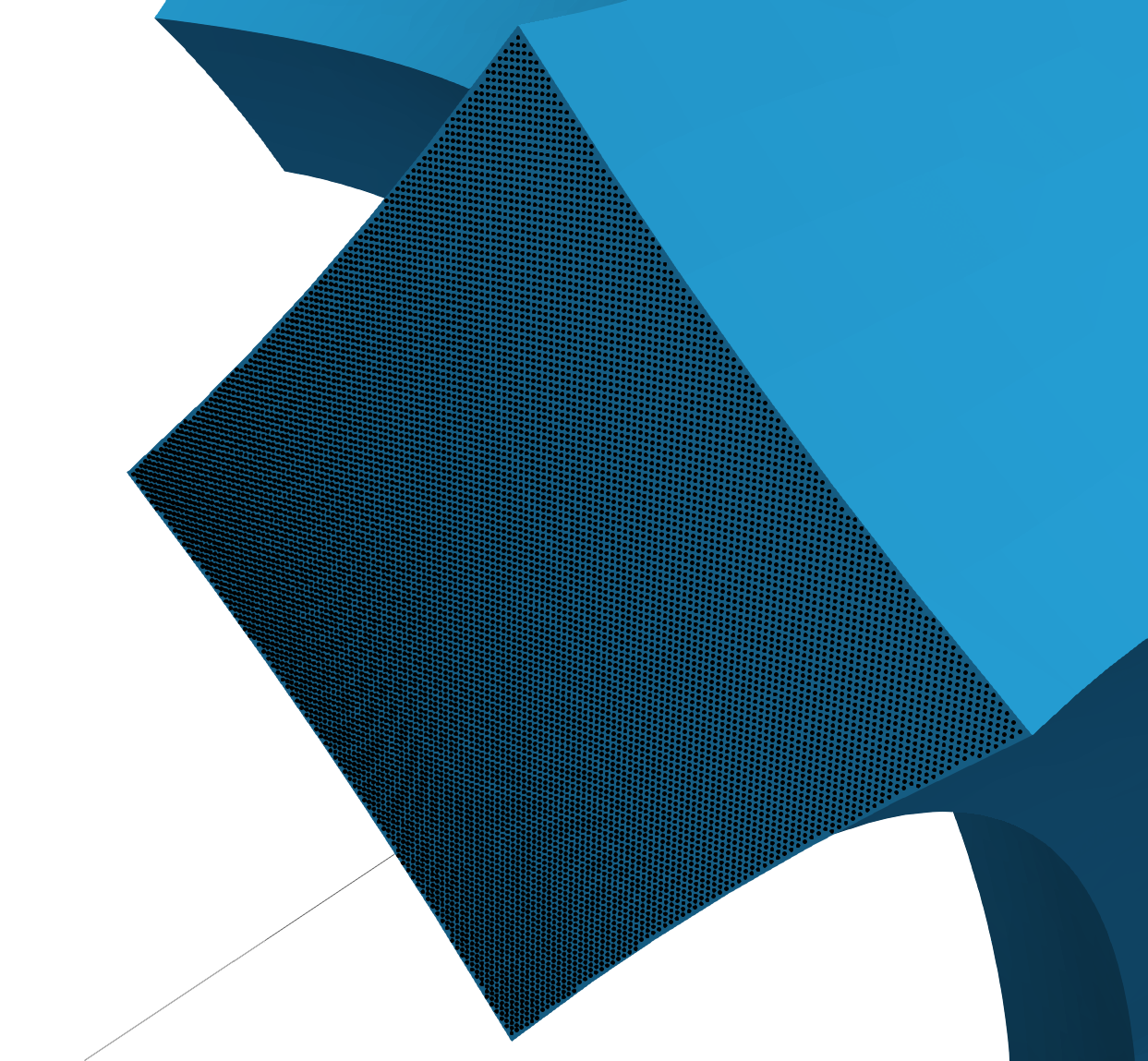}
		\includegraphics[width=.45\linewidth]{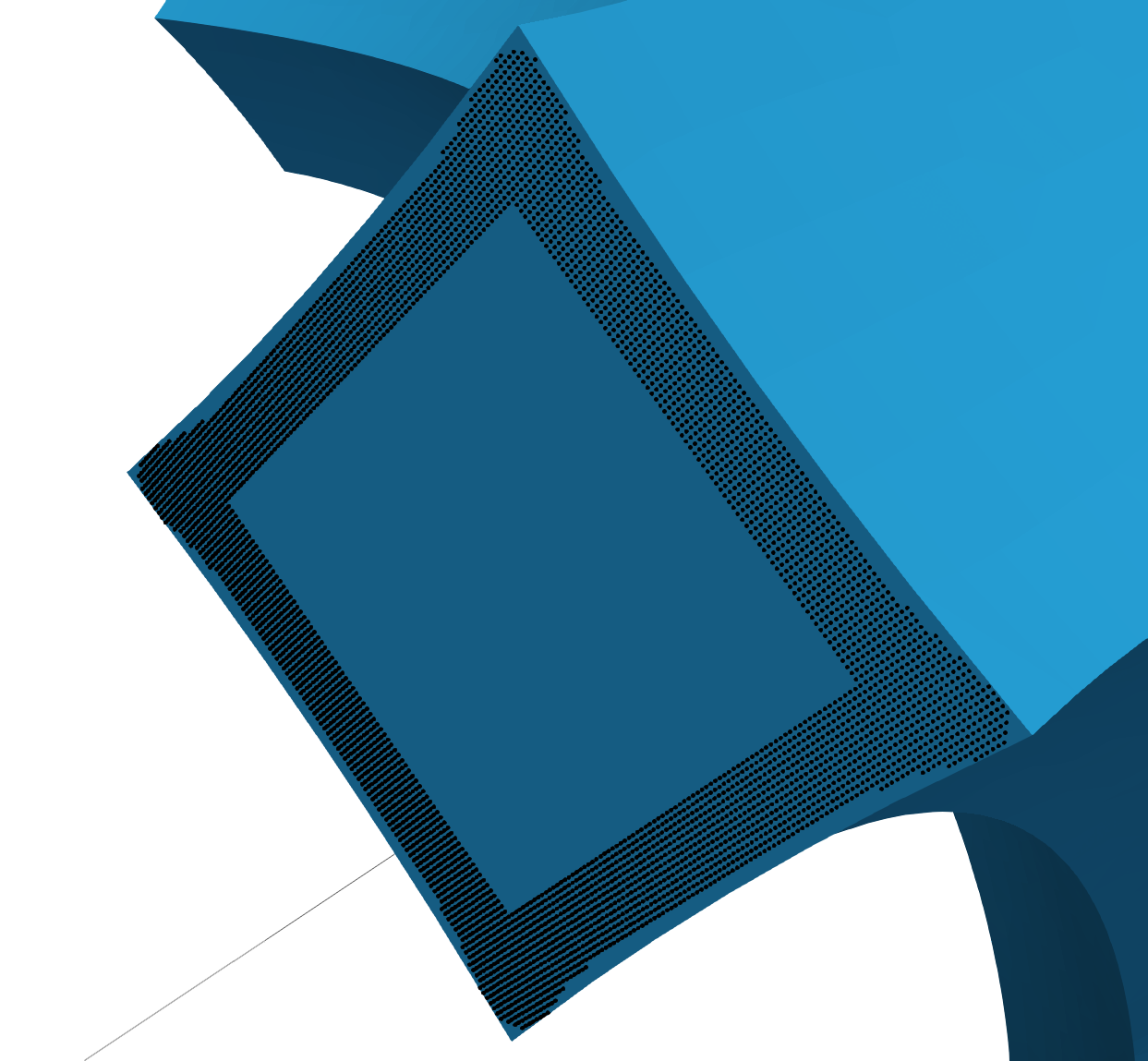}
		\includegraphics[width=.5\linewidth]{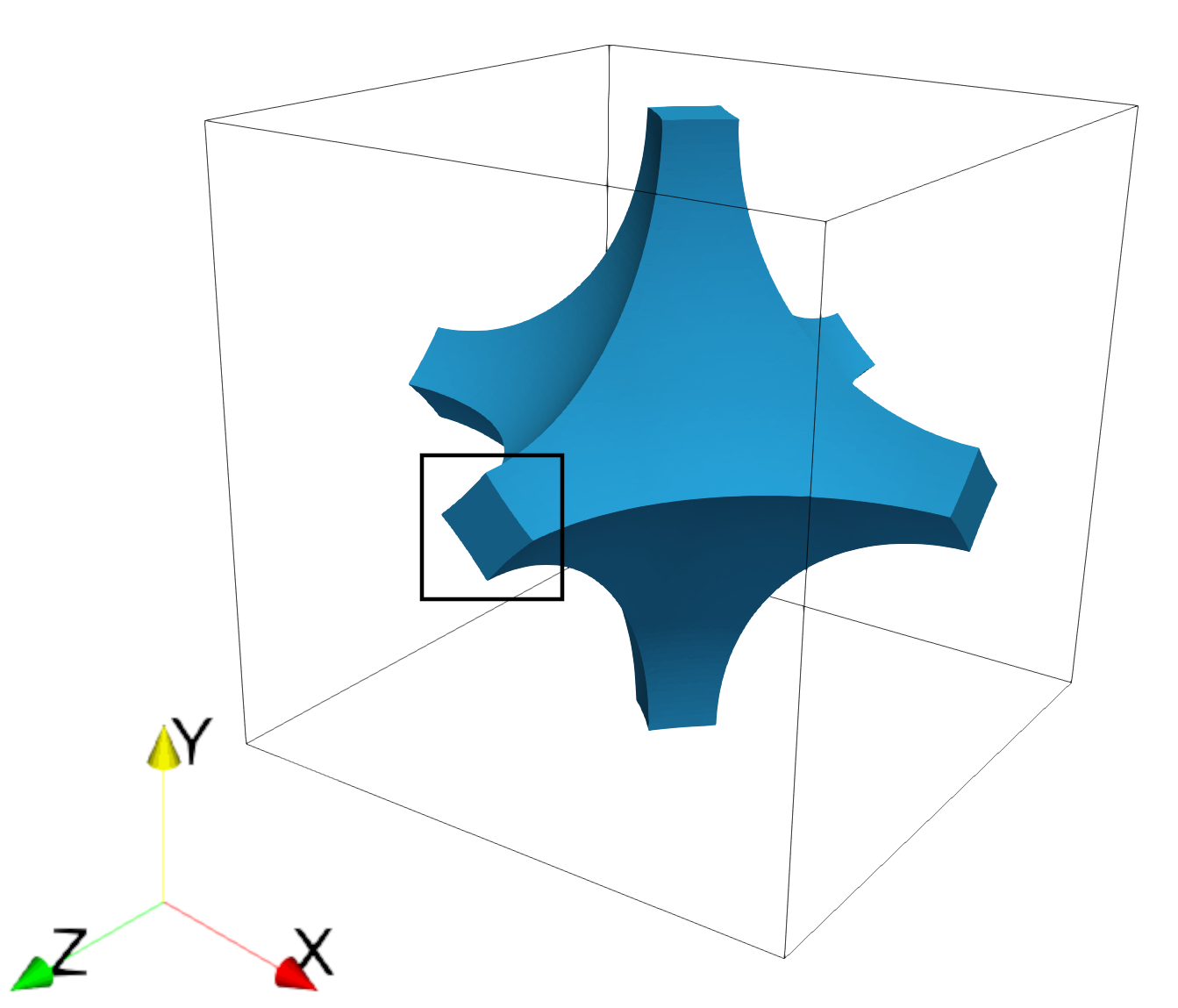}
		\caption{Initial positions of the streamlines for the calculation of $T_s$ (\textit{top left}) and the identification of the recirculation/percolating volume (\textit{top right}) for \del{the low-porosity} \add{$\phi=0.1$} SC sample. The bottom-left image shows which part of the domain is shown in the remaining ones. The fluid volume $\Omega$ is depicted in blue.}
		\label{fig:r0-6526_sl_seeds}
	\end{figure}

	The initial positions of the tracers used to generate streamlines depend on the task at hand. Due to the very different geometries and steady-state velocity fields in each of the SC samples, we tailor-cut the streamline generation for each case. For $\phi=0.1$ SC sample, for the calculation of the streamline-based tortuosity $T_s$, we place them uniformly at the fluid part of the $x_0=0$ plane, with the distance $h_s=10^{-3}$ between the tracers along $x_1$- and $x_2$-direction (see Fig.~\ref{fig:r0-6526_sl_seeds}). To find the recirculation/percolating volumes, we place the tracers only in the vicinity of the solid boundaries at $x_0=0$ plane, with the distance $h_s=10^{-3}$ between the tracers. For this task, the grid from which the positions of the tracers are taken is rotated by $\pi/4$ about $x_0$-axis for a better compliance with the solid boundaries. For both tasks, we use the timestep length $\Delta t_s=10^{-3}$ during the integration along characteristics. Considering the initial positions of the streamlines for both tasks, theoretically, none of them should hit the solid wall, have a negative increment in \del{$x$} \add{$x_0$}-direction, and all the streamlines should connect the inlet and the outlet (should be percolating). If any streamline fails to satisfy the mentioned conditions, we assume this is due to numerical errors and exclude it from further analysis.
	
\begin{figure}[!ht]
		\raggedright
		\includegraphics[height=.45\linewidth]{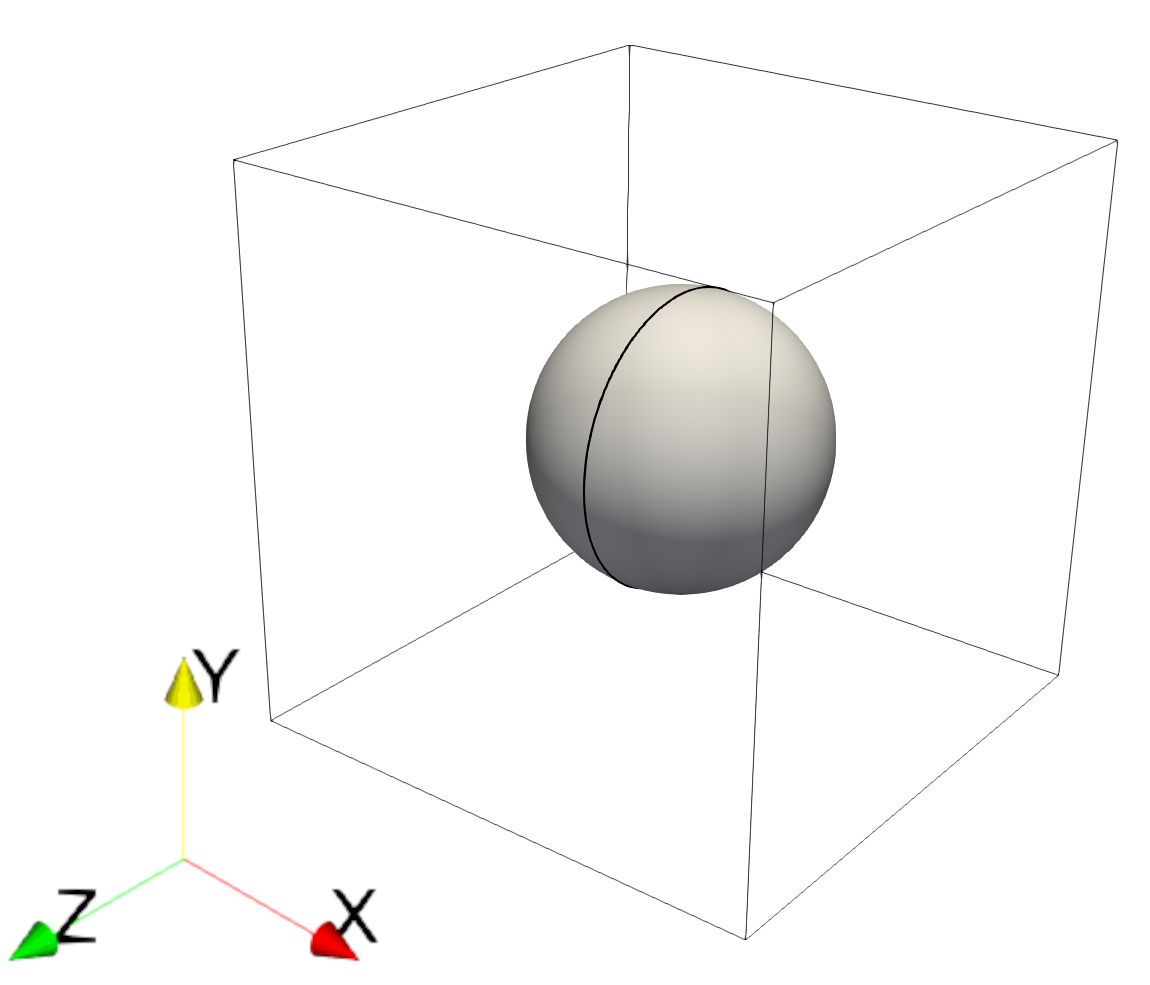}
		\includegraphics[height=.45\linewidth]{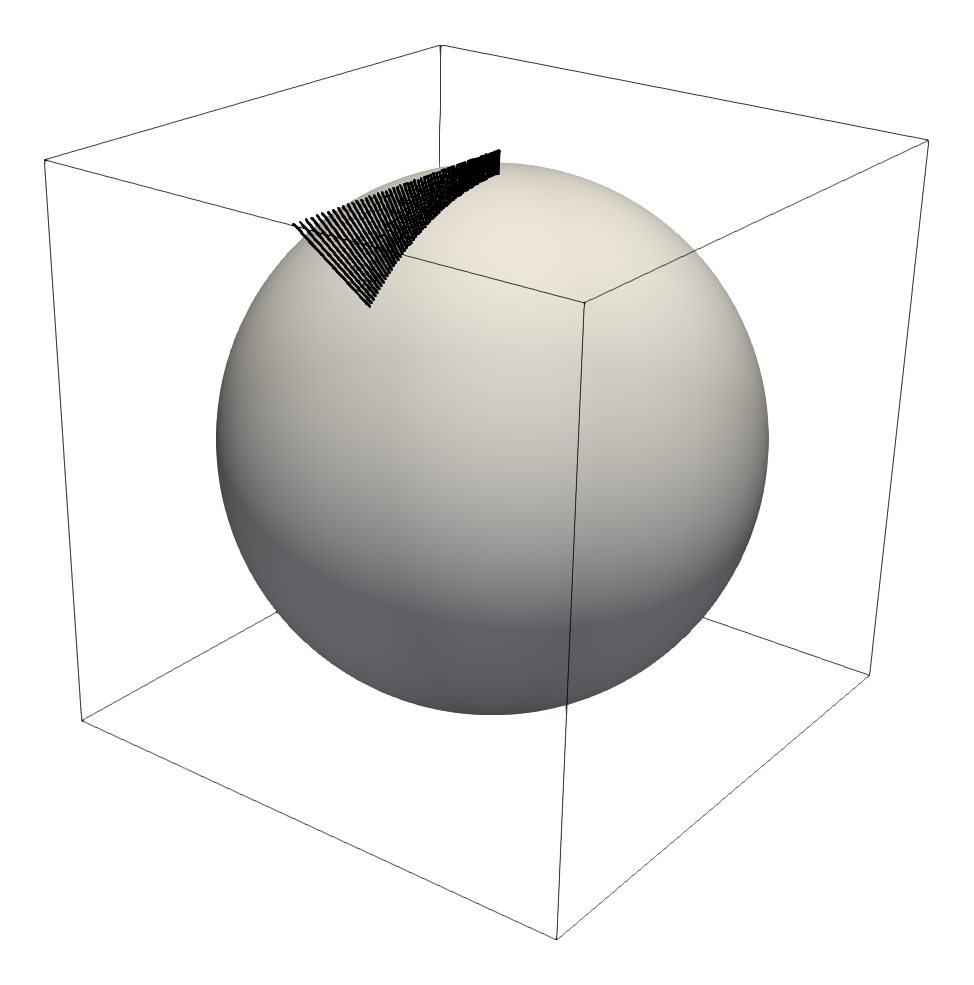}
		\caption{\add{Initial positions of the streamlines for the identification of the recirculation/percolating volume for $\phi=0.93$ (\textit{left}) and $\phi=0.59$ (\textit{right}) SC sample.}}
		\label{fig:r0-46_26_sl_seeds}
	\end{figure}

	\add{In case of $\phi=0.93$ and $\phi=0.59$ SC samples, for the calculation of streamline-based tortuosity, we place the starting points of the streamlines on the intersection between the fluid volume $\Omega$ and the plane $x_0=0.5$. The positions are taken from a square grid lying on this intersection with steps $h_s = 1.25 \cdot 10^{-3}$ and $h_s = 2.5 \cdot 10^{-3}$, respectively. The timestep lengths used for the backward and forward integration of streamlines are the same as the corresponding $h_s$ values. For the identification of the percolating/recirculation volumes, we use the same intersection of $\Omega$ and $x_0=0.5$ plane, but this time the positions of the starting points are taken from a polar grid, concentric with the spherical obstacle. The angular steps are chosen to be $2\pi R/ \tilde{h}_s$, where $R$ is the radius of the sphere and $\tilde{h}_s=2.5 \cdot 10^{-4}$ for $\phi=0.93$ and $\tilde{h}_s= 10^{-3}$ for $\phi=0.59$. In the case of $\phi=0.93$ SC sample, we place only a few layers of points directly adjacent to the sphere's surface, see Fig.~\ref{fig:r0-46_26_sl_seeds}, left. In the case of $\phi=0.59$ sample, the set of such generated streamlines is not enough to identify the percolating/recirculation volumes with enough precision, so we place the points along the whole segments of the radii of the polar grid that fell into the periodic cell. To speed up the integration of the streamlines, we limit ourselves only to the angular coordinates of the polar system within $[0;\pi/4]$ range, see Fig.~\ref{fig:r0-46_26_sl_seeds}, right. For $\phi=0.59$ sample, we use the third-order explicit time integration scheme for both streamline-related tasks. The same exclusion criteria for the streamlines are applied for $\phi=0.59$ and $\phi=0.93$ systems, as for the lowest-porosity one.} \add{Due to numerical errors in the solution of the governing equations, the mean values of $x_1$- and $x_2$-velocities in $\phi=0.59$ and $\phi=0.93$ cases are not identically zero, although they are orders of magnitude lower than the mean $x_0$-velocity. It does not impact the calculation of $T_\Omega$ significantly, however, it causes the nearly-straight streamlines to experience a drift in a direction perpendicular to $x_0$. This results in their elongation from $L$ by a certain value, the magnitude of which is comparable to the value of $T_s-1$. In turn, the streamline-based tortuosity is artificially increased to a non-negligible extent. Due to this, for those two SC samples, in Eq.~\eqref{eq:single_streamline_tortuosity}, we use the distance between the streamline's start and end as the normalization length, rather than the porous sample's side length $L$.}
	
\begin{figure}[!ht]
		\centering
		\includegraphics[width=\linewidth]{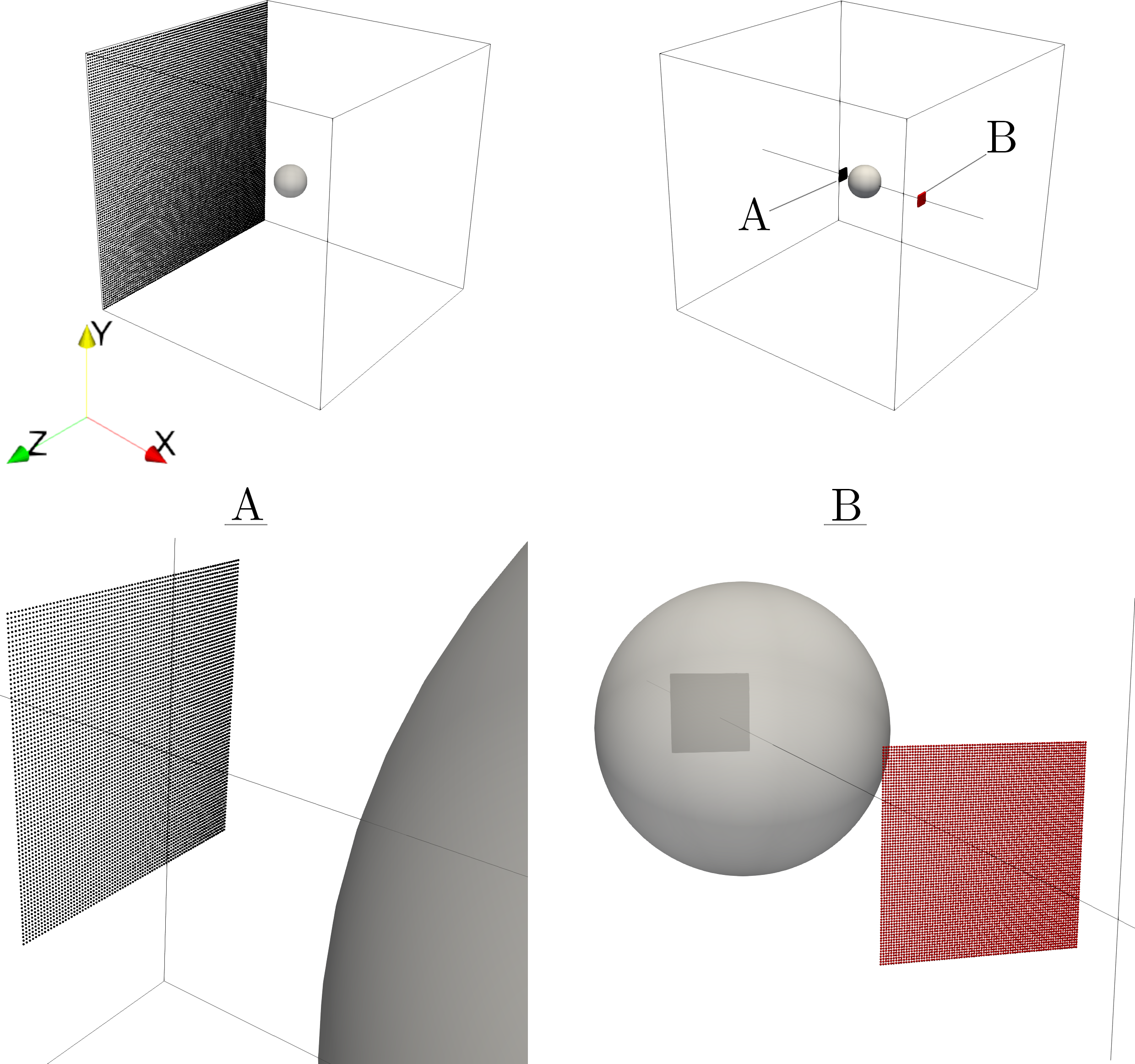}
		\caption{Initial positions of the streamlines for the calculations in \del{the high-porosity} \add{$\phi=0.999$} SC sample. \textit{Top-left}: for the calculation of $T_s$. \textit{Top-right}: for the identification of recirculation/percolating volumes. \textit{Bottom row}: magnified views at the initial positions of the forward- (\textit{A}) and backward-integrated (\textit{B}) streamlines. The $x_1=x_2=0.5$ line is shown for clarity.}
		\label{fig:r0-062_sl_seeds}
	\end{figure}

	In case of $\phi=0.999$ SC sample, for the calculation of $T_s$ we place the tracers on a square grid of step $h_s=10^{-2}$ at the $x_0=0$ plane (see Fig.~\ref{fig:r0-062_sl_seeds}, top-left) and use the timestep length $\Delta t_s=10^{-2}$. We dismiss the streamlines that do not fulfill the previously stated criteria. \del{Due to numerical errors in the solution of the governing equations, the mean values of $x_1$- and $x_2$-velocities in this case are not identically zero, although they are two orders of magnitude lower than the mean $x_0$-velocity. It does not impact the calculation of $T_\Omega$ significantly, however, it causes the nearly-straight streamlines to experience a drift in a direction perpendicular to $x_0$. This results in their elongation form $L$ by a certain value, the magnitude of which is comparable to the value of $T_s-1$. In turn, the streamline-based tortuosity is artificially increased to a non-negligible extent. Due to this, for the higher-porosity SC sample, in Eq.~\eqref{eq:single_streamline_tortuosity} we use the distance between the streamline's start and end as the normalization length, rather than the porous sample's side length $L$.} \add{For the same reasons as in the intermediate porosity SC samples, we use the distance between the streamlines endpoints as the normalization length while calculating the tortuosity $T_s$.} For the determination of the recirculation/percolating volume, we use two sets of the initial positions of the tracers (see Fig.~\ref{fig:r0-062_sl_seeds}, top-right). To properly capture the flow separation on the rear part of the sphere's surface, we place the tracers upwind from the sphere, in the vicinity of the domain's centerline $x_1=x_2=0.5$ (see Fig.~\ref{fig:r0-062_sl_seeds}, bottom-left). After the integration, we apply the same criteria to dismiss certain streamlines as previously. To properly capture the saddle point at the end of the recirculation zone in the wake of the sphere, we place the tracers downwind from the sphere (also in the vicinity of the centerline) and integrate them in the upwind direction (see Fig.~\ref{fig:r0-062_sl_seeds}, bottom-right). We do not apply the previously described streamline exclusion criteria; rather, we stop the integration before the streamlines reach the separation ring on the sphere's surface, to avoid them falling into the recirculation zone due to the numerical errors. We use the separation $h_s=5\cdot 10^{-4}$ between the streamline starting positions along $x_1$- and $x_2$-direction and the timestep length $\Delta t_s=10^{-3}$ for the identification of the recirculation/percolating volume.
	
	\subsection{Determining the recirculation and percolating volumes of fluid}\label{subapp:volumes_histogram}
	
\begin{figure}[!ht]
		\centering
		\includegraphics[width=\linewidth]{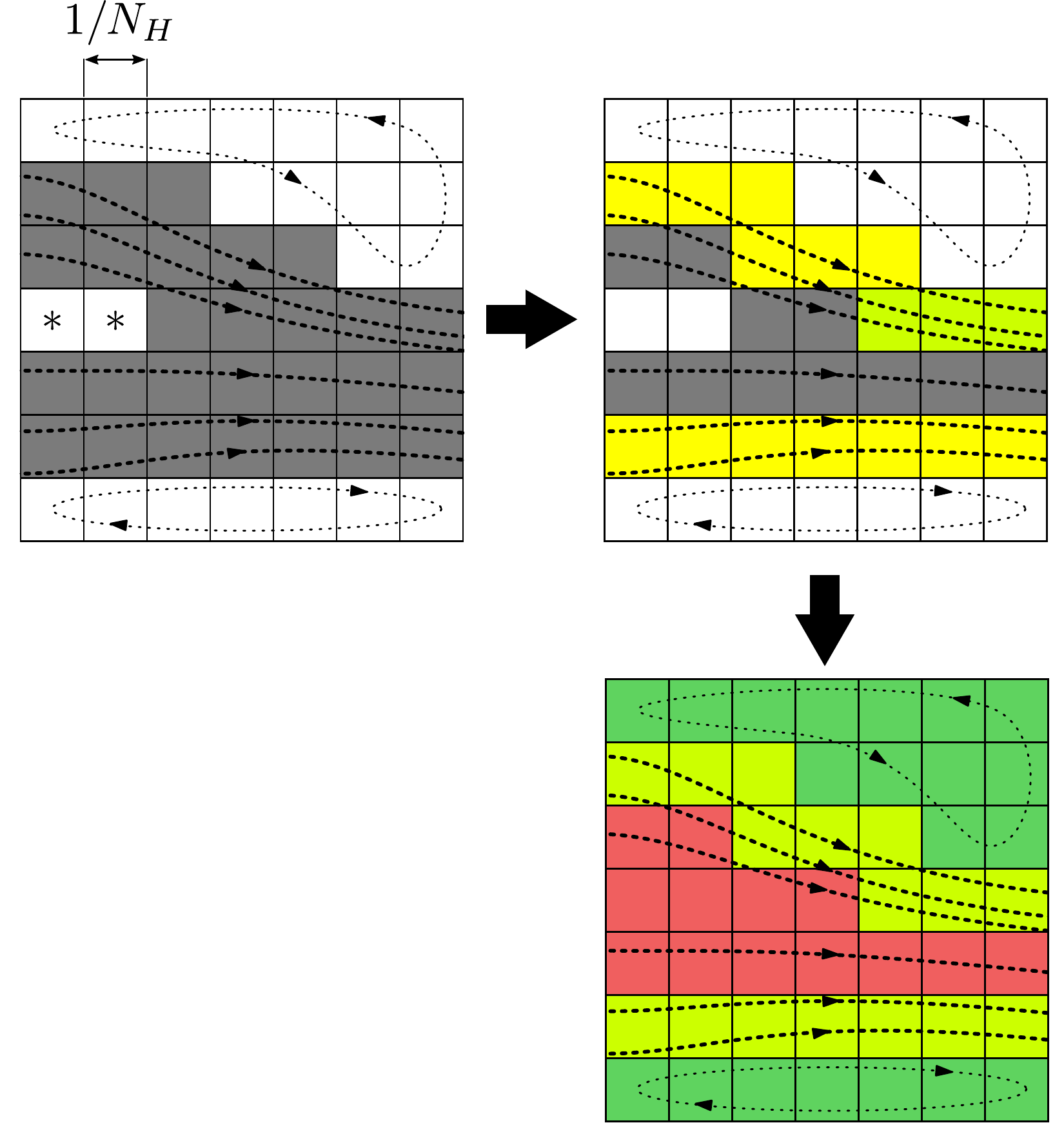}
		\caption{Schematic representation of the procedure for identifying the percolating ($\Omega_p$) and recirculation ($\Omega_v$) volumes. Each square represents a cell into which we divide the fluid volume. We note that those cells do not comply with the discretization. The percolating streamlines are plotted with thick dashed lines.}
		\label{fig:streamilne_histogram}
	\end{figure}

	To determine which points of the fluid domain $\Omega$ of the SC samples belong to the recirculation or percolating volume, the crucial part is to approximate the boundary between the two, $\partial\Omega_{pv}$. To do this, first we divide the domain into $N_H^3$ cubic cells and consider only those occupied by the fluid. We use $N_H=40$ for $\phi=0.59$ system and $N_H=160$ for the remaining SC systems. We then mark those of the cells which contain at least one point of the streamlines generated for the purpose of identification of $\Omega_p$ and $\Omega_v$ (Fig.~\ref{fig:streamilne_histogram}, top-left). As mentioned above, we generate only the percolating streamlines (although not all of them). Due to this, holes may appear in the clusters of marked cells, denoted by an asterisk. Having obtained the cluster of the marked cells, we consider those lying on the cluster's boundary (i.e. having at least one non-marked neighbor), near the expected position of $\partial \Omega_{pv}$, as the approximation of $\partial \Omega_{pv}$ (Fig.~\ref{fig:streamilne_histogram}, top-right). Lastly, all the cells lying on either side of the approximate boundary are marked as belonging to the recirculation or the percolating volume (Fig.~\ref{fig:streamilne_histogram}, bottom-right). The last step of this procedure automatically takes care of the proper coloring of the cells which do not contain any streamlines but should do so, if sufficiently many percolating streamlines were generated (the ones marked with an asterisk in Fig.~\ref{fig:streamilne_histogram}). This way, one needs to generate the streamlines only in the vicinity of $\partial\Omega_{pv}$.
	
	\bibliography{main}
	
\end{document}